\documentclass[11pt]{article}%\pdfoutput=1
\usepackage{cite}
\usepackage{scalerel}
\usepackage{shuffle}
\usepackage{amsmath,amsfonts,amssymb}
\usepackage{latexsym,epsfig}
\usepackage{dsfont}
\usepackage{hyperref}
\usepackage{extarrows}
\usepackage{comment} % To comment several lines at once.
\usepackage{tikz-cd}

\def\hybrid{
        \topmargin -20pt
        \oddsidemargin 0pt
        \headheight 0pt \headsep 0pt
        \textwidth 6.25in % A4 paper
        \textheight 9.5in % A4 paper
        \marginparwidth .875in
        \parskip 5pt plus 1pt \jot = 1.5ex}

\hybrid

 \csname
@addtoreset\endcsname{equation}{section}

\def\BV{{\rm BV}_\infty}
\def\cB{{\cal B}}

\def\cO{{\cal O}}
\def\cA{{\cal A}}
\def\cE{{\cal E}}
\def\cM{{\cal M}}
\def\cN{{\cal N}}

\def\cP{{\cal P}}

\def\cV{{\cal V}}
\def\cZ{{\cal Z}}
\def\cK{{\cal K}}

\def\cX{{\cal X}}

\def\cT{{\cal T}}
\def\cW{{\cal W}}
\def\del{\partial}
\def\l{\langle}
\def\r{\rangle}

\def\B{\square}

\def\Pperp#1{\big\{ #1 \big\}_{\perp}}

\allowdisplaybreaks

\def\bpm{\begin{pmatrix}}
\def\epm{\end{pmatrix}}

\begin{document}

\begin{titlepage}
\rightline{}
%\rightline\today
\rightline{September  2023}
\rightline{HU-EP-23/50-RTG}  
\begin{center}
\vskip 1.5cm
{\Large \bf{Weakly Constrained Double Field Theory as the\\[1ex]  
Double Copy of Yang-Mills Theory}}
\vskip 1.7cm

{\large\bf {Roberto Bonezzi, Christoph Chiaffrino, Felipe D\'iaz-Jaramillo \\ [1.5ex] 
and %\\[1ex] 
Olaf Hohm}}
\vskip 1.6cm

{\it  Institute for Physics, Humboldt University Berlin,\\
 Zum Gro\ss en Windkanal 6, D-12489 Berlin, Germany}\\[1.5ex] 
 ohohm@physik.hu-berlin.de, 
roberto.bonezzi@physik.hu-berlin.de, 
felipe.diaz-jaramillo@hu-berlin.de, chiaffrc@hu-berlin.de
\vskip .1cm

\vskip .2cm

\end{center}

\bigskip\bigskip
\begin{center} 
\textbf{Abstract}

\end{center} 
\begin{quote}

Weakly constrained double field theory, in the sense of Hull and Zwiebach, captures  the subsector of string theory on toroidal backgrounds 
that includes  gravity, $B$-field and dilaton together with all of their massive Kaluza-Klein and winding modes, which are 
encoded in 
doubled coordinates subject to the `weak constraint'. 
Due to the complications of the weak constraint, this theory was only known to cubic order. 
Here we construct the quartic interactions for the case that all 
dimensions are toroidal and doubled. Starting  from the kinematic $C_{\infty}$ algebra ${\cal K}$
of pure Yang-Mills theory and its hidden Lie-type algebra, we construct  the $L_{\infty}$ algebra 
of weakly constrained double field theory on a subspace of the `double copied' tensor product 
space ${\cal K}\otimes\bar{\cal K}$, by doing 
homotopy transfer to the weakly constrained subspace and performing a non-local shift that is well-defined 
on the torus. We test the resulting three-brackets, and establish their uniqueness up to cohomologically 
trivial terms, by verifying the Jacobi identities up to homotopy for the gauge sector.

\end{quote} 
\vfill
\setcounter{footnote}{0}
\end{titlepage}

\tableofcontents
%\newpage

\section{Introduction}

String theory is often  thought of  as something quite different from  quantum field theory. 
There are, however, formulations of string theory  as an `ordinary'  field theory, known as string field theory, 
with the only somewhat unusual feature being that it 
carries  an infinite number of  component fields (see \cite{Erbin:2021smf,Doubek:2020rbg} for  modern reviews). 
These component fields include familiar fields such as vector spin-1  
gauge fields (in open string theory) or 
the tensor spin-2 fluctuations of gravity (in closed string field theory). 
Since string theory is UV-finite, we thus have with  closed string field theory   
a quantum field theory of gravity that is at least perturbatively  well-defined. 

However, string theory and string field theory are technically infamously  involved and also include numerous exotic ingredients such as extra dimensions and 
infinite towers of massive fields of ever-increasing spin. Undoubtedly, this state of affairs is 
part of the reason that so far no compelling scenario has emerged of how to connect string theory to real-world  observations. 
At the same time, general qualitative aspects of the real-world physics encoded in the standard model of particle physics 
are naturally found in string theory. Yang-Mills theory, for instance, which governs all interactions in the realm of particle 
physics, can be obtained from open string field theory by eliminating (or rather integrating out) all massive string modes. 
Since Yang-Mills theory defines a perfectly good quantum field theory, without any need to pass to the full open string field theory, 
one may thus wonder, by analogy,  whether there are consistent theories of quantum gravity that are `smaller' than the full closed string field theory, 
perhaps consistent duality-invariant subsectors of string theory that include only  some of the massive string modes.  
(Since general relativity and the low-energy supergravity actions of string theory are certainly non-renormalizable and are unlikely to be  UV-finite 
it is clear that, in contrast to Yang-Mills theory, any putative quantum gravity theory has to include some, and most likely  infinite towers of, 
extra states in order to improve the UV-behavior.) 

In this paper, we explicitly construct a gravity  theory, known as weakly constrained double field theory \cite{Hull:2009mi},  
that  includes some infinite towers of  
massive string modes, which provide  a promising subsector for two reasons: First,  at the 
level of scattering amplitudes there are deep relations between open and closed string theory, the Kawai-Lewellen-Tye (KLT) relations \cite{Kawai:1985xq},  
which have been shown by Bern, Carrasco and Johansson (BCJ) to have field theory analogues \cite{Bern:2008qj,Bern:2019prr}.  
Such so-called   `double copy' constructions  relate pure Yang-Mills theory to `${\cal N}=0$ supergravity', i.e.~Einstein-Hilbert gravity coupled 
to a two-form (B-field) and a scalar (dilaton). In view  of the double copy this theory  is most efficiently formulated as a (strongly constrained) 
double field theory (DFT) \cite{Siegel:1993th,Hohm:2010jy,Hohm:2010pp,Hohm:2011dz}. 
Second, DFT is believed to exist  also in a weakly constrained version that features genuine massive string modes and is 
expected to exhibit an improved UV behavior. Concretely, weakly constrained DFT 
is defined on toroidal backgrounds and includes the massless fields of  ${\cal N}=0$ supergravity together with all of their massive Kaluza-Klein 
and winding modes. Such a theory can in principle be derived from the full closed string field theory by integrating out all fields that do 
not belong to the DFT sector \cite{Sen:2016qap,Arvanitakis:2020rrk,Arvanitakis:2021ecw}. 
As argued by Sen, the weakly constrained DFT so obtained would inherit the UV finiteness of string theory \cite{Sen:2016qap}. 
Therefore, `bootstrapping' such a theory directly from Yang-Mills theory via double copy, 
plausibly upon  also   including $\alpha'$ corrections \cite{Hohm:2013jaa,Hohm:2014xsa},  appears to be a promising  
path towards quantum gravity.

The construction of weakly constrained DFT to be presented here, which  was announced and outlined in \cite{Bonezzi:2023ced}, 
is based on homotopy algebras such as homotopy Lie or $L_{\infty}$ algebras. In theoretical physics, such structures were first 
discovered in string field theory \cite{Zwiebach:1992ie} and only later realized  to govern also conventional field theories such as Yang-Mills theory,  see 
in particular the early work of A.~Zeitlin \cite{Zeitlin:2007vv,Zeitlin:2008cc} (which remarkably anticipated already aspects of double copy  \cite{Zeitlin:2009tj,Zeitlin:2014xma}). 
Homotopy algebras also play a role in the formulation of 
quantum field theory due to Costello \cite{costellorenormalization} and Gwilliam-Costello \cite{Costello:2016vjw}. We refer to \cite{Hohm:2017pnh} for a self-contained introduction 
to  $L_{\infty}$ algebras and the general dictionary between field theories and $L_{\infty}$ algebras.

Using this framework one may start from Yang-Mills theory, viewed as an $L_{\infty}$ algebra, and give a perfectly precise meaning 
to the notion of `stripping off' color factors. While there is no such thing as a field theory of `color-stripped Yang-Mills fields', 
there is a homotopy algebra of such `fields', a $C_{\infty}$ rather than an $L_{\infty}$ algebra.  
Specifically, the vector space $X_{\rm YM}$,  on which the $L_{\infty}$ algebra  of Yang-Mills theory is defined, 
can be decomposed as the tensor product $X_{\rm YM}={\cal K}\otimes \frak{g}$, where $\frak{g}$ is the `color' Lie algebra 
of the gauge group, and  the $C_{\infty}$  algebra ${\cal K}$ encodes the `kinematics' of Yang-Mills theory \cite{Zeitlin:2008cc}. 
$C_{\infty}$  algebras are 
homotopy versions of \textit{commutative associative} algebras, which means that the (graded commutative) product is only associative `up to homotopy', 
with the failure of associativity being governed by a `3-product'.  
The kinematic algebra ${\cal K}$ of Yang-Mills theory is thus  an `associative-type' algebra. However,  the study of scattering amplitudes  
underlying the double copy indicates that there is also a hidden `Lie-type' algebra.  
At the level of any local  off-shell Lagrangian formulation there is no such Lie algebra in the strict sense, not even `up to homotopy', 
but a further relaxation of the Lie algebra axioms was proposed by Reiterer in \cite{Reiterer:2019dys} and proved to be realized in 
Yang-Mills theory in four (Euclidean) dimensions (with complexified fields and momenta). 
This algebra goes under the forbidding yet fitting name ${\rm BV}_\infty^\B$ and is a generalization of a Batalin-Vilkovisky (BV) algebra. 
Here  $\B$ refers to the flat space wave operator that, being of second order, 
is the origin of the obstructions that prevent ${\cal K}$ from carrying  a homotopy Lie algebra.  Nevertheless, the ${\rm BV}_\infty^\B$ algebra 
seems to be at the core of the so-called color-kinematics duality of Yang-Mills scattering amplitudes. 
More recently,  for Yang-Mills theory in arbitrary dimensions (and spacetime signature), we displayed this algebra 
up to trilinear maps \cite{Bonezzi:2022bse}.

In order to double copy Yang-Mills theory  one considers the tensor product ${\cal K}\otimes \bar{\cal K}$ 
with a second copy $\bar{\cK}$ of the kinematic algebra. This total space  consists of functions of doubled (unconstrained) coordinates (coordinates $x$ 
associated to $\cK$ and coordinates $\bar{x}$ associated to $\bar{\cK}$). This space inherits an algebra of precisely the same kind:  
a ${\rm BV}_\infty^\Delta$ algebra, where $\Delta=\frac{1}{2}(\B-\bar{\B})$ is the difference of the respective wave operators. 
This unconstrained space does not carry an unobstructed $L_{\infty}$ algebra, and hence no consistent field theory,  
due to $\Delta$ being second order, but by restricting to a suitable subspace one can eliminate the obstructions. Identifying 
coordinates $x$ with coordinates $\bar{x}$ implies   $\Delta\equiv 0$, which yields the  \textit{strongly constrained}  DFT that can be viewed as a 
duality invariant formulation  of $\cN=0$ supergravity. More precisely, so far this  was established  to quartic order in fields \cite{Bonezzi:2022bse}.  
(See \cite{Bonezzi:2023ciu} for a quick  and polemic introduction, \cite{Diaz-Jaramillo:2021wtl, Lee:2018gxc,Cho:2019ype, Berman:2020xvs,Lescano:2022nhp,Lescano:2023pai} for DFT and double copy, and \cite{Bonezzi:2023pox,Borsten:2021hua,Borsten:2022vtg, Borsten:2023reb,Borsten:2023ned, Borsten:2023paw} for homotopy algebras and double copy.)

In this paper we will show how to construct \textit{weakly constrained} DFT for toroidal and hence Euclidean backgrounds  in which $\Delta=0$ is imposed as a constraint on fields, 
which then still genuinely depend on doubled coordinates and hence encode both physical winding and momentum modes. 
(Since all dimensions are toroidal and doubled this theory does not yet include a non-compact time direction, which at least in conventional 
thinking should remain undoubled. We leave the construction of the full Lorentzian theory for future work.)
To this end, one performs homotopy transfer to the subspace 
with $\Delta=0$ (see, e.g., \cite{Arvanitakis:2020rrk} for a self-contained introduction to homotopy transfer).  
This still does not give an unobstructed $L_{\infty}$ algebra, but by further imposing an algebraic constraint 
known from the level-matching constraints of string theory one can redefine  the desired  3-bracket by a non-local but perfectly 
well-defined shift so that one obtains a genuine 
$L_{\infty}$ algebra to the order relevant for the quartic theory. This solves a  problem that was outstanding since the modern inception of 
DFT by Hull and Zwiebach in 2009 \cite{Hull:2009mi}. It should be emphasized that this solution of the problem is unique, 
up to cohomologically trivial redefinitions,  given the `initial data' of the 
differential $B_1$ and 2-bracket $B_2$ of weakly constrained DFT encoded in the cubic theory  of \cite{Hull:2009mi}.

What is perhaps the most striking aspect of this solution of the long-open problem of constructing weakly constrained DFT 
is that it relates to, and in some ways  is almost identical with, deep hidden structures \textit{that are present in Yang-Mills theory proper}, 
without any reference to gravity. While the conventional Lagrangian formulation of Yang-Mills theory 
relies only on the color Lie algebra $\frak{g}$ and the kinematic $C_{\infty}$ algebra, the computation of scattering amplitudes 
requires more structures, as for instance exhibited  in gauge conditions. Given these extra structures, the kinematic 
vector space comes close to be a (homotopy) Lie algebra, but this is obstructed by the wave operator $\B$ being of second order. 
When computing scattering amplitudes one goes on-shell, so that $\B$ gives zero when acting on single fields (polarization vectors), 
but even then the algebraic structure is obstructed since the product of two on-shell fields is generally not on-shell. 
Thus, even on the subspace with $\B=0$  the  ${\rm BV}_\infty^\B$ algebra does not yield 
an unobstructed homotopy Lie algebra. In  the amplitude literature it has been shown how to shift the  kinematic numerators so that 
these obey Jacobi-type identities, a property known as color-kinematic duality. 
The problem of constructing weakly constrained DFT is therefore technically 
analogous to the problem of making color-kinematics manifest in Yang-Mills theory proper, 
just with ${\rm BV}_\infty^\Delta$ instead of ${\rm BV}_\infty^\B$. We hope to further explore this intriguing connection in the future.

The remainder of this paper is organized as follows. 
In sec.~2 we introduce  the $C_{\infty}$ algebra of the kinematic space  ${\cal K}$ of Yang-Mills theory, 
to the order of tri-linear maps, and we introduce the  ${\rm BV}_\infty^\B$ algebra. While to a large part this is  a review 
of results presented  in \cite{Bonezzi:2022bse}, we also introduce a more streamlined notation for objects of ${\cal K}$ and its 
multilinear maps, which 
is instrumental in order to efficiently compute the double copied maps in later sections. These results are useful additions to 
\cite{Bonezzi:2022bse}, even just for strongly constrained DFT. In sec.~3 we prove, again to the order of tri-linear maps relevant for 
the quartic theory, that the (unconstrained) doubled space ${\cal K}\otimes \bar{\cK}$ carries a ${\rm BV}_\infty^\Delta$ algebra. 
Finally, in sec.~4 we construct weakly constrained DFT to quartic order, by first doing  homotopy transfer to the subspace 
with $\Delta=0$ and then performing a non-local but well-defined shift. We verify the inevitability of this non-local shift 
by computing   the  3-brackets of the gauge sector and by verifying  the generalized Jacobi identities. 
We close with a summary and outlook in sec.~5. where we discuss possible applications and generalizations. 
In two appendices we collect all maps for Yang-Mills theory and we give some of the  technically challenging proofs.

\section{The kinematic algebra of Yang-Mills}

Here we start by reviewing the $\BV^\B$ algebra of Yang-Mills theory, up to its trilinear maps. In doing so we will introduce the necessary formalism and fix our conventions and notation. We follow closely the discussion in \cite{Bonezzi:2022yuh,Bonezzi:2022bse}, although with some differences in the notation that, we believe, lead to a more streamlined treatment. 

We employ a formulation of Yang-Mills theory with an auxiliary scalar field $\varphi$, whose  action is given by   \cite{Bonezzi:2022yuh}
\begin{equation}\label{YMaction}
S=\int d^dx\,\Big[\tfrac12\,A^\mu_a\B A_\mu^a-\tfrac12\,\varphi_a\varphi^a+\varphi_a\,\del^\mu A_\mu^a-f_{abc}\,\del_\mu A_\nu^a A^{\mu b}A^{\nu c}-\tfrac14\,f^e{}_{ab}f_{ecd}\,A_\mu^a A_\nu^b A^{\mu c}A^{\nu d}\Big]\;,    
\end{equation}
where $f_{abc}$ are the structure constants of the color Lie algebra $\mathfrak{g}$.
The cubic and quartic vertices are standard, and one recovers the usual action upon integrating out $\varphi$.
In Yang-Mills theory all objects including  gauge parameters, fields, field equations, etc., take values in the Lie algebra $\mathfrak{g}$ of 
the gauge group. It is thus natural to view the space of Yang-Mills theory as the tensor product $\cK\otimes\mathfrak{g}$, where elements of $\cK$ are color-stripped local spacetime fields, which we still refer to as gauge parameters, fields and so on.

\subsection{The graded vector space $\cK$}

Let us describe in more detail the structure of the kinematic vector space $\cK$. It is a graded vector space given by the direct sum of subspaces $\cK_i$ of homogenous degree: $\cK=\bigoplus_{i=0}^3\cK_i$. Elements in each $\cK_i$ are identified as gauge parameters $\lambda$, fields $\cA$, field equations $\cE$ and Noether identities $\cN$ according to the following diagram:
\begin{equation}\label{K}
\begin{tikzcd}[row sep=2mm]
\cK_{0} & \cK_{1}& \cK_2 & \cK_3\\
\lambda& \cA & \cE &\cN
\end{tikzcd}\;.
\end{equation}
We take the field $\cA$ to contain both the (color-stripped) gauge vector field $A_\mu$ and the scalar $\varphi$, and similarly the equations of motion $\cE$ have a vector and a scalar component.

In order to display explicitly its degree structure, we find it useful to view $\cK$ as the tensor product of a finite-dimensional graded vector space $\cZ=\bigoplus_{i=0}^3\cZ_i$ with the space of smooth   spacetime  functions of degree zero: $\cK=\cZ\otimes C^\infty(\cM)$. Here $\cM$ is flat $d-$dimensional Minkowski spacetime, but the signature is immaterial for the following discussion. 
The vector space  $\cZ$ is defined by giving a basis. To this end 
let us introduce a $(d+2)$--component graded vector $\theta_M=(\theta_+,\theta_\mu,\theta_-)$, where $\mu=0,1,\ldots, d-1$ is a Lorentz vector index. 
The degrees of the components are given by 
\begin{equation}
|\theta_+|=0\;,\quad |\theta_\mu|=1\;,\quad |\theta_-|=2\;, 
\end{equation}
which we sometimes summarize by writing $|\theta_M|=1-M$, where $M$ means $+1$, $0$ or $-1$, depending on the index. 
Next, we take a second copy of these vectors with degrees shifted by one, which we denote by $c\,\theta_M$, with $|c\,\theta_M|=2-M$, or
\begin{equation}
|c\,\theta_+|=1\;,\quad |c\,\theta_\mu|=2\;,\quad |c\,\theta_-|=3\;.    
\end{equation}
A basis of $\cZ$ is then given by  
 \begin{equation}
   Z_A=(\theta_M, c\,\theta_M)\;. 
 \end{equation}

 The above characterization of $\cZ$  exhibits the manifest  $\mathbb{Z}_2$ symmetry that exchanges $\theta_M$ and $c\, \theta_M$.  
 This isomorphism between the subspaces generated by $\theta_M$ and $c\,\theta_M$, respectively, 
 can be implemented by nilpotent operators $b$ and $c$ defined by their action on the basis $Z_A$:
\begin{equation}\label{bc}
\begin{split}
c\,(\theta_M)&:=c\,\theta_M\;,\quad c\,(c\,\theta_M):=0\;,\\    
b\,(\theta_M)&:=0\;,\quad\hspace{5mm} b\,(c\,\theta_M):=\theta_M\;.    
\end{split}
\end{equation}
The degrees of $b$ and $c$ are thus fixed to be $|c|=+1$ and $|b|=-1$, and from their definition one can see that they obey the algebra
\begin{equation}
c^2=0\;,\quad b^2=0\;,\quad bc+cb=1\;.  \end{equation}
The basis elements of $\cZ$ can be displayed according to their degree in a way that emphasizes the $\mathbb{Z}_2$ symmetry:
\begin{equation}\label{Zdiagram}
\begin{tikzcd}[row sep=2mm]
\cZ_{0} & \cZ_{1} & \cZ_2 & \cZ_3\\
\theta_+& \theta_\mu & \theta_-\\[2mm]
 &\arrow{ul}{b}c\,\theta_+&\arrow{ul}{b}c\,\theta_\mu &\arrow{ul}{b}c\,\theta_-
\end{tikzcd}\;,
\end{equation}
where we have indicated the action of $b$ ($c$ acts by reversing the arrows).

In addition to the above $\mathbb{Z}_2$ symmetry, 
$\cZ$ can be equipped with an odd symplectic bilinear form\footnote{This is closely related to the field theoretic odd symplectic structure of the Batalin-Vilkovisky formalism, see \emph{e.g.} \cite{Zwiebach:1992ie,Barnich:2003wj,Barnich:2004cr,Grigoriev:2023lcc}.} $\omega$ of degree $|\omega|=-3$, satisfying 
 \begin{equation}
  \omega(Z_1, Z_2) = (-1)^{Z_1Z_2} \omega(Z_2,Z_1)\;, 
 \end{equation} 
which is symmetric since it always pairs odd with even elements.  
We specify $\omega$ by giving its components   $\omega(Z_A, Z_B)$ in the above basis:
\begin{equation}\label{sympomega}
\begin{split}
\omega(\theta_+,c\,\theta_-)&=\omega(c\,\theta_-,\theta_+)=-1\;,\\
\omega(\theta_-,c\,\theta_+)&=\omega(c\,\theta_+,\theta_-)=+1\;,\\
\omega(\theta_\mu,c\,\theta_\nu)&=\omega(c\,\theta_\nu,\theta_\mu)=\eta_{\mu\nu}\;,
\end{split}    
\end{equation}
where $\eta_{\mu\nu}$ is the $d-$dimensional Minkowski metric, and all other pairings vanish.

Upon tensoring $\cZ$ with smooth functions, we obtain the kinematic space $\cK$ of Yang-Mills theory. The degree in $\cK$ coincides with the one in $\cZ$, meaning that for an homogeneous element $\psi=Z\,f(x)$ one has $|\psi|=|Z|$. An arbitrary element in $\cK$ can thus be expanded as
\begin{equation}\label{psi expansion}
\psi=Z_A\,\psi^A(x)\;.  \end{equation}
Comparing the degree structure \eqref{K} of $\cK$ with \eqref{Zdiagram}, one infers  that the Yang-Mills fields, parameters and so on are given by the following vectors in $\cK$ with homogeneous degrees:
\begin{equation}
\begin{split}
\lambda&=\theta_+\,\lambda(x)\,\in\cK_0\;,\\
\cA&=\theta_\mu\,A^\mu(x)+c\,\theta_+\,\varphi(x)\,\in\cK_1\;,\\
\cE&=\theta_-\,E(x)+c\,\theta_\mu\,E^\mu(x)\,\in\cK_2\;,\\
\cN&=c\,\theta_-\,\cN(x)\,\in\cK_3\;.
\end{split}    
\end{equation}
The $\mathbb{Z}_2$ structure and the action of $b$ and $c$ are inherited from $\cZ$. One can indeed draw the same diagram \eqref{Zdiagram} in $\cK$ to display this:
\begin{equation}\label{Kdiagram}
\begin{tikzcd}[row sep=2mm]
\cK_{0} & \cK_{1} & \cK_2 & \cK_3\\
\lambda& A^\mu & E\\[2mm]
 &\arrow{ul}{b}\varphi&\arrow{ul}{b}E^\mu&\arrow{ul}{b}\cN
\end{tikzcd}\;,
\end{equation}
where we omitted the $Z_A$ and only wrote the component fields. The odd symplectic pairing $\omega$ induces a degree $-3$ inner product $\l\,,\,\r$ in $\cK$, defined by
\begin{equation}
\l\psi_1,\psi_2\r=\int d^dx\,\psi_1^A(x)\,\psi_2^B(x)\,\omega(Z_A,Z_B)\;.    
\end{equation}
More specifically, using \eqref{sympomega} one can see that the non-vanishing pairings are between fields $\cA$ and field equations $\cE$:
\begin{equation}
 \l\cA,\cE\r=\int d^dx\,\Big[A_\mu(x)\,E^\mu(x)+\varphi(x)\,E(x)\Big]\;,  \end{equation}
and between gauge parameters $\lambda$ and Noether identities $\cN$:
\begin{equation}
\l\lambda,\cN\r=-\int d^dx\,\lambda(x)\,\cN(x)\;.
\end{equation}

\subsection{$C_\infty$ algebra on $\cK$}

Having described $\cK$ as a graded vector space, we now turn to reviewing the algebraic structures that can be defined on it. The consistency of Yang-Mills theory as a field theory (this includes, for instance, gauge covariance of the field equations and closure of the gauge algebra) is encoded, upon factoring out color, by a $C_\infty$ algebra structure on $\cK$ \cite{Zeitlin:2008cc,Borsten:2021hua,Bonezzi:2022yuh}. This is a homotopy generalization of a commutative associative algebra where a graded vector space ($\cK$ in the case at hand), is equipped with multilinear maps or products $m_n$ obeying a set of quadratic relations.
For the case of Yang-Mills theory, the only non-vanishing maps  are an operator $m_1$ of degree $+1$, a bilinear product $m_2$ of degree zero and a trilinear product $m_3$ of degree $-1$, summarized as $|m_n|=2-n$.

The nontrivial $C_\infty$ relations to be satisfied then consist of 
\begin{itemize}
\item Nilpotency of $m_1$:
\begin{equation}
m_1^2(\psi)=0\;,    
\end{equation}
stating that $m_1$ is a differential, which makes $\cK$ into a chain complex. Physically, $m_1^2=0$ encodes, in particular, 
gauge invariance of the free theory under linearized gauge transformations.
\item The differential $m_1$ acts as a derivation on $m_2$ (Leibniz rule):
\begin{equation}
m_1\,m_2(\psi_1,\psi_2)=m_2(m_1\psi_1,\psi_2)+(-1)^{\psi_1}m_2(\psi_1,m_1\psi_2)\;,
\end{equation}
where $\psi_i$ in exponents always denotes  the degree $|\psi_i|$. This requirement ensures, upon tensoring with color, consistency of Yang-Mills theory up to cubic order.
\item The product $m_2$ is associative up to homotopy:
\begin{equation}
\begin{split}
&m_2\big(m_2(\psi_1,\psi_2),\psi_3\big)-m_2\big(\psi_1,m_2(\psi_2,\psi_3)\big)=m_1\,m_3(\psi_1,\psi_2,\psi_3)+m_3(m_1\psi_1,\psi_2,\psi_3)\\&+(-1)^{\psi_1}m_3(\psi_1,m_1\psi_2,\psi_3)+(-1)^{\psi_1+\psi_2}m_3(\psi_1,\psi_2,m_1\psi_3)\;,   
\end{split}    
\end{equation}
which is responsible for consistency of the theory up to quartic order and thus fully, given that Yang-Mills theory has no higher vertices.
\end{itemize}

In a $C_\infty$ algebra, the products $m_n$ have to further obey symmetry constraints under permutations of arguments. Specifically, the $m_n$ have to vanish under so-called shuffle permutations. For our purposes, the relevant symmetry properties are
\begin{equation}
\begin{split}
m_2(\psi_1,\psi_2)-(-1)^{\psi_1\psi_2}m_2(\psi_2,\psi_1)&=0\;,\\
m_3(\psi_1,\psi_2,\psi_3)-(-1)^{\psi_1\psi_2}m_3(\psi_2,\psi_1,\psi_3)+(-1)^{\psi_1(\psi_2+\psi_3)}m_3(\psi_2,\psi_3,\psi_1)&=0\;,   
\end{split}    
\end{equation}
which, for $m_2$, is the same as graded symmetry. For the following discussion we find it more convenient to work with a different representation of $m_3$, which we denoted $m_{3\rm h}$ in \cite{Bonezzi:2022bse}, defined as
\begin{equation}
\begin{split}
m_{3\rm h}(\psi_1,\psi_2,\psi_3)&:=\tfrac13\,\Big(m_3(\psi_1,\psi_2,\psi_3)+(-1)^{\psi_1\psi_2}m_3(\psi_2,\psi_1,\psi_3)\Big)\;,\\
m_{3}(\psi_1,\psi_2,\psi_3)&=m_{3\rm h}(\psi_1,\psi_2,\psi_3)-(-1)^{\psi_1(\psi_2+\psi_3)}m_{3\rm h}(\psi_2,\psi_3,\psi_1)\;.
\end{split}    
\end{equation}
This is just a redefinition of $m_3$, not a projection, as it can be inverted explicitly by using the above formula. The redefined $m_{3\rm h}$ is graded symmetric in its first two arguments and vanishes upon total graded symmetrization. Otherwise stated, it is a graded hook representation in terms of Young diagrams.

Given the tensor product structure of $\cK=\cZ\otimes C^\infty(\cM)$ and the expansion \eqref{psi expansion} of arbitrary vectors, the $m_n$ products act on elements of $\cK$ as follows:
\begin{equation}\label{mhats}
\begin{split}
m_1(\psi)&=\hat m_1(Z_A)\,\psi^A(x)\;,\\
m_2(\psi_1,\psi_2)&=\mu\left[\hat m_2(Z_A,Z_B)\,\Big(\psi_1^A(x)\otimes\psi_2^B(x)\Big)\right]\;,\\
m_{3\rm h}(\psi_1,\psi_2,\psi_3)&=\mu\left[\hat m_{3\rm h}(Z_A,Z_B,Z_C)\,\Big(\psi_1^A(x)\otimes\psi^B_2(x)\otimes\psi^C_3(x)\Big)\right]\;, 
\end{split}    
\end{equation}
where the operations on the right are defined as follows: 
First, the operators $\hat m_n(Z_{A_1},\ldots, Z_{A_n})$ are $\cZ-$valued multidifferential operators acting 
on the component fields as 
\begin{equation}
\begin{split}
\hat m_1(Z)&:C^\infty(\cM)\;\rightarrow\;\cZ\otimes C^\infty(\cM)\;,\\ 
\hat m_2(Z_1,Z_2)&:C^\infty(\cM)\otimes C^\infty(\cM)\;\rightarrow\;\cZ\otimes\Big(C^\infty(\cM)\otimes C^\infty(\cM)\Big)\;,\\
\hat m_{3\rm h}(Z_1,Z_2,Z_3)&:C^\infty(\cM)\otimes C^\infty(\cM)\otimes C^\infty(\cM)\;\rightarrow\;\cZ\otimes\Big(C^\infty(\cM)\otimes C^\infty(\cM)\otimes C^\infty(\cM)\Big)\;. 
\end{split}
\end{equation}
Second, $\mu$ just denotes the local pointwise product:
\begin{equation}
\mu\left[f_1(x)\otimes\ldots\otimes f_n(x)\right]=f_1(x)\cdots f_n(x)\;.    
\end{equation}
To clarify this notation, let us give some explicit examples (the explicit form of all $m_n$ products can be found in \cite{Bonezzi:2022yuh}). Acting on the basis vectors of $\cZ_1$ (corresponding to fields), we have
\begin{equation}
\begin{split}
\hat m_1(\theta_\mu)\ &= \ c\,\theta_\mu\,\B \ + \ \theta_-\,\del_\mu\;,\\
\hat m_1(c\,\theta_+)\ &= \ -c\,\theta_\mu \del^\mu \ - \ \theta_-\;.
\end{split}    
\end{equation}
Using \eqref{mhats} one computes the action of  the differential $m_1$ on a field 
$\cA=\cZ_A\psi^A = \theta_{\mu}A^{\mu} + c\,\theta_+\varphi\in \cK_1$ as
\begin{equation}
\begin{split}
m_1(\cA)&=\hat m_1(Z_A)\,\psi^A  \\
&= \hat m_1(\theta_\mu)\,A^\mu+\hat m_1(c\,\theta_+)\,\varphi\\
&=c\,\theta_\mu\,\Big(\B A^\mu-\del^\mu\varphi\Big)+\theta_-\,\Big(\del\cdot A-\varphi\Big)\;,  \end{split}
\end{equation}
where we omitted the explicit spacetime dependence of the component fields, and denoted the contraction of Lorentz indices with a dot. One can see that setting $m_1(\cA)=0$ corresponds to the free Maxwell equations upon solving for $\varphi$.

For the next example, the non-vanishing part of $m_2$ between fields $\cA_1$ and $\cA_2$ is encoded in the bi-differential operator
\begin{equation}
\begin{split}
\hat  m_2(\theta_\mu,\theta_\nu)&=c\,\theta_\nu\Big[\big(\del_\mu\otimes \mathds{1}\big)+2\,\big(\mathds{1}\otimes\del_\mu\big)\Big]-c\,\theta_\mu\Big[\big(\mathds{1}\otimes\del_\nu\big)+2\,\big(\del_\nu\otimes \mathds{1}\big)\Big]\\
&+c\,\theta_\rho\,\eta_{\mu\nu}\,\Big[\big(\del^\rho\otimes \mathds{1}\big)-\big(\mathds{1}\otimes\del^\rho\big)\Big]\;.    
\end{split}    
\end{equation}
This acts on $\big(A_1^\mu\otimes A_2^\nu\big)$ as
\begin{equation}
\begin{split}
\hat  m_2(\theta_\mu,\theta_\nu)\,\big(A_1^\mu\otimes A_2^\nu\big)&=c\,\theta_\nu\Big[\big(\del\cdot A_1\otimes A_2^\nu\big)+2\,\big(A_1^\mu\otimes\del_\mu A_2^\nu\big)\Big]\\
&-c\,\theta_\mu\Big[\big(A_1^\mu\otimes\del\cdot A_2\big)+2\,\big(\del_\nu A_1^\mu\otimes A_2^\nu\big)\Big]\\
&+c\,\theta_\rho\,\Big[\big(\del^\rho A_1^\mu\otimes A_{2\mu}\big)-\big(A_1^\mu\otimes \del^\rho A_{2\mu}\big)\Big]\;. 
\end{split}    
\end{equation}
The pointwise multiplication implemented by  $\mu$ then  yields (with a dot denoting contraction of Lorentz indices)
\begin{equation}\label{cubicvert}
\begin{split}
m_2(\cA_1,\cA_2)&=\mu\left[\hat  m_2(\theta_\mu,\theta_\nu)\,\big(A_1^\mu\otimes A_2^\nu\big)\right]\\
&=c\,\theta_\mu\,\Big(\del\cdot A_1\,A_2^\mu+2\,A_1\cdot\del A_2^\mu+\del^\mu A_1\cdot A_2-(1\leftrightarrow2)\Big)\\
&\equiv c\, \theta_{\mu}\, \big(A_{1}\bullet A_{2}\big)^{\mu}\;, 
\end{split}
\end{equation}
which gives the color-stripped cubic vertex of Yang-Mills as $\big\l\cA_3,m_2(\cA_1,\cA_2)\big\r$. Similarly, the only non-vanishing component of $m_{3\rm h}$ comes from the operator
\begin{equation}
\hat m_{3\rm h}(\theta_\mu,\theta_\nu,\theta_\rho)=\Big(c\,\theta_\mu\,\eta_{\nu\rho}-c\,\theta_\nu\,\eta_{\mu\rho}\Big)\,\big(\mathds{1}\otimes\mathds{1}\otimes\mathds{1}\big)\;,   
\end{equation}
yielding
\begin{equation}
\begin{split}
m_{3\rm h}(\cA_1,\cA_2,\cA_3)&=\mu\left[
\hat m_{3\rm h}(\theta_\mu,\theta_\nu,\theta_\rho)\,\big(A_1^\mu\otimes A_2^\nu\otimes A_3^\rho\big)\right]\\
&=c\,\theta_\mu\,\Big(A_1^\mu\,A_2\cdot A_3-(1\leftrightarrow2)\Big)\;,    
\end{split}
\end{equation}
corresponding to the color-stripped quartic vertex. A complete list of the operators $\hat m_n$ can be found in appendix \ref{App:YM}.

\subsection{$\BV^\B$ algebra on $\cK$}

While it is rather straightforward to determine that $\cK$ carries a $C_\infty$ structure (this is `just' rephrasing its usual consistency conditions), the next algebraic layer on $\cK$ is highly nontrivial and plays a crucial role in the double copy construction. To see how this deeper structure arises on $\cK$, one has to look at the interplay of the $C_\infty$ algebra, given by the products $m_n$, with the $b$ operator introduced before. From its definition in \eqref{bc} and the expression of the differential $m_1$ one may verify that  it obeys
\begin{equation}\label{bproperties}
 b^2=0\;,\quad b\,m_1+m_1b=\B\;,\quad |b|=-1\;,   
\end{equation}
where $\B=\del^\mu\del_\mu$ is the wave operator.  
It turns out that
\eqref{bproperties} should be viewed as the general defining property of $b$, with our realization \eqref{bc}, \eqref{Kdiagram} being a particular case.\footnote{For different realizations of the $b$ operator in various gauge theories, see \emph{e.g.} \cite{Reiterer:2019dys,Ben-Shahar:2021doh,Ben-Shahar:2021zww,Borsten:2022vtg,Borsten:2023reb}}  
Although $b$ does not play a role in the consistency relations encoded  in the $C_\infty$ algebra 
of the theory, it can be viewed as providing a gauge fixing condition as $b(\cA)=0$, as well as the related propagator as $\frac{b}{\B}$ acting on the space of equations. The peculiar property of our realization \eqref{bc} is that it acts on $\cK$ as
\begin{equation}
b(\psi)=(b\,Z_A)\,\psi^A(x)\;,   
\end{equation}
implying in particular that it is local and does not contain spacetime derivatives. This will be instrumental in order to construct a local theory from double copy.

With this second differential at our disposal, one can study its compatibility with the $C_\infty$ products. Its graded commutator with $m_1$ is given in \eqref{bproperties}. Going one step further, $b$ does not act as a derivation on $m_2$. Rather, the failure to do so defines a bracket $b_2$:
\begin{equation}\label{b2}
b_2(\psi_1,\psi_2):=b\,m_2(\psi_1,\psi_2)-m_2(b\,\psi_1,\psi_2)-(-1)^{\psi_1}m_2(\psi_1,b\,\psi_2)\;,    
\end{equation}
on which $b$ acts as a derivation by construction. Given a product $m_2$ and a bracket $b_2$, one can ask if they are mutually compatible, \emph{i.e.} if they obey the graded Poisson identity
\begin{equation}\label{Poisson}
b_2\big(\psi,m_2(\psi_1,\psi_2)\big)=m_2\big(b_2(\psi,\psi_1),\psi_2\big)+(-1)^{\psi_1\psi_2}m_2\big(b_2(\psi,\psi_2),\psi_1\big)\;.    
\end{equation}
If this were the case (which also requires $m_2$ to be associative), $b_2$ would be a graded Lie bracket, and the triplet $(b,m_2,b_2)$ would form a BV algebra. While this happens for Chern-Simons theory \cite{Ben-Shahar:2021zww,Borsten:2022vtg}, it is not the case for Yang-Mills theory (at least in standard formulations). The compatibility \eqref{Poisson} holds only up to an homotopy $\theta_3$ and further $\B$ deformations originating from \eqref{bproperties}. This prompts a cascade of higher relations, defined as a $\BV^\B$ algebra in \cite{Reiterer:2019dys}. 

In order to give all the relevant relations of the resulting $\BV^\B$ algebra (up to trilinear maps), we shall review a convenient input-free notation introduced in \cite{Bonezzi:2022bse}, which will allow us to establish the results of the forthcoming sections.
In this part we will denote by $\cO$ any linear operator in $\cK$, of degree $|\cO|$. Generic bilinear and trilinear maps will be denoted by $\cM$ and $\cT$, respectively, with arbitrary intrinsic degrees $|\cM|$ and $|\cT|$.
Similarly to \eqref{mhats}, these generic maps act on elements of $\cK$ as
\begin{equation}\label{general mhats}
\begin{split}
\cO(\psi)&=\hat \cO(Z_A)\,\psi^A(x)\;,\\
\cM(\psi_1,\psi_2)&=\mu\left[\hat \cM(Z_A,Z_B)\,\Big(\psi_1^A(x)\otimes\psi_2^B(x)\Big)\right]\;,\\
\cT(\psi_1,\psi_2,\psi_3)&=\mu\left[\hat \cT(Z_A,Z_B,Z_C)\,\Big(\psi_1^A(x)\otimes\psi^B_2(x)\otimes\psi^C_3(x)\Big)\right]\;,
\end{split}    
\end{equation}
where $\hat\cO$, $\hat\cM$ and $\hat\cT$ are $\cZ-$valued multidifferential operators acting on the component functions.
We define the graded commutator of operators $\cO_1$, $\cO_2$ by
\begin{equation}\label{commO1O2}
[\cO_1,\cO_2](\psi):=\cO_1\big(\cO_2 \psi\big)-(-1)^{\cO_1\cO_2}\cO_2\big(\cO_1 \psi\big)\;,    
\end{equation}
where every symbol in exponents refers to the degree of a map or element.
The commutators of an operator $\cO$ with bilinear and trilinear maps $\cM$ and $\cT$ are the bilinear map $[\cO,\cM]$ and trilinear map $[\cO,\cT]$ given by
\begin{equation}\label{OMOT}
\begin{split}
[\cO,\cM](\psi_1, \psi_2)&:=\cO\cM (\psi_1,\psi_2)-(-1)^{\cO\cM}\Big[\cM(\cO \psi_1,\psi_2)+(-1)^{\psi_1\cO}\cM(\psi_1,\cO \psi_2)\Big]\;,\\
[\cO,\cT](\psi_1, \psi_2, \psi_3)&:=\cO\cT(\psi_1, \psi_2,\psi_3)-(-1)^{\cO\cT}\Big[\cT(\cO \psi_1,\psi_2,\psi_3)\\
&\hspace{15mm}+(-1)^{\cO\psi_1}\cT(\psi_1,\cO \psi_2,\psi_3)+(-1)^{\cO(\psi_1+\psi_2)}\cT(\psi_1,\psi_2,\cO \psi_3)\Big]\;.    
\end{split}    
\end{equation}
The action of an operator $\cO$ on a map (be it another operator, a bilinear or trilinear map) gives a map of the same kind, \emph{e.g.}
\begin{equation}
\big(\cO\cM\big)(\psi_1,\psi_2):=\cO\big(\cM(\psi_1,\psi_2)\big)\;. \end{equation}
Finally, composition of bilinear maps is defined from the left and denoted by juxtaposition:
\begin{equation}\label{leftnest}
\cM_1\cM_2(\psi_1, \psi_2, \psi_3):=\cM_1\big(\cM_2(\psi_1, \psi_2), \psi_3\big)\;.    
\end{equation} 
This is sufficient for our purposes, since all bilinear maps involved are graded symmetric.
With this notation one can check that $[\cO,-]$ is a derivation on commutators and compositions, in the sense that it obeys
\begin{equation}
\begin{split}
[\cO_1,[\cO_2,\cM]]&=[[\cO_1,\cO_2],\cM]+(-1)^{\cO_1\cO_2}[\cO_2,[\cO_1,\cM]]\;,\\
[\cO_1,\cO_2\cM]&=[\cO_1,\cO_2]\cM+(-1)^{\cO_1\cO_2}\cO_2[\cO_1,\cM]\;,\\
[\cO,\cM_1\cM_2]&=[\cO,\cM_1]\cM_2+(-1)^{\cO\cM_1}\cM_1[\cO,\cM_2]\;.
\end{split}    
\end{equation}

We now turn to discuss the symmetry properties of trilinear maps $\cT$. Since they are all graded symmetric in the first two arguments (this is the reason we chose to work with $m_{3\rm h}$ rather than $m_3$), they can be decomposed into a totally graded symmmetric part, which we denote by $\cT_{\rm s}$, and a graded hook part $\cT_{\rm h}:=\cT-\cT_{\rm s}$.
In terms of $\cT$, the symmetrized map $\cT_{\rm s}$ acts on three inputs as
\begin{equation}\label{Tpinputs}
\begin{split}
\cT_{\rm s}(\psi_1,\psi_2,\psi_3)&=\frac13\,\Big\{\cT(\psi_1,\psi_2,\psi_3)+(-1)^{\psi_1(\psi_2+\psi_3)}\cT(\psi_2,\psi_3,\psi_1)\\
&\hspace{10mm}+(-1)^{\psi_3(\psi_1+\psi_2)}\cT(\psi_3,\psi_1,\psi_2)\Big\}\;.
\end{split}    
\end{equation}
In line with  \eqref{general mhats}  we want to associate a multidifferential operator $\hat\cT_{\rm s}$ to $\cT_{\rm s}$, such that
\begin{equation}\label{Tpi}
\cT_{\rm s}(\psi_1,\psi_2,\psi_3)=\mu\left[\hat{\cT}_{\rm s}(Z_A,Z_B,Z_C)\,\Big(\psi_1^A\otimes\psi_2^B\otimes\psi_3^C\Big)\right]\;.    
\end{equation}
To do so, we start by introducing a permutation operator $\Sigma$, which acts on trilinear operators as
\begin{equation}
\big(\cO_1\otimes\cO_2\otimes\cO_3\big)\Sigma:=\big(\cO_3\otimes\cO_1\otimes\cO_2\big)\;,    
\end{equation}
and thus obeys 
\begin{equation}
\begin{split}
\mu\Big[\hat\cT(Z_A,Z_B,Z_C)\Sigma\,\big(f_1\otimes f_2\otimes f_3\big)\Big]&=\mu\Big[\hat\cT(Z_A,Z_B,Z_C)\big(f_2\otimes f_3\otimes f_1\big)\Big]\;,\\    
\mu\Big[\hat\cT(Z_A,Z_B,Z_C)\Sigma^2\,\big(f_1\otimes f_2\otimes f_3\big)\Big]&=\mu\Big[\hat\cT(Z_A,Z_B,Z_C)\big(f_3\otimes f_1\otimes f_2\big)\Big]\;,
\end{split}    
\end{equation}
and $\Sigma^3=1$. We then use this to define a projector $\pi$, obeying $\pi^2=\pi$, so that the symmetrized and hook operators $\hat\cT_{\rm s}$ and $\hat\cT_{\rm h}$ are defined via
\begin{equation}
\hat\cT_{\rm s}:=\hat\cT\pi\;,\quad \hat\cT_{\rm h}:=\hat\cT(1-\pi)\;,\quad\hat\cT=\hat\cT_{\rm s}+\hat\cT_{\rm h}\;.    
\end{equation}
In terms of the permutation operator $\Sigma$, $\pi$ is explicitly given by
\begin{equation}\label{Tpisigma}
\begin{split}
\hat{\cT}\pi(Z_A,Z_B,Z_C)
&=\frac13\,\Big\{\hat\cT(Z_A,Z_B,Z_C)+(-1)^{Z_A(Z_B+Z_C)}\hat\cT(Z_B,Z_C,Z_A)\,\Sigma\\
&\hspace{10mm}+(-1)^{Z_C(Z_A+Z_B)}\hat\cT(Z_C,Z_A,Z_B)\,\Sigma^2\Big\}\;,
\end{split}    
\end{equation}
which reproduces, upon using \eqref{Tpi},  the expression \eqref{Tpinputs} for the map $\cT_{\rm s}$. 
We will then use interchangeably the notation $\cT_{\rm s}\equiv\cT\pi$ for the symmetrized map as well.
The operator $\hat\cT_{\rm s}$ obeys the graded symmetry property
\begin{equation}\label{Tpiproperty}
\hat\cT_{\rm s}(Z_A,Z_B,Z_C)=(-1)^{Z_A(Z_B+Z_C)}\hat\cT_{\rm s}(Z_B,Z_C,Z_A)\Sigma=(-1)^{Z_C(Z_A+Z_B)}\hat\cT_{\rm s}(Z_C,Z_A,Z_B)\Sigma^2\;,   
\end{equation}
which implies the standard graded symmetry of the map $\cT_{\rm s}(\psi_1,\psi_2,\psi_3)$ upon permutations of the inputs.

Let us illustrate the action of $\pi$ with a concrete example. We consider the trilinear map $\cT$ associated with the operator
\begin{equation}\label{Ex:T3}
\hat\cT(\theta_\mu,\theta_\nu,\theta_\rho)=c\,\theta_-\,\eta_{\mu\nu}\,\Big[\big(\mathds{1}\otimes\del_\rho\otimes\mathds{1}\big)-\big(\del_\rho\otimes\mathds{1}\otimes\mathds{1}\big)\Big]\;,    
\end{equation}
which is part of the actual map $m_2m_2(\cA_1,\cA_2,\cA_3)$. Acting with \eqref{Ex:T3} on $(A_1^\mu\otimes A_2^\nu\otimes A_3^\rho)$ and 
evaluating the pointwise product one obtains
\begin{equation}
\cT(A_1,A_2,A_3)=c\,\theta_-\,\Big(A_{1\mu}\del_\rho A_2^\mu A_3^\rho-A_{2\mu}\del_\rho A_1^\mu A_3^\rho \Big)\;,    
\end{equation}
where we abbreviated $A_i=\theta_\mu A_i^\mu$.
According to the definition \eqref{Tpisigma}, the symmetrized operator $\hat\cT\pi$ is given by
\begin{equation}
\begin{split}
\hat{\cT}\pi(\theta_\mu,\theta_\nu,\theta_\rho)&=\frac13\,c\,\theta_-\Big(\eta_{\mu\nu}\,\big(\mathds{1}\otimes\del_\rho\otimes\mathds{1}\big)-\eta_{\mu\nu}\,\big(\del_\rho\otimes\mathds{1}\otimes\mathds{1}\big)+\eta_{\nu\rho}\,\big(\mathds{1}\otimes\mathds{1}\otimes\del_\mu\big)\\
&\hspace{18mm}-\eta_{\nu\rho}\,\big(\mathds{1}\otimes\del_\mu\otimes\mathds{1}\big)+\eta_{\mu\rho}\,\big(\del_\nu\otimes\mathds{1}\otimes\mathds{1}\big)-\eta_{\mu\rho}\,\big(\mathds{1}\otimes\mathds{1}\otimes\del_\nu\big)\Big) \;,  
\end{split}    
\end{equation}
yielding the symmetrized map
\begin{equation}
\begin{split}
\cT\pi(A_1,A_2,A_3)&=\frac13\,c\,\theta_-\Big(A_{1\mu}\del_\rho A_2^\mu A_3^\rho-A_{2\mu}\del_\rho A_1^\mu A_3^\rho+A_{2\mu}\del_\rho A_3^\mu A_1^\rho-A_{3\mu}\del_\rho A_2^\mu A_1^\rho\\
&\hspace{17mm}+A_{3\mu}\del_\rho A_1^\mu A_2^\rho-A_{1\mu}\del_\rho A_3^\mu A_2^\rho\Big)\;.
\end{split}    
\end{equation}
From the definition \eqref{OMOT} of the commutator $[\cO,\cT]$, one can check that the action of $\cO$ preserves the symmetry property of the map $\cT$ in the sense  that
\begin{equation}
[\cO,\cT]\pi=[\cO,\cT\pi]\;.   
\end{equation}

We conclude this review of the input-free formulation by focusing on the possible $\B$ obstructions. Since we are working on flat spacetime, the wave operator $\B$ commutes with all the multidifferential operators $\hat\cO$, $\hat\cM$ and $\hat\cT$ in \eqref{general mhats}. Its commutators with the maps $\cO$, $\cM$ and $\cT$ are thus entirely determined by the commutator of $\B$ on the pointwise product of functions. We thus define the following operators, acting on three local functions:
\begin{equation}\label{d's}
\begin{split}
d_s(f_1\otimes f_2\otimes f_3)&:=2\,(\del^\mu f_1\otimes\del_\mu f_2\otimes f_3)\;,\\
d_\B(f_1\otimes f_2\otimes f_3)&:=2\,(\del^\mu f_1\otimes\del_\mu f_2\otimes f_3)+2\,(f_1\otimes\del^\mu f_2\otimes\del_\mu f_3)+2\,(\del^\mu f_1\otimes f_2\otimes\del_\mu f_3)\;.
\end{split}    
\end{equation}
The subscript in $d_s$ alludes to the Mandelstam variable $s$, and should not be confused with the symmetrization $\cT_{\rm s}$. One can compose a $\cZ-$valued tri-differential operator $\hat\cT$ with $d_s$ and $d_\B$, which we denote by juxtaposition: 
\begin{equation}
\begin{split}
\hat\cT d_s&:=2\,\hat\cT\circ(\del^\mu\otimes\del_\mu\otimes\mathds{1})\;,\\
\hat\cT d_\B&:=2\,\hat\cT\circ\Big\{(\del^\mu\otimes\del_\mu\otimes\mathds{1})+(\mathds{1}\otimes\del^\mu\otimes\del_\mu)+(\del^\mu\otimes\mathds{1}\otimes\del_\mu)\Big\}\;.
\end{split}    
\end{equation}
These are also $\cZ-$valued tri-differential operators which generate the corresponding maps $\cT d_s$ and $\cT d_\B$. For instance, one has $\cT d_s(\psi_1,\psi_2,\psi_3)=2\,\cT(\del^\mu\psi_1,\del_\mu\psi_2,\psi_3)$ and so on. Under projection by $\pi$, $d_s$ and $d_\B$ obey
\begin{equation}\label{dboxrel}
\cT d_\B \,\pi=\cT\pi\, d_\B=3\,\cT\pi\, d_s\,\pi\;.   
\end{equation}
The $d_\B$ operator is always related to a total commutator with $\B$, in the sense that
\begin{equation}
[\B,\cT]=\cT\,d_\B\;,
\end{equation}
while $\cT\,d_s$ is not.
Lastly, from the definition of $[\cO,\cT]$ it follows that $d_s$ and $d_\B$ commute with linear operators $\cO$:
\begin{equation}\label{Odscommutator}
[\cO,\cT]d_s=[\cO,\cT d_s]\;,\quad [\cO,\cT]d_\B=[\cO,\cT d_\B]\;.
\end{equation}
With this notation at hand, we can summarize all the relevant ${\rm BV}_\infty^\B$ relations up to trilinear maps \cite{Bonezzi:2022bse}:
\begin{equation}\label{BVcrazy}
\begin{array}{ll}
m_1^2=0\;,\quad b^2=0\;,\quad [m_1,b]=\B\;,&\text{differentials and central obstruction,}\\[3mm]
[m_1,m_2]=0\;,\quad m_2m_2(1-\pi)=[m_1,m_{3\rm h}]\;,& C_\infty\;{\rm structure},\\[3mm]
b_2=[b,m_2]\;,\quad[m_1,b_2]=[\B ,m_2]\;,& \text{two-bracket and deformed Leibniz},\\[3mm]
b_2m_2+m_2b_2\,(1-3\,\pi)=[m_1,\theta_3]+m_{3\rm h}(d_\B-3\,d_s\,\pi)\;,&\text{deformed homotopy Poisson},\\[3mm]
3\,b_2b_2\pi+[m_1,b_3]+3\,\theta_{3}d_s\pi=0\;,&\text{deformed homotopy Jacobi},\\[3mm]
\theta_{3\rm h}+[b,m_{3\rm h}]=0\;,\quad b_3+[b,\theta_{3\rm s}]=0\;, & \text{compatibility of homotopies.}
\end{array}  
\end{equation}
The explicit maps for $m_n$ and $\theta_3$ can be found in the appendix of \cite{Bonezzi:2022bse}, while $b_2$ and $b_3$ are easily derived from these by taking $b-$commutators. The corresponding differential operators $\hat m_n$ and $\hat\theta_3$ are listed in appendix \ref{App:YM}.

From the above table one can see that the only consistent subsector is given by the original $C_\infty$ algebra $(m_1,m_2,m_{3\rm h})$. On the other hand, the brackets $(b_1\equiv m_1, b_2,b_3)$ form an $L_\infty$ algebra only up to $\B-$deformations, governed at this level by $m_2$ and $\theta_3$. 
Armed with this structure on $\cK$, in the next section we will show that a natural $\BV^\Delta$ algebra exists on the tensor product $\cK\otimes\bar\cK$ of two copies of $\cK$.

\section{$\rm{BV}_{\infty}^{\Delta}$ algebra on $\mathcal{K}\otimes \bar{\mathcal{K}}$}\label{sec:BVDel}

In this section we will consider two copies of the kinematic algebra $\cK$ and show that the respective ${\rm BV}_\infty^\B$ algebras give rise to a $\BV^\Delta$ algebra 
on the tensor product $\cX:=\cK\otimes\bar\cK$, 
where $\Delta:=\frac{1}{2}(\B-\bar{\B})$.
This will be used in the next sections to derive the three-brackets of double field theory (DFT), both on arbitrary flat backgrounds in the strongly constrained sense and  on a torus 
in the weakly constrained sense. At this point we should mention that the space $\cX$ is \emph{not} the $L_\infty$ graded vector space of DFT,  which we refer to as $\cV_{\rm DFT}$ and which  is a linear subspace of $\cX$ to  be described below.

\subsection{Grading  and maps on $\cK\otimes\bar\cK$}

We start by spelling out the structure of the tensor product $\cX=\cK\otimes\bar\cK$ as a graded vector space. From now on, we will denote all elements and maps of $\bar\cK$ with a bar on the same symbols used for $\cK$. Recalling that $\cK=\cZ\otimes C^\infty(\cM)$, one obtains that $\cX$ similarly factorizes as a finite-dimensional graded vector space tensored with functions on a doubled spacetime:
\begin{equation}\label{X}
\begin{split}
\cX=\cK\otimes\bar\cK&=\big(\cZ\otimes C^\infty(\cM)\big)\otimes\big(\bar\cZ\otimes C^\infty(\bar\cM)\big)\\
&\simeq \big(\cZ\otimes\bar\cZ\big)\otimes C^\infty(\cM\times\bar\cM) \;,   
\end{split}    
\end{equation}
using  that  $ C^\infty(\cM) \, \otimes \, C^\infty(\bar\cM)\simeq C^\infty(\cM\times\bar\cM)$, which under a 
certain topological completion  holds for  the tensor product.    
Throughout this section, we will take $\cM$ and $\bar{\cM}$ to be $d-$dimensional flat spaces with unspecified signature, but later on we will specialize to Euclidean signature. 
Given the structure \eqref{X} of $\cX$ and two copies of the finite-dimensional basis, i.e.~$\{Z_A\}$ of $\cZ$ and $\{\bar{Z}_{\bar{A}}\}$ 
of $\bar{\cZ}$,  
we can expand an arbitrary element $\Psi\in\cX$ as
\begin{equation}\label{Psi expansion}
\Psi(x,\bar x) =Z_A\bar Z_{\bar B}\,\Psi^{A\bar B}(x,\bar x)\;,    
\end{equation}
where we denote coordinates of the doubled space by $(x^\mu,\bar x^{\bar\mu})$, and from now on we will use capital $\Psi$ for elements in $\cX$.
The degree in $\cX$ is defined as the sum of the degrees in $\cK$ and $\bar\cK$, with an additional shift by $2$. Specifically, for a homogenous element $\Psi=Z\bar Z\,F(x,\bar x)$ we set
\begin{equation}\label{Xdegree}
|\Psi|=|Z|+|\bar Z|-2\;,    
\end{equation}
where we recall that degrees in $\cZ$ (the same for $\bar\cZ$) are displayed in \eqref{Zdiagram}. The shift in degree is by an even amount, so it is immaterial for sign factors like $(-1)^{|\Psi|}$ and thus strictly not necessary. However, 
the definition \eqref{Xdegree} complies with standard $L_\infty$ degrees in the resulting double field theory. 

Given the definition \eqref{general mhats} and the expansion \eqref{Psi expansion} we now proceed to lift the action of operators $\cO:\cK\to\cK$ and $\bar\cO:\bar\cK\to\bar\cK$ to $\cX$ by defining
\begin{equation}
\cO(\Psi):=\hat \cO(Z_A)\bar Z_{\bar B}\,\Psi^{A\bar B}(x,\bar x)\;,\qquad \bar\cO(\Psi):=(-1)^{Z_A\bar\cO}Z_A\,\hat{\bar \cO}(\bar Z_{\bar B})\,\Psi^{A\bar B}(x,\bar x)  \;,  
\end{equation}
where the differential operators $\hat\cO$ and $\hat{\bar\cO}$ act on functions of $x$ and $\bar x$ by taking $\del_\mu=\frac{\del}{\del x^\mu}$ and $\bar\del_{\bar\mu}=\frac{\del}{\del\bar x^{\bar\mu}}$ derivatives, respectively.
This allows us to sum operators from $\cK$ and $\bar\cK$ and yield well-defined operators on $\cX$, such as $\cO+\bar\cO$. Similarly, tensor products of bilinear maps $\cM$ and $\bar\cM$ are defined to act on elements of $\cX$ as follows:
\begin{equation}\label{MMbar}
\begin{split}
&\big(\cM\otimes\bar\cM\big)(\Psi_1,\Psi_2)=\big(\cM\otimes\bar\cM\big)\big(Z_A\bar Z_{\bar B}\,\Psi_1^{A\bar B}, Z_C\bar Z_{\bar D}\,\Psi_2^{C\bar D}\big)
\\&:=(-1)^{Z_C\bar Z_{\bar B}+\bar\cM(Z_A+Z_C)}\mu\left[\hat\cM(Z_A, Z_C)\,\hat{\bar\cM}(\bar Z_{\bar B},\bar Z_{\bar D})\,\Big(\Psi_1^{A\bar B}(x,\bar x)\otimes\Psi_2^{C\bar D}(x,\bar x)\Big)\right]\;,     
\end{split}
\end{equation}
with a completely analogous expression for $\big(\cT\otimes\bar\cT\big)(\Psi_1,\Psi_2,\Psi_3)$. With these definitions we can extend the input-free notation of the previous section to $\cX$. It turns out that operators $\cO$ and $\bar\cO$ commute (in the graded sense). To show this we compute
\begin{equation}\label{oobarcomm}
\begin{split}
\cO(\bar\cO\,\Psi)&=(-1)^{Z_A\bar\cO}\cO\big(Z_A\,\hat{\bar \cO}(\bar Z_{\bar B})\,\Psi^{A\bar B}\big)\\
&=(-1)^{Z_A\bar\cO}\hat\cO(Z_A)\,\hat{\bar \cO}(\bar Z_{\bar B})\,\Psi^{A\bar B}(x,\bar x)\\
&=(-1)^{Z_A\bar\cO+\bar\cO(Z_A+\cO)}{\bar \cO}\big(\hat\cO(Z_A)\,\bar Z_{\bar B}\,\Psi^{A\bar B}\big)\\
&=(-1)^{\cO\bar \cO}\bar\cO(\cO\,\Psi)\;,
\end{split}    
\end{equation}
where we omitted the explicit dependence on $(x,\bar x)$ in intermediate steps. This can be written as the input-free relation $[\cO,\bar\cO]=0$, where the graded commutator is defined as in \eqref{commO1O2}, albeit acting on elements of $\cK\otimes\bar\cK$. A similar computation using the definition \eqref{MMbar} shows that operators of $\cK$ commute with bilinear and trilinear maps of $\bar\cK$ and viceversa, in the sense
\begin{equation}\label{OMbarcomm}
[\cO,\cM\otimes\bar\cM]=[\cO,\cM]\otimes\bar\cM\;, \qquad [\bar\cO,\cM\otimes\bar\cM]=(-1)^{\cM\bar\cO}\cM\otimes[\bar\cO,\bar\cM]\;,       
\end{equation}
with analogous formulas for commutators with $\cT\otimes\bar\cT$.
Nesting of bilinear maps can also be extended naturally by defining
\begin{equation}
\big(\cM_1\otimes\bar\cM_1\big)\,\big(\cM_2\otimes\bar\cM_2\big):=(-1)^{\bar\cM_1\cM_2}\cM_1\cM_2\otimes\bar\cM_1\bar\cM_2\;,   
\end{equation}
where the composition $\cM_1\cM_2$ (same for the barred ones) is defined by \eqref{leftnest}.

Finally, one can introduce on $\cX$ a symmetric projector $\Pi$, obeying $\Pi^2=\Pi$, via
\begin{equation}\label{Pi}
\begin{split}
&\big(\cT\otimes\bar\cT\big)\Pi\,\big(\Psi_1,\Psi_2,\Psi_3\big)=\frac13\,(-1)^{\cE}\mu\Big[\Big(\hat\cT(Z_A,Z_B,Z_C)\,\hat{\bar\cT}(\bar Z_{\bar A},\bar Z_{\bar B},\bar Z_{\bar C})\\
&+(-1)^{Z_A(Z_B+Z_C)+\bar Z_{\bar A}(\bar Z_{\bar B}+\bar Z_{\bar C})}\hat\cT(Z_B,Z_C,Z_A)\Sigma\,\hat{\bar\cT}(\bar Z_{\bar B},\bar Z_{\bar C},\bar Z_{\bar A})\bar\Sigma\\
&+(-1)^{Z_C(Z_A+Z_B)+\bar Z_{\bar C}(\bar Z_{\bar A}+\bar Z_{\bar B})}\hat\cT(Z_C,Z_A,Z_B)\Sigma^2\,\hat{\bar\cT}(\bar Z_{\bar C},\bar Z_{\bar A},\bar Z_{\bar B})\bar\Sigma^2\Big)\Big(\Psi_1^{A\bar A}\otimes\Psi_2^{B\bar B}\otimes\Psi_3^{C\bar C}\Big)\Big]\;,    
\end{split}    
\end{equation}
where the global phase is $\cE=Z_B\bar Z_{\bar A}+Z_C(\bar Z_{\bar A}+\bar Z_{\bar B})+(Z_A+Z_B+Z_C)\bar\cT$. In terms of the map $\cT\otimes\bar\cT$, this results in
\begin{equation}
\begin{split}
\big(\cT\otimes\bar\cT\big)\Pi(\Psi_1,\Psi_2,\Psi_3)&:=\frac13\,\Big\{\big(\cT\otimes\bar\cT\big)(\Psi_1,\Psi_2,\Psi_3)+(-1)^{\Psi_1(\Psi_2+\Psi_3)}\big(\cT\otimes\bar\cT\big)(\Psi_2,\Psi_3,\Psi_1)\\
&+(-1)^{\Psi_3(\Psi_1+\Psi_2)}\big(\cT\otimes\bar\cT\big)(\Psi_3,\Psi_1,\Psi_2)\Big\}\;,
\end{split}
\end{equation}
which makes the graded symmetry manifest.
One can lift the definition of the single copy $\pi$ or $\bar\pi$ to a trilinear map $\cT\otimes\bar\cT$ on $\cX$ by
\begin{equation}
(\cT\otimes\bar\cT)\pi:=(\cT\pi)\otimes\bar\cT\;,\qquad (\cT\otimes\bar\cT)\bar\pi:=\cT\otimes(\bar\cT\bar\pi)\;,
\end{equation}
and using \eqref{Tpi}, \eqref{Tpisigma} for the single copy symmetrized maps.
From this it follows that $\pi\Pi=\bar\pi\Pi=\pi\bar\pi$, which further implies the decomposition
\begin{equation}\label{Pipi}
\begin{split}
\Pi&=\Big[\pi\bar\pi+(1-\pi)(1-\bar\pi)\Big]\Pi\;,\\
1-\Pi&=\Big[\pi(1-\bar\pi)+(1-\pi)\bar\pi+(1-\pi)(1-\bar\pi)\Big](1-\Pi)\;.
\end{split}    
\end{equation}

\subsection{$\BV^\Delta$ algebra on $\cX$}

We are now ready to show that, given the $\BV^\B$ algebras on $\cK$ and $\bar\cK$, a natural $\BV^\Delta$ algebra arises on $\cX$. As for the single copies, for the moment we will work this out up to trilinear maps. The starting point is the $C_\infty$ sector of \eqref{BVcrazy}. With the differentials $m_1$, $\bar m_1$ and the two-products $m_2$ and $\bar m_2$ we define a differential $M_1$ and two-product $M_2$ on $\cX$ by
\begin{equation}\label{M1M2}
\begin{split}
M_1&:=m_1+\bar m_1\;,\\
M_2&:=m_2\otimes\bar m_2\;.
\end{split}    
\end{equation}
From the graded symmetry of $m_2$ and $\bar m_2$ one can easily show that $M_2$ is graded symmetric in $\cX$. Due to the degree shift \eqref{Xdegree}, one has $|M_1|=+1$ and $|M_2|=+2$.
Upon using \eqref{oobarcomm} and \eqref{OMbarcomm} it is immediate to see that $M_1$ is nilpotent and acts as a derivation on $M_2$:
\begin{equation}
\begin{split}
M_1^2&=(m_1+\bar m_1)^2=m_1\bar m_1+\bar m_1 m_1=0\;,\\
[M_1,M_2]&=[m_1+\bar m_1, m_2\otimes\bar m_2]\\
&=[m_1, m_2]\otimes\bar m_2+ m_2\otimes[\bar m_1,\bar m_2]=0\;.
\end{split}    
\end{equation}

We next study the associativity of $M_2$ by computing its associator. Due to  the graded symmetry of $M_2$, the latter is equivalent to the hook projection $M_2M_2(1-\Pi)$.
Using the definition \eqref{M1M2}, the property \eqref{Pipi} of projectors and the homotopy associativity of $m_2$ and $\bar m_2$ we can compute
\begin{equation}\label{M2ass}
\begin{split}
M_2M_2(1-\Pi)&=m_2m_2\otimes\bar m_2\bar m_2\,\Big(\pi(1-\bar\pi)+(1-\pi)\bar\pi+(1-\pi)(1-\bar\pi)\Big)(1-\Pi)\\
&=\Big\{m_2m_2\pi\otimes[\bar m_1,\bar m_{3\rm h}]+[m_1, m_{3\rm h}]\otimes\bar m_2\bar m_2\bar\pi+[m_1, m_{3\rm h}]\otimes[\bar m_1,\bar m_{3\rm h}]\Big\}(1-\Pi)\\
&=\big[M_1,m_2m_2\pi\otimes\bar m_{3\rm h}+ m_{3\rm h}\otimes\bar m_2\bar m_2\bar\pi\big](1-\Pi)+[m_1, m_{3\rm h}]\otimes[\bar m_1,\bar m_{3\rm h}](1-\Pi)\;,
\end{split}    
\end{equation}
where in the last step we used $[m_1,m_2m_2]=0$ (and the barred relation) to extract a total differential $M_1$.
At this stage an ambiguity arises on how to treat the last term, since
\begin{equation}
[m_1, m_{3\rm h}]\otimes[\bar m_1,\bar m_{3\rm h}]=\big[M_1,(\tfrac12-\xi)\,m_{3\rm h}\otimes[\bar m_1,\bar m_{3\rm h}]+(\tfrac12+\xi)\,[ m_1, m_{3\rm h}]\otimes\bar m_{3\rm h}\big]    
\end{equation}
for arbitrary $\xi$.
Keeping $\xi$ arbitrary leads to a one-parameter family of three-products, differing by an $M_1-$exact term (which for maps means a total $M_1$ commutator). This is expected, since in a homotopy associative algebra the three-product is defined only up to an $M_1-$closed quantity. For simplicity we choose $\xi=0$ and obtain
\begin{equation}\label{M3h}
M_{3\rm h}=\frac12\,\Big\{m_{3\rm h}\otimes \bar m_2\bar m_2(1+\bar\pi)+ m_2 m_2(1+\pi)\otimes \bar m_{3\rm h}\Big\}(1-\Pi)\;,    
\end{equation}
obeying homotopy associativity in the form $M_2M_2(1-\Pi)=[M_1,M_{3\rm h}]$. Even though the original $C_\infty$ algebras on $\cK$ and $\bar\cK$ have no higher products than $m_{3\rm h}$, the tensor algebra $\cX$ are expected to have  infinitely many higher $M_n$, which we will not explore further.

In order to go beyond the $C_\infty$ structure, one has to identify the analogue of the $b$ differential on the tensor space. Two natural candidates are the linear combinations
\begin{equation}
b^\pm:=\tfrac12 (b\pm\bar b)\;.    
\end{equation}
Both $b^\pm$ are nilpotent and (anti)commute with each other. Their commutators with $M_1$, which determine the obstruction $\Delta$, are given by
\begin{equation}
[M_1,b^\pm]=\tfrac12\,\big(\B\pm\bar\B\big) \;.   
\end{equation}
For establishing a $\BV^\Delta$ algebra on $\cX$, both choices $b^\pm$ for the second differential are equivalent. Our choice is dictated by the goal of constructing double field theory on a suitable subspace of $\cX$ which, in particular, requires an unobstructed $L_\infty$ algebra. In view of the fact that the subspace $\cV_{\rm DFT}$ is partly determined by constraining $\big(\B-\bar\B\big)\Psi=0$, the natural choice for the $b$ operator is $b^-$, yielding
\begin{equation}
[M_1,b^-]=\Delta\;,\quad \Delta:=\tfrac12  \big(\B-\bar\B\big)  \;.  
\end{equation}
Since $\Delta$ arises as the above commutator, it is guaranteed to commute with both $M_1$ and $b^-$: $[M_1,\Delta]=[b^-,\Delta]=0$.
With this choice one can construct a degree $+1$ two-bracket $B_2$, in perfect analogy with the single copy version \eqref{b2}:
\begin{equation}
\begin{split}
B_2&:=[b^-, M_2] 
=\tfrac12 [b-\bar b,m_2\otimes\bar m_2]\\ 
&=\tfrac12\,\Big(b_2\otimes\bar m_2-m_2\otimes\bar b_2\Big)\;,  
\end{split}
\end{equation}
where in the second line we emphasized its tensor product structure that is of the schematic form  ``Lie$\,\otimes\,$Commutative". 
While $b^-$ is trivially a derivation for $B_2$, $M_1$ is not. The obstruction is easily computed and takes the same form as in \eqref{BVcrazy}:
\begin{equation}
\begin{split}
[M_1, B_2]&=[M_1, [b^-,M_2]]\\
&=[[M_1,b^-],M_2]-[b^-, [M_1,M_2]]\\
&=[\Delta, M_2]\;,
\end{split}   
\end{equation}
where we used $[M_1,M_2]=0$. For the $\Delta-$obstructions on trilinear maps we define $D_\Delta:=\frac12 (d_\B- \bar d_{\bar\B})$ and $D_s:=\frac12 (d_s- \bar d_s)$ which act on three functions $F_i(x,\bar x)$ as
\begin{equation}\label{Ds}
\begin{split}
D_s(F_1\otimes F_2\otimes F_3)&=(\del^\mu F_1\otimes\del_\mu F_2\otimes F_3)-(\bar\del^{\bar\mu}F_1\otimes\bar\del_{\bar\mu}F_2\otimes F_3)\;,\\[2mm]
D_\Delta(F_1\otimes F_2\otimes F_3)&=(\del^\mu F_1\otimes\del_\mu F_2\otimes F_3)-(\bar\del^{\bar\mu}F_1\otimes\bar\del_{\bar\mu}F_2\otimes F_3)\\
&+(F_1\otimes\del^\mu F_2\otimes\del_\mu  F_3)-(F_1\otimes\bar\del^{\bar\mu}F_2\otimes\bar\del_{\bar\mu} F_3)\\
&+(\del^\mu  F_1\otimes F_2\otimes\del_\mu  F_3)-(\bar\del^{\bar\mu} F_1\otimes F_2\otimes\bar\del_{\bar\mu} F_3)\;.
\end{split}
\end{equation}
Given the definition of $D_s$ and $D_\Delta$ and the projector \eqref{Pi}, they obey
\begin{equation}\label{Ddeltarel}
\big(\cT\otimes\bar\cT\big)\,D_\Delta\,\Pi=\big(\cT\otimes\bar\cT\big)\,\Pi\,D_\Delta=3\,\big(\cT\otimes\bar\cT\big)\,\Pi\,D_s\,\Pi\;,  
\end{equation}
similarly to \eqref{dboxrel} for $d_s$ and $d_\B$.

Given the product $M_2$ and the bracket $B_2=[b^-, M_2]$, the Poisson compatibility condition (defined as in \eqref{Poisson} upon replacing $m_2\rightarrow M_2$ and $b_2\rightarrow B_2$) can be formulated as
\begin{equation}\label{Input free Poisson}
\begin{split}
B_2M_2+M_2B_2\,(1-3\,\Pi)\equiv[b^-,M_2M_2]-3\,M_2[b^-,M_2]\,\Pi \stackrel{?}{=}0\;.
\end{split}    
\end{equation}
The first form of \eqref{Input free Poisson} emphasizes the Poisson relation between $B_2$ and $M_2$, while the second form shows that this is equivalent to $b^-$ being second order with respect to $M_2$. Since the single copy $b_2$ and $m_2$ obey
\begin{equation}\label{Poiss reminder}
[b,m_2m_2]-3\,m_2b_2\,\pi=[m_1,\theta_3]+m_{3\rm h}(d_\B-3\,d_s\,\pi)\;,    
\end{equation}
one does not expect \eqref{Input free Poisson} to vanish, but rather to obey a similar relation in terms of $M_{3\rm h}$ and a $\Theta_3$ yet to be determined. To show that this is indeed the case, 
it is convenient to split the computation of \eqref{Input free Poisson} into its totally symmetric and hook parts. We start from the hook, which is simple to determine:
\begin{equation}\label{Phook}
\begin{split}
\Big\{[b^-,M_2M_2]-3\,M_2B_2\,\Pi\Big\}(1-\Pi)&=[b^-,M_2M_2](1-\Pi)\\
&=[b^-,M_2M_2(1-\Pi)]=\big[b^-,[M_1,M_{3\rm h}]\big]\\
&=-\big[M_1,[b^-,M_{3\rm h}]\big]+\big[[M_1,b^-],M_{3\rm h}\big]\\
&=-\big[M_1,[b^-,M_{3\rm h}]\big]+[\Delta,M_{3\rm h}]\\
&=[M_1,\Theta_{3\rm h}]+M_{3\rm h}\,D_\Delta\;,
\end{split}    
\end{equation}
where we identified $\Theta_{3\rm h}=-[b^-,M_{3\rm h}]$.
One can see that \eqref{Phook} has the same form as the hook projection of \eqref{Poiss reminder}, and the relation between $\Theta_{3\rm h}$ and $M_{3\rm h}$ is the same as in \eqref{BVcrazy} for compatibility of the homotopies. Computing the symmetric projection of \eqref{Input free Poisson} is considerably more involved. We spell out the computation in detail in appendix \ref{sec:Theta3}, which results in
\begin{equation}
\Big\{[b^-,M_2M_2]-3\,M_2B_2\,\Pi\Big\}\Pi=[M_1,\Theta_{3\rm s}]-3\,M_{3\rm h}\,D_s\,\Pi\;, 
\end{equation}
where $\Theta_{3\rm s}$ is determined in terms of Yang-Mills maps as
\begin{equation}
\begin{split}
\Theta_{3\rm s}&:=\tfrac12\,\Big(\theta_{3\rm s}\otimes \bar m_2\bar m_2- m_2 m_2\otimes\bar\theta_{3\rm s}+3\,m_{3\rm h}\otimes\bar m_2\bar b_2+3\, m_2 b_2\otimes\bar m_{3\rm h} \\
&\hspace{12mm}-\tfrac32\,m_{3\rm h}\otimes\bar m_{3\rm h}\,(d_s+\bar d_s)-[b^-, m_2m_2\otimes\bar m_{3\rm h}+m_{3\rm h}\otimes\bar m_2\bar m_2]\Big)\,\Pi\;.    
\end{split}    
\end{equation}
Given the Poisson relation, one can determine the jacobiator $3\,B_2B_2\Pi$ of the two bracket $B_2$ by taking a $b^-$ commutator. This is also presented in appendix \ref{sec:Theta3} and yields the deformed $L_\infty$ relation
\begin{equation}
3\,B_2B_2\Pi+[M_1,B_3]+3\,\Theta_{3}\,D_s\,\Pi=0\;,    
\end{equation}
where the three-bracket is given by $B_3=-[b^-,\Theta_{3\rm s}]$.

This concludes the $\BV^\Delta$ relations up to trilinear maps, which we summarize in the following table, analogous to the single copy one \eqref{BVcrazy}:
\begin{equation}\label{BVXcrazy}
\begin{array}{ll}
M_1^2=0\;,\quad (b^-)^2=0\;,\quad [M_1,b^-]=\Delta\;,&\text{differentials and central obstruction,}\\[3mm]
[M_1,M_2]=0\;,\quad M_2M_2(1-\Pi)=[M_1,M_{3\rm h}]\;,& C_\infty\;{\rm structure},\\[3mm]
B_2=[b^-,M_2]\;,\quad[M_1,B_2]=[\Delta,M_2]\;,& \text{two-bracket and deformed Leibniz},\\[3mm]
B_2M_2+M_2B_2\,(1-3\,\Pi)=[M_1,\Theta_3]+M_{3\rm h}(D_\Delta-3\,D_s\,\Pi)\;,&\text{deformed homotopy Poisson},\\[3mm]
3\,B_2B_2\Pi+[M_1,B_3]+3\,\Theta_{3}\,D_s\,\Pi=0\;,&\text{deformed homotopy Jacobi},\\[3mm]
\Theta_{3\rm h}+[b^-,M_{3\rm h}]=0\;,\quad B_3+[b^-,\Theta_{3\rm s}]=0\;, & \text{compatibility of homotopies.}
\end{array}  
\end{equation}
In particular, notice that the brackets of the $L_\infty$ sector (albeit obstructed) are all determined in terms of other structures as $B_1\equiv M_1$, $B_2=[b^-,M_2]$, $B_3=-[b^-, \Theta_{3\rm s}]$. This is in close analogy with closed string field theory, where all the string brackets, apart from $B_1$, are $b^--$exact \cite{Zwiebach:1992ie}.

We conclude this section by collecting the expressions of the relevant maps in terms of Yang-Mills building blocks:
\begin{equation}\label{YMtoDFT}
\begin{split}
M_1&:=m_1+\bar m_1\;,\quad b^\pm:=\tfrac12\,(b\pm\bar b)\;,\\
M_2&:=m_2\otimes\bar m_2\;,\\
M_{3\rm h}&:=\tfrac12\,\Big(m_{3\rm h}\otimes\bar m_2\bar m_2\,(1+\bar\pi)+m_2 m_2\,(1+\bar\pi)\otimes\bar m_{3\rm h}\Big)\,(1-\Pi)\;,\\
\Theta_{3\rm s}&:=\tfrac12\,\Big(\theta_{3\rm s}\otimes \bar m_2\bar m_2- m_2 m_2\otimes\bar\theta_{3\rm s}+3\,m_{3\rm h}\otimes\bar m_2\bar b_2+3\, m_2 b_2\otimes\bar m_{3\rm h} \\
&\hspace{12mm}-\tfrac32\,m_{3\rm h}\otimes\bar m_{3\rm h}\,(d_s+\bar d_s)-[b^-, m_2m_2\otimes\bar m_{3\rm h}+m_{3\rm h}\otimes\bar m_2\bar m_2]\Big)\,\Pi\;.
\end{split}    
\end{equation}

\section{Weakly constrained double field theory}

In this section we will start from the $\BV^\Delta$ algebra on $\cX$ and construct the $L_\infty$ algebra of weakly constrained DFT on a spatial torus, up to and including the three-bracket, which encodes all the quartic structures of the theory. The obvious issue is that the $L_\infty$ sector of \eqref{BVXcrazy} is obstructed on $\cX$, due to $\Delta$. The idea is that these obstructions should be milder when considering the relevant subspace of weakly constrained fields, obeying $\Delta\Psi=0$. 

Let us start by giving the precise definition of the graded vector space $\cV_{\rm DFT}$ carrying the $L_\infty$ algebra of double field theory. This is given by the following linear subspace of $\cX=\cK\otimes\bar\cK$:
\begin{equation}\label{VDFT}
\cV_{\rm DFT}:=\Big\{\Psi\in\cX,\; {\rm s.t.}\;\Delta\Psi=0\;,\; b^-\Psi=0\Big\}=\Big({\rm ker}\Delta\cap{\rm ker}b^-\Big)\subset\cX\;.    
\end{equation}
Given that an arbitrary element of $\cX$ can be expanded as in \eqref{Psi expansion}, with the graded vectors $Z_A=(\theta_M, c\,\theta_M)$ and $\bar Z_{\bar A}=(\bar\theta_{\bar M}, \bar c\,\bar\theta_{\bar M})$, one can explicitly characterize the elements of $\cV_{\rm DFT}$ as
\begin{equation}\label{Psi DFT expansion}
\Psi=\theta_M\bar\theta_{\bar N}\,\psi^{M\bar N}(x, \bar x)+c^+\theta_M\bar\theta_{\bar N}\,\chi^{M\bar N}(x, \bar x)\;,\quad\Delta\psi^{M\bar N}=\Delta\chi^{M\bar N}=0\;,    
\end{equation}
where $c^+:=c+\bar c$, so that $c^+\theta_M\bar\theta_{\bar N}=c\,\theta_M\bar\theta_{\bar N}+(-1)^{|\theta_M|}\theta_M\,\bar c\,\bar\theta_{\bar N}$. From the degree assignment \eqref{Xdegree}, one can see that the degrees in $\cV_{\rm DFT}$ are given by
\begin{equation}
|\theta_M\bar\theta_{\bar N}|=-(M+\bar N)\;,\quad |c^+\theta_M\bar\theta_{\bar N}|=1-(M+\bar N)\;,   
\end{equation}
where we recall that we associate degrees $(+1,0,-1)$ to $M=(+,\mu,-)$. One thus finds, for instance,  the DFT gauge parameters in degree $-1$  to be given by
\begin{equation}\label{gengaugeparam}
\Lambda=\theta_+\bar\theta_{\bar\mu}\,\bar\lambda^{\bar\mu}-\theta_\mu\bar\theta_+\,\lambda^\mu-2\,c^+\theta_+\bar\theta_+\,\eta \;, 
\end{equation}
which thus consist  of two vector parameters, related to generalized diffeomorphisms, and a St\"uckelberg scalar parameter. Similarly, one has the fields in degree zero:
\begin{equation}
\psi=\theta_\mu\bar\theta_{\bar\nu}\,e^{\mu\bar\nu}+2\,\theta_+\bar\theta_-\,\bar e+2\,\theta_-\bar\theta_+\, e+2\,c^+\theta_\mu\bar\theta_{+}\,f^\mu+2\,c^+\theta_+\bar\theta_{\bar\mu}\,\bar f^{\bar\mu}\;,
\end{equation}
comprising the tensor fluctuation $e_{\mu\bar\nu}$, two scalars (one combination is related to the dilaton, the other is pure gauge), and two auxiliary vectors. This is precisely the field content of DFT as first introduced in \cite{Hull:2009mi}.

In order to construct an $L_\infty$ algebra on $\cV_{\rm DFT}$, we will proceed in two steps: we will first transport the $\BV^\Delta$ structure of $\cX$ to the subspace $\overline{\cX}:={\rm ker}\Delta$ via homotopy transfer, to be discussed momentarily, and in the second step we will restrict the maps to act on ${\rm ker}b^-$. In this last step the $\BV^\Delta$ structure will be lost, leaving an unobstructed $L_\infty$ algebra on $\cV_{\rm DFT}$.

\subsection{Homotopy transfer to ${\rm ker}\Delta$}

In order to perform homotopy transfer, we shall find a suitable projector $\cP_\Delta$ to ${\rm ker}\Delta$, together with an homotopy operator $h$ of degree $|h|=-1$, obeying
\begin{equation}\label{homotopy relation}
\big[M_1,h\big]=1-\cP_\Delta \;.   
\end{equation}
Since $\overline{\cX}={\rm ker}\Delta\subset\cX$ is a subspace of $\cX$, we consider the projector $\cP_\Delta:\cX\rightarrow\overline{\cX}$ as an operator in $\cX$, implicitly assuming a trivial inclusion map $\iota:\overline{\cX}\rightarrow\cX$ (see \emph{e.g.}~\cite{Arvanitakis:2020rrk} for an introduction to homotopy transfer). 
The projector has to be properly normalized and is required to be a chain map, meaning that it should obey
\begin{equation}\label{Pprop}
\cP_\Delta^2=\cP_\Delta\;,\quad \cP_\Delta\,M_1=M_1\,\cP_\Delta\;.    
\end{equation}
We further require that the homotopy $h$ obeys the so-called side conditions:
\begin{equation}\label{hprop}
h\,\cP_\Delta=\cP_\Delta\,h=0\;,\quad h^2=0\;.    
\end{equation}

In order to define $\cP_\Delta$, we specialize the signature and topology of our underlying doubled space to be a doubled Euclidean square torus. In particular, we have two copies of the Euclidean metric $\delta_{\mu\nu}$ and $\delta_{\bar\mu\bar\nu}$ and we identify coordinates with periodicity $2\pi$. Any function on the doubled torus can be expanded in discrete Fourier modes as
\begin{equation}\label{Fourier}
f(x,\bar x)=\sum_{k,\bar k}\tilde f(k,\bar k)\,e^{ik\cdot x+i\bar k\cdot\bar x}\;,    
\end{equation}
with an unconstrained sum over discrete  momenta $(k^\mu, \bar k^{\bar\mu})\in\mathbb{Z}^{2d}$.
The obstruction $\Delta$ acts on \eqref{Fourier} as 
\begin{equation}
\Delta\,f(x,\bar x)=-\frac12\sum_{k,\bar k}(k^2-\bar k^2)\,\tilde f(k,\bar k)\,e^{ik\cdot x+i\bar k\cdot\bar x}\;. 
\end{equation}
This allows us to write the projectors $\cP_\Delta$ and $(1-\cP_\Delta)$ explicitly in terms of the Fourier expansion by inserting 
suitable Kronecker deltas:
\begin{equation}
\begin{split}
\Big(\cP_\Delta f\Big)(x,\bar x)&=\sum_{k,\bar k}\delta_{k^2,\bar k^2}\tilde f(k,\bar k)\,e^{ik\cdot x+i\bar k\cdot\bar x}=\sum_{k^2=\bar k^2}\tilde f(k,\bar k)\,e^{ik\cdot x+i\bar k\cdot\bar x}\;,\\
\Big((1-\cP_\Delta)f\Big)(x,\bar x)&=\sum_{k,\bar k}(1-\delta_{k^2,\bar k^2})\tilde f(k,\bar k)\,e^{ik\cdot x+i\bar k\cdot\bar x}=\sum_{k^2\neq\bar k^2}\tilde f(k,\bar k)\,e^{ik\cdot x+i\bar k\cdot\bar x}\;.
\end{split}    
\end{equation}
This operator clearly projects to ${\rm ker}\Delta$, since
\begin{equation}
\Delta\,\cP_\Delta=\cP_\Delta\,\Delta=0\;,    
\end{equation}
and squares to itself: $\cP_\Delta^2=\cP_\Delta$. Moreover, being a function of $\Delta$, it commutes with $M_1$, thus complying with the properties \eqref{Pprop}.
To construct the homotopy operator, we shall first introduce the ``propagator'' $G$. Given a function orthogonal to ${\rm ker}\Delta$, meaning it obeys $\cP_\Delta\,f=0$, one can invert $\Delta$ by means of the propagator $G$, defined as
\begin{equation}
\big(Gf\big) (x,\bar x)=-\sum_{k,\bar k}\frac{2}{k^2-\bar k^2}\,\tilde f(k,\bar k)\,e^{ik\cdot x+i\bar k\cdot\bar x}   \;,\quad \tilde f(k,\bar k)\equiv0\;\; \forall\;\; k^2=\bar k^2\;. 
\end{equation}
Such operator is clearly not defined on ${\rm ker}\Delta$. 
Nevertheless, the following operator relations hold  on the full space $\cX$:  
\begin{equation}
G\,\Delta=1-\cP_\Delta\;,\quad \Delta\,G\,(1-\cP_\Delta)=1-\cP_\Delta\;,     
\end{equation}
as one may quickly verify. 
Since $\Delta=[M_1,b^-]$, it is now straightforward to find the homotopy:
\begin{equation}\label{h}
h:=b^-G\,(1-\cP_\Delta)\;,    
\end{equation}
which indeed obeys the fundamental relation \eqref{homotopy relation} and the side conditions \eqref{hprop}.  
To verify this one uses that $b^-$ is nilpotent and commutes with $\cP_\Delta$ 
and $G$.

Equipped with the projector and homotopy maps we are now ready to transport the $\BV^\Delta$ structure to $\overline{\cX}$. Homotopy transfer is well-established for $L_\infty$ and $A_\infty$ algebras but, to the best of our knowledge, it has not been discussed for $\BV^\B$ or $\BV^\Delta$ algebras. We will thus proceed step by step in a constructive way. From now on, we will denote all transferred maps in $\overline{\cX}$ with an overline, which should not be confused with the second copy $\bar\cK$, since the underlying Yang-Mills maps will not play a role anymore.
Here all inputs are intended as $\overline{\Psi}_i$, living in ${\rm ker}\Delta$. Since we do not display any input, we shall denote by $\cM\rvert_{\overline{\cX}}$ the restriction of any multilinear map $\cM$ (including operators) to act on elements of  the subspace $\overline{\cX}\subset \cX$.

We begin with the differential, which is unchanged thanks  to $[M_1,\cP_\Delta]=0$:  
\begin{equation}
\overline{M}_1:=M_1\big\rvert_{\overline{\cX}}\;.    
\end{equation}
Next,  the transferred 2-product $M_2$  is simply obtained by restricting the map to the subspace 
and projecting the output back to the subspace:  
\begin{equation}
\overline{M}_2:=\cP_\Delta M_2\big\rvert_{\overline{\cX}} \;, 
\end{equation}
which obeys $\Delta \overline{M}_2=0$ by definition. 
Similarly, $\overline{B}_2$ is given by projection, and it retains its BV relation with $\overline{M}_2$:
\begin{equation}
\overline{B}_2:=\cP_\Delta\, B_2\rvert_{\overline{\cX}}=\cP_\Delta\,[b^-,M_2]\rvert_{\overline{\cX}}=[b^-,\cP_\Delta\, M_2\rvert_{\overline{\cX}}]=[b^-,\overline{M}_2]\;.    
\end{equation}
The original failure of $M_1$ to be a derivation of $B_2$ is now cured for $\overline{B}_2$:
\begin{equation}
\begin{split}
[\overline{M}_1,\overline{B}_2]&=[M_1,[b^-,\overline{M}_2]]=[\Delta,\overline{M}_2]=0 \;,   
\end{split}    
\end{equation}
where we used the fact that acting on elements of ${\rm ker}\Delta$ one has $[\Delta,\overline{M}_2]=\Delta\overline{M}_2=0$.

We move on to the first trilinear map, which requires the homotopy operator. In order to find $\overline{M}_{3\rm h}$, we compute the associator of $\overline{M}_2$ (we leave the restriction $(\cdots)\rvert_{\overline{\cX}}$ implicit):
\begin{equation}
\begin{split}
\overline{M}_2\overline{M}_2\,(1-\Pi)&=\cP_\Delta M_2\,\cP_\Delta M_2 \,(1-\Pi)=\cP_\Delta\,\Big(M_2M_2-M_2[M_1,h]M_2\Big)\,(1-\Pi) \\
&=\cP_\Delta\,\Big([M_1,M_{3\rm h}]-[M_1,M_2hM_2]\Big)\,(1-\Pi)\\
&=[\overline{M}_1,\overline{M}_{3\rm h}]\;,
\end{split}    
\end{equation}
with the transferred three-product
\begin{equation}\label{M3bar}
\overline{M}_{3\rm h}=\cP_\Delta\,\Big({M}_{3\rm h}-M_2hM_2\,(1-\Pi)\Big)\Big\rvert_{\overline{\cX}}\;.
\end{equation}
Notice that we used $[M_1,M_2]=0$ to compute
\begin{equation}
[M_1, M_2hM_2]=M_2[M_1,hM_2]=M_2[M_1,h]M_2\;.    
\end{equation}
We now define the Poisson compatibility of the transferred $\overline{M}_2$ and $\overline{B}_2$ as the one in $\cX$ (see \eqref{Input free Poisson}) by
\begin{equation}\label{Poiss transfer}
\overline{B}_2\overline{M}_2+\overline{M}_2\overline{B}_2\,(1-3\,\Pi)\equiv[b^-,\overline{M}_2\overline{M}_2]-3\,\overline{M}_2\overline{B}_2\,\Pi \;.   
\end{equation}
Since now both $\overline{M}_2$ and $\overline{B}_2$ commute with $M_1$, one immediately has that the expression \eqref{Poiss transfer} is $M_1-$closed. Its hook part is also exact, since
\begin{equation}
\begin{split}
\Big\{[b^-,\overline{M}_2\overline{M}_2]-3\,\overline{M}_2\overline{B}_2\,\Pi\Big\}\,(1-\Pi)&=[b^-,\overline{M}_2\overline{M}_2\,(1-\Pi)]=[b^-,[M_1,\overline{M}_{3\rm h}]]\\
&=-[M_1,[b^-,\overline{M}_{3\rm h}]]+[\Delta,\overline{M}_{3\rm h}]\\
&=[M_1,\overline{\Theta}_{3\rm h}]\;,
\end{split}
\end{equation}
for $\overline{\Theta}_{3\rm h}=-[b^-,\overline{M}_{3\rm h}]$. The $\Delta-$obstruction above vanishes, since $[\Delta,\overline{M}_{3\rm h}]=\Delta \overline{M}_{3\rm h}=0$ when acting on inputs in $\overline{\cX}$.
Computing the symmetric part of \eqref{Poiss transfer} is, as usual, more complicated. We obtain
\begin{equation}\label{K3s}
\begin{split}
&\Big\{[b^-,\overline{M}_{2}\overline{M}_{2}]-3\,\overline{M}_{2}\overline{B}_{2}\,\Pi\Big\}\,\Pi=\cP_\Delta\,
\Big([b^-, {M}_{2}\cP_\Delta{M}_{2}]-3\,{M}_{2}\cP_\Delta{B}_{2}\Big)\,\Pi\\
&=\cP_\Delta\,
\Big([b^-, {M}_{2}{M}_{2}]-3\,{M}_{2}{B}_{2}-\big[b^-, {M}_{2}[M_1,h]{M}_{2}\big]+3\,{M}_{2}[M_1,h]{B}_{2}\Big)\,\Pi\;.
\end{split}    
\end{equation}
Now we can use
\begin{equation}
\begin{split}
M_2[M_1,h]M_2&=[M_1,M_2hM_2]\;,\\
[M_1,hB_2]&=[M_1,h]B_2-h[M_1, B_2]=[M_1,h]B_2-h[\Delta, M_2]\;, 
\end{split}    
\end{equation}
in order to pull out some $M_1-$commutators:
\begin{equation}\label{barK3s}
\begin{split}
\eqref{K3s}&=\cP_\Delta\,\Big([b^-, {M}_{2}{M}_{2}]-3\,{M}_{2}{B}_{2}-[b^-, [M_1,{M}_{2}h{M}_{2}]]+3\,[M_1,{M}_{2}h{B}_{2}]+3\,M_2h[\Delta,M_2]\Big)\Pi\\
&=\cP_\Delta\,\Big([b^-, {M}_{2}{M}_{2}]-3\,{M}_{2}{B}_{2}+\big[M_1,[b^-, {M}_{2}h{M}_{2}]+3\,{M}_{2}h{B}_{2}\big]+3\,M_2h[\Delta,M_2]\Big)\Pi\;,
\end{split}    
\end{equation}
where we used $\cP_\Delta\,[\Delta, M_2hM_2]\rvert_{\overline{\cX}}=\cP_\Delta\,\Delta M_2hM_2\rvert_{\overline{\cX}}=0$. The last term above can be further manipulated as follows:
\begin{equation}
\begin{split}
3\,\cP_\Delta\,M_2h[\Delta,M_2]\Pi&=3\,\cP_\Delta\,M_2hM_2D_s\Pi=3\,p\,M_2hM_2\big(\Pi+1-\Pi\big)D_s\Pi\\
&=\cP_\Delta\,M_2hM_2D_\Delta \Pi+3\,\cP_\Delta\,M_2hM_2(1-\Pi)D_s\Pi\\
&=\cP_\Delta\,[\Delta,M_2hM_2]\Pi+3\,\cP_\Delta\,M_2hM_2(1-\Pi)D_s\Pi\\
&=3\,\cP_\Delta\,M_2hM_2(1-\Pi)D_s\Pi\;,
\end{split}
\end{equation}
where we used \eqref{Ddeltarel} and the vanishing of total $\Delta-$commutators under $\cP_\Delta$. We can now use the homotopy Poisson relation \eqref{Poissfinal} on $\cX$ to rewrite \eqref{barK3s} as
\begin{equation}
\begin{split}
\eqref{barK3s}&=\cP_\Delta\,\Big([M_1, \Theta_{3\rm s}]-3\,M_{3\rm h}D_s+\big[M_1,[b^-, {M}_{2}h{M}_{2}]+3\,{M}_{2}h{B}_{2}\big]+3\,M_2hM_2(1-\Pi)D_s\Big)\Pi\\
&=\Big[M_1,\cP_\Delta\,\big(\Theta_{3\rm s}+[b^-, {M}_{2}h{M}_{2}]+3\,{M}_{2}h{B}_{2}\big)\Pi \Big]-3\,\cP_\Delta\,\big(M_{3\rm h}-M_2hM_2\,(1-\Pi)\big)D_s\Pi\\
&=[\overline{M}_1, \overline{\Theta}_{3\rm s}]-3\,\overline{M}_{3\rm h}D_s\Pi\;,
\end{split}    
\end{equation}
with the transferred $\overline{\Theta}_{3}$ given by
\begin{equation}\label{Theta3transfer}
\overline{\Theta}_{3}=\cP_\Delta\,\Big(\Theta_{3}+[b^-, {M}_{2}h{M}_{2}]+3\,{M}_{2}h{B}_{2}\,\Pi\Big)\Big\rvert_{\overline{\cX}}\;.    
\end{equation}
The full homotopy Poisson relation thus reads
\begin{equation}\label{Poiss transfer full}
[b^-,\overline{M}_{2}\overline{M}_{2}]-3\,\overline{M}_{2}\overline{B}_{2}\,\Pi=[\overline{M}_1, \overline{\Theta}_{3}]-3\,\overline{M}_{3\rm h}D_s\Pi\;.    
\end{equation}
This is telling us that the ${\rm BV}_\infty^\Delta$ algebra is well-behaved under homotopy transfer, but this is not enough to completely remove the obstructions. The reason is that $\Delta$ is not zero as an operator on $\overline{\cX}$. Rather, only total $\Delta-$commutators are transferred to zero. 

The main difference of \eqref{Poiss transfer full} compared to the identity \eqref{Poissfinal} is that both sides of \eqref{Poiss transfer full} are $M_1-$closed. This is obvious for the left-hand side, since $\overline{M}_1$ commutes with both 
$\overline{M}_2$ and $\overline{B}_2$, while checking it for the right-hand side requires some computation:
\begin{equation}
\begin{split}
\Big[\overline{M}_1,\overline{M}_{3\rm h}D_s\Pi\Big]&=\big[\overline{M}_1,\overline{M}_{3\rm h}]D_s\Pi\\
&=\overline{M}_2\overline{M}_2(1-\Pi)D_s\Pi\\
&=-\tfrac13\,\overline{M}_2\overline{M}_2D_\Delta\Pi+\overline{M}_2\overline{M}_2D_s\Pi\\
&=-\tfrac13\,[\Delta, \overline{M}_2\overline{M}_2]\Pi+\overline{M}_2[\Delta, \overline{M}_2]\Pi=0\;.
\end{split}    
\end{equation}
We thus see that the $\Delta-$obstruction in \eqref{Poiss transfer full} is $M_1-$closed. If it were exact, one could shift $\overline{\Theta}_3$ to obtain a genuine homotopy Poisson identity. If this is not the case, on the other hand, $\overline{M}_{3\rm h}D_s\Pi$  would be a genuine cohomological obstruction.

For the last step, let us determine the homotopy Jacobi relation. Due to $\overline{B}_2=[b^-, \overline{M}_2]$, the Jacobiator is given by taking a $b^--$commutator of the left-hand side of \eqref{Poiss transfer full}. The computation is entirely analogous to the one leading to \eqref{JacX}, and we obtain
\begin{equation}\label{homJactransfer}
3\,\overline{B}_2\overline{B}_2\Pi+[\overline{M}_1,\overline{B}_3]+3\,\overline{\Theta}_3D_s\Pi=0\;,    
\end{equation}
including the deformation. It is interesting to note that only the hook part of $\overline{\Theta}_3$ (and thus $\overline{M}_{3\rm h}$) contributes to \eqref{homJactransfer}, since 
\begin{equation}\label{thetasno}
\overline{\Theta}_{3\rm s}D_s\Pi=\overline{\Theta}_{3}\Pi D_s\Pi=\tfrac13\,\overline{\Theta}_{3}D_\Delta\Pi=0 \;.     
\end{equation}
The transferred three-bracket is given by $\overline{B}_3=-[b^-,\overline{\Theta}_{3\rm s}]$, as dictated by ${\rm BV}_\infty^\Delta$. Upon using the expression \eqref{Theta3transfer}, one can see that $\overline{B}_3$ has also the standard form in terms of homotopy transfer of $L_\infty$ algebras:
\begin{equation}\label{B3bar}
\overline{B}_3=\cP_\Delta\,\Big(B_3-3\,B_2hB_2\,\Pi\Big)\Big\rvert_{\overline{\cX}}\;.    
\end{equation}

In the following subsection we will study the obstruction to the homotopy Jacobi identity encoded in the last term of 
\eqref{homJactransfer}  upon restricting the inputs to ${\rm ker}b^-$. Only the $L_\infty$ brackets
$\overline{B}_n$ will restrict to $\cV_{\rm DFT}$, and we will show that a well-defined albeit non-local deformation of $\overline{B}_3$ can be used to remove the obstruction.

\subsection{Restriction to ${\rm ker}\,b^-$ and three-bracket}

In the last section we have shown that the space $\overline{\cX}$ of weakly constrained fields still carries a $\BV^\Delta$ structure with milder, but non-vanishing, $\Delta$ deformations. The DFT vector space \eqref{VDFT} can be viewed, in terms of $\overline{\cX}$, as the subspace   $\cV_{\rm DFT}={\rm ker}b^-\subset\overline{\cX}$. In the following we will denote the restriction of maps $\overline{\cM}$ (which already act on inputs in $\overline{\cX}$) to act on elements in ${\rm ker}b^-$  
by $\overline{\cM}\rvert_{{\rm ker}b^-}$. Let us stress that here we are merely restricting the inputs to lie in ${\rm ker}b^-$, but no homotopy transfer is involved. While generic operators and maps of the $\BV^\Delta$ algebra on $\overline{\cX}$ are not well-defined on $\cV_{\rm DFT}$ upon restricting the inputs, the brackets of the $L_\infty$ sector are, as we will show in a moment.

From now on we will focus on the (obstructed) $L_\infty$ sector on $\overline{\cX}$, with brackets $\overline{B}_n$ given by
\begin{equation}
\begin{split}
\overline{B}_1&=\overline{M}_1\;,\\ \overline{B}_2&=[b^-,\overline{M}_2]\;,\\
\overline{B}_3&=-[b^-,\overline{\Theta}_{3\rm s}]\;,
\end{split}    
\end{equation}
obeying the following
quadratic relations:
\begin{equation}\label{LooDelta}
\begin{split}
\overline{B}_1^2&=0\;,\\
\big[\overline{B}_1,\overline{B}_2\big]&=0\;,\\
3\,\overline{B}_2\overline{B}_2\,\Pi+[\overline{B}_1,\overline{B}_3]&=-3\,\overline{\Theta}_{3\rm h}\,D_s\,\Pi\;.
\end{split}    
\end{equation}
Before studying the obstruction of the homotopy Jacobi identity, we recall that restricting the inputs to ${\rm ker}b^-$ gives us a well-defined differential and two-bracket on $\cV_{\rm DFT}$ \cite{Bonezzi:2022yuh}, which we denote by $\cB_1$ and $\cB_2$, respectively:
\begin{equation}\label{calB12}
\begin{split}
\cB_1&:=\overline{B}_1\rvert_{{\rm ker}b^-}\;,\\
\cB_2&:=\overline{B}_2\rvert_{{\rm ker}b^-}=\big[b^-,\overline{M}_2\big]\rvert_{{\rm ker}b^-}=b^-\overline{M}_2\rvert_{{\rm ker}b^-}\;.
\end{split}    
\end{equation}
As we have mentioned, $\cB_1$ and $\cB_2$ are well-defined on ${\rm ker}b^-$, since they obey $b^-\cB_i=0$. While it is trivial to see this for $\cB_2$ from \eqref{calB12}, for the differential it can be shown by computing
\begin{equation}
\begin{split}
b^-\cB_1&=b^-\overline{B}_1\rvert_{{\rm ker}b^-}=\big[b^-,\overline{B}_1\big]\rvert_{{\rm ker}b^-}\\
&=\big[b^-,M_1\big]\rvert_{{\rm ker}\Delta}\rvert_{{\rm ker}b^-}=\Delta\rvert_{{\rm ker}\Delta}\rvert_{{\rm ker}b^-}=0\;.   
\end{split}    
\end{equation}
This confirms that $\cB_1:\cV_{\rm DFT}\rightarrow\cV_{\rm DFT}$ and $\cB_2:\cV_{\rm DFT}^{\otimes 2}\rightarrow\cV_{\rm DFT}$ restrict correctly and obey nilpotency and the Leibniz relation, thus defining a consistent field theory (DFT) to cubic order. 

We now move on to the Jacobi identity of $\cB_2$, which is given by restriction of the corresponding relation \eqref{LooDelta}:
\begin{equation}\label{Looobstructed}
3\,\cB_2\cB_2\,\Pi+\big[\cB_1,\overline{B}_3\rvert_{{\rm ker}b^-}\big]=\cO\;,\quad  {\rm where} 
\qquad \cO:=-3\,\overline{\Theta}_{3\rm h}\,D_s\,\Pi\rvert_{{\rm ker}b^-}\;,   
\end{equation}
with  the obstruction denoted by $\cO$. Taking into account the restriction to ${\rm ker}b^-$, one can express the obstruction as
\begin{equation}\label{deltaJac}
\begin{split}
\cO&=-3\,\overline{\Theta}_{3\rm h}\,D_s\,\Pi\rvert_{{\rm ker}b^-}=3\,[b^-,\overline{M}_{3\rm h}]\,D_s\,\Pi\rvert_{{\rm ker}b^-}\\
&=3\,b^-\overline{M}_{3\rm h}\,D_s\,\Pi\rvert_{{\rm ker}b^-}\;,
\end{split}    
\end{equation}
where we used that $[b^-,\cT]\rvert_{{\rm ker}b^-}=b^-\cT\rvert_{{\rm ker}b^-}$. This expression can be further manipulated by using the definition \eqref{M3bar} of the transferred $\overline{M}_{3\rm h}$:
\begin{equation}
\begin{split}
\cO&=3\,b^-\overline{M}_{3\rm h}\,D_s\,\Pi\rvert_{{\rm ker}b^-}\\
&=3\,\cP_\Delta\,b^-\Big({M}_{3\rm h}-M_2hM_2\,(1-\Pi))\Big)\,D_s\,\Pi\Big\rvert_{{\rm ker}b^-\cap\,{\rm ker}\Delta}\\
&=3\,\cP_\Delta\,b^-\Big({M}_{3\rm h}\,D_s\,\Pi-M_2\Delta hM_2\,\Pi\Big)\Big\rvert_{{\rm ker}b^-\cap\,{\rm ker}\Delta}\;,
\end{split}    
\end{equation}
where we used $M_2\,hM_2\,D_s\rvert_{{\rm ker}\Delta}=M_2\,[\Delta,hM_2]\rvert_{{\rm ker}\Delta}=M_2\,\Delta h\,M_2\rvert_{{\rm ker}\Delta}$.
The homotopy \eqref{h} obeys $\Delta h=b^-(1-\cP_\Delta)$ and, for inputs in ${\rm ker}b^-$, we can also write
\begin{equation}
b^-M_2\,b^-M_2\rvert_{{\rm ker}b^-}=[b^-,M_2]\,b^-M_2\rvert_{{\rm ker}b^-}=B_2\,b^-M_2\rvert_{{\rm ker}b^-}=B_2B_2\rvert_{{\rm ker}b^-}\;,  
\end{equation}
yielding a simpler expression for the obstruction:
\begin{equation}\label{deltaJacfinal}
\cO=3\,\cP_\Delta\,\Big(b^-{M}_{3\rm h}\,D_s-B_2\,(1-\cP_\Delta)\,B_2\Big)\,\Pi\Big\rvert_{{\rm ker}b^-\cap\,{\rm ker}\Delta}\;,    
\end{equation}
which we will use to determine whether it can be removed.

First of all, the obstruction is closed: $[\cB_1,\cO]=0$, as  can be seen by taking a $\cB_1-$commutator of \eqref{Looobstructed}. However, $\cO$ is given by projection $\cP_\Delta$ of an otherwise not closed quantity:
\begin{equation}
\cO=\cP_\Delta\,\widetilde{\cO}\;,\quad\big[\cB_1,\widetilde{\cO}\big]=\Delta\,\cW\quad\Rightarrow\quad \big[\cB_1,{\cO}\big]=0\;,   
\end{equation}
with explicit $\widetilde{\cO}$ and $\cW$ given by
\begin{equation}
\begin{split}
\widetilde{\cO}&=3\,\Big(b^-{M}_{3\rm h}\,D_s-B_2\,(1-\cP_\Delta)\,B_2\Big)\Pi\Big\rvert_{{\rm ker}b^-\cap\,{\rm ker}\Delta}\;,\\ 
\cW&=\Big(3\,M_{3\rm h}\,D_s+b^-M_2M_2-3\,M_2\,(1-\cP_\Delta)\,B_2\Big)\,\Pi\Big\rvert_{{\rm ker}b^-\cap\,{\rm ker}\Delta}\;.
\end{split}
\end{equation}
Since $\widetilde{\cO}$ is not closed, it certainly cannot be exact. It is thus hard to expect that one can extract a $\cB_1-$commutator from $\cO$ in a simple way.

In order to prove that $\cO$ is, in fact, exact, we shall consider the Laplacian corresponding to Euclidean signature: 
\begin{equation}\label{Delta+}
\Delta_+:=\frac12\,\big(\del^\mu\del_\mu+\bar\del^{\bar\mu}\bar\del_{\bar\mu}\big)  \;,\quad\big[\cB_1,b^+\big]=\Delta_+\;,  
\end{equation}
which acts on the Fourier expansion \eqref{Fourier} as
\begin{equation}
\Delta_+\,f(x,\bar x)=-\frac12\sum_{k,\bar k}(k^2+\bar k^2)\,\tilde f(k,\bar k)\,e^{ik\cdot x+i\bar k\cdot\bar x}\;.    
\end{equation}
Since the metric in both $k^2$ and $\bar k^2$ is Euclidean, $\Delta_+$ is almost invertible. The only solution to $\Delta_+f(x,\bar x)=0$ is the doubled zero mode $\tilde f(0,0)$, which is allowed on the doubled torus due to its nontrivial topology. We can thus associate a zero mode projector $\cP_0$ to ${\rm ker}\Delta_+$, with a corresponding homotopy
\begin{equation}
h_0:=b^+\frac{1}{\Delta_+}\,(1-\cP_0) \;,\quad \big[\cB_1,h_0\big]=1-\cP_0\;.  
\end{equation}
At this stage, it is important to notice that nonlinear combinations of fields of the form
\begin{equation}\label{goodones}
\Big((1-\cP_\Delta)f\,\cP_\Delta g\Big)(x,\bar x)\sum_{k^2\neq\bar k^2\;,\;l^2=\bar l^2}\tilde f(k,\bar k)\,\tilde g(l,\bar l)\,e^{i(k+l)\cdot x+i(\bar k+\bar l)\cdot\bar x}\;,    
\end{equation}
do not contain zero modes. This is easily seen from the fact that in the sum above $(k^\mu, \bar k^{\bar\mu})\neq-(l^\mu, \bar l^{\bar\mu})$, given that $k^2\neq\bar k^2$, while $l^2=\bar l^2$. The total momentum above is thus $(k^\mu+l^\mu, \bar k^{\bar\mu}+\bar l^{\bar\mu})\neq(0,0)$. On such combinations one has $(1-\cP_0)=1$, yielding
\begin{equation}
(1-\cP_\Delta)f\,\cP_\Delta g=\big[\cB_1,h_0\big]\Big((1-\cP_\Delta)f\,\cP_\Delta g\Big)\;.   
\end{equation}
In order to see that we can apply this argument to our obstruction, let us act with $\cO$ in \eqref{deltaJacfinal} on three arbitrary inputs $(\Psi_1,\Psi_2,\Psi_3)$ in $\cV_{\rm DFT}$:
\begin{equation}
\begin{split}
\cO(\Psi_1,\Psi_2,\Psi_3)&\stackrel{(123)}{=}3\,\cP_\Delta\,b^-\Big({M}_{3\rm h}\big(\del^\mu\Psi_1,\del_\mu\Psi_2,\Psi_3\big)-{M}_{3\rm h}\big(\bar\del^{\bar\mu}\Psi_1,\bar\del_{\bar\mu}\Psi_2,\Psi_3\big)\Big)\\
&\hspace{5mm}-3\,\cP_\Delta\,B_2\big((1-\cP_\Delta)B_2(\Psi_1,\Psi_2),\Psi_3\big)\;,    
\end{split}    
\end{equation}
where by $(123)$ we denote graded symmetrization in the labels. The $B_2B_2$ term above has momenta of the form \eqref{goodones}, given the explicit projector $(1-\cP_\Delta)$ and 
recalling $\cP_\Delta\Psi_i=\Psi_i$. The $M_{3\rm h}$ term falls in the same category, since $\cP_\Delta$ only acts on input functions, and one has
\begin{equation}
\begin{split}
\mu\Big[D_s(F_1\otimes F_2\otimes F_3 )\Big]&=\mu\Big[(\del^\mu F_1\otimes\del_\mu F_2\otimes F_3)-(\bar\del^{\bar\mu}F_1\otimes \bar\del_{\bar\mu}F_2\otimes F_3)\Big]\\
&=\del^\mu F_1\,\del_\mu F_2\,F_3-\bar\del^{\bar\mu}F_1\, \bar\del_{\bar\mu}F_2\, F_3=\Delta(F_1F_2)\,F_3\\
&=(1-\cP_\Delta)\big[\Delta(F_1F_2)]\,\cP_\Delta F_3\;,
\end{split}
\end{equation}
for weakly constrained functions $F_i(x,\bar x)$ obeying $\Delta F_i=0$.

Having shown that $\Delta_+$ is invertible on $\cO$, we can prove that
$\cO$ is exact:
\begin{equation}
\begin{split}
\cO&=\Big[\cB_1,\frac{b^+}{\Delta_+}\Big]\cO=\Big[\cB_1,\frac{b^+}{\Delta_+}\cO\Big]+\frac{b^+}{\Delta_+}\Big[\cB_1,\cO\Big]\\
&=\Big[\cB_1,\frac{b^+}{\Delta_+}\cO\Big]\;,
\end{split}    
\end{equation}
where we used $[\cB_1,\cO]=0$.

Since we have shown  that the obstruction is exact, we can shift the original $\overline{B}_3\rvert_{{\rm ker}b^-}$ appearing in \eqref{Looobstructed} by $\frac{b^+}{\Delta_+}\cO$ and obtain a genuine $L_\infty$ relation on $\cV_{\rm DFT}$:
\begin{equation}
3\,\cB_2\cB_2\,\Pi+[\cB_1,\cB_3]=0\;,   
\end{equation}
where $\cB_1$ and $\cB_2$ are given by \eqref{calB12}, and the final three-bracket reads
\begin{equation}\label{finalB3}
\cB_3=\cP_\Delta\,\Big(B_3-3\,\frac{b^+}{\Delta_+}\big(b^-{M}_{3\rm h}\,D_s-B_2\,(1-\cP_\Delta)\,B_2\big)\,\Pi\Big)\Big\rvert_{{\rm ker}b^-\cap\,{\rm ker}\Delta}\;.
\end{equation}
Notice that the standard homotopy part $B_2hB_2$ in the definition \eqref{B3bar} of the transported $\overline{B}_3$ drops on ${\rm ker}b^-$, due to $h\propto b^-$ and ${B}_2\propto b^-$.
We have thus succeeded in constructing the three-bracket of weakly constrained DFT on a purely spatial torus. Since the whole construction is fairly abstract and intricate, in the next section we will provide an explicit check of the above results by computing the gauge algebra.

\subsection{Gauge algebra}

We now compute explicitly a consistent subsector of the gauge algebra of weakly constrained DFT, which is encoded in the homotopy Jacobi relation
\begin{equation}\label{HomJacLam}
    \text{Jac}(\Lambda_{1},\Lambda_{2},\Lambda_{2})+[\mathcal{B}_{1}, \mathcal{B}_{3}](\Lambda_{1},\Lambda_{2},\Lambda_{3})=0\; ,
\end{equation}
with the Jacobiator $\text{Jac}(\Lambda_{1},\Lambda_{2},\Lambda_{3})$ defined as
\begin{equation}
    \text{Jac}(\Lambda_{1},\Lambda_{2},\Lambda_{3}):= 3\, \mathcal{B}_{2}(\mathcal{B}_{2}(\Lambda_{[1},\Lambda_{2}),\Lambda_{3]})\stackrel{[123]}{=} 3\, \mathcal{B}_{2}(\mathcal{B}_{2}(\Lambda_{1},\Lambda_{2}),\Lambda_{3})\; .
\end{equation}
The input labels inside of the square brackets $[123]$ on top of the the last equal sign denote antisymmetrization of the labels. 
In the above equation, as in the remainder of the paper,  we employ  the convention
\begin{equation}
    3\, \mathcal{B}_{2}(\mathcal{B}_{2}(\Lambda_{[1},\Lambda_{2}),\Lambda_{3]})=\mathcal{B}_{2}(\mathcal{B}_{2}(\Lambda_{1},\Lambda_{2}),\Lambda_{3})+\mathcal{B}_{2}(\mathcal{B}_{2}(\Lambda_{2},\Lambda_{3}),\Lambda_{1})+\mathcal{B}_{2}(\mathcal{B}_{2}(\Lambda_{3},\Lambda_{1}),\Lambda_{2})\; ,
\end{equation}
where we used the antisymmetry of $\mathcal{B}_{2}$ when acting on gauge parameters: $\mathcal{B}_{2}(\Lambda_{1},\Lambda_{2})=-\mathcal{B}_{2}(\Lambda_{2},\Lambda_{1})$. 

In order to check the identity \eqref{HomJacLam}, it will be convenient to rewrite the individual terms of the Jacobiator in a different but equivalent way. One can rewrite the inner projector of the nested brackets in the Jacobiator as $\mathcal{P}_{\Delta}=1-(1-\mathcal{P}_{\Delta})$ while keeping the external projector untouched. Doing so yields
\begin{equation}
\begin{split}
    \text{Jac}(\Lambda_{1},\Lambda_{2},\Lambda_{3})&\stackrel{[123]}{=}3\, \mathcal{P}_{\Delta}\, B_{2}(B_{2}(\Lambda_{1},\Lambda_{2}),\Lambda_{3})-3\, \mathcal{P}_{\Delta}\, B_{2}(\, (1-\mathcal{P}_{\Delta})\, B_{2}(\Lambda_{1},\Lambda_{2}),\Lambda_{3})\\
    &\stackrel{[123]}{=}3\, \mathcal{P}_{\Delta}\, B_{2}(B_{2}(\Lambda_{1},\Lambda_{2}),\Lambda_{3})-3\, \mathcal{P}_{\Delta}\, B_{2}\big(\Pperp{B_{2}(\Lambda_{1},\Lambda_{2})},\Lambda_{3}\big)\; ,
\end{split}
\end{equation}
where here and in what follows it is understood that all the maps are acting on ${\rm ker}b^-\cap\,{\rm ker}\Delta$ and in order to simplify our notation we introduced the perpendicular projector $\Pperp{B_{2}}$ in the second term in the last line which denotes $(1-\mathcal{P}_{\Delta})\, B_{2}$. The above split will be useful once we compute the part of the homotopy Jacobi relation that contains the three-bracket $\mathcal{B}_{3}$ because it contains a term with $\mathcal{P}_{\Delta}\, B_{2}(1-\mathcal{P}_{\Delta})B_{2}$.

As presented in equation \eqref{gengaugeparam}, a generic gauge parameter in double field theory has three components: two vector components and one scalar component. However, in order to simplify the computation we will restrict our attention to vanishing $\bar\lambda^{\bar\nu}$ and $\eta$ while only keeping $\lambda^{\mu}$. For this reason from now on we consider the consistent subsector of the gauge algebra with parameters of the form
\begin{equation}\label{paramremain}
    \Lambda=-\theta_{\mu}\bar \theta_{+}\, \lambda^{\mu}
    \; .
\end{equation}
The homotopy Jacobi relation \eqref{HomJacLam} takes values in the space of gauge parameters, and hence consists of three components. For this reason we will check the gauge algebra explicitly displaying the basis elements of the DFT space $Z_{A}\bar Z_{\bar B}$. This will allow us to keep track of the different components of the relation.

We now turn to finding the two-bracket between gauge parameters using the technology developed in section \ref{sec:BVDel}. For gauge parameters defined as in \eqref{paramremain}, we have
\begin{equation}
\begin{split}
    \mathcal{B}_{2}(\Lambda_{1},\Lambda_{2})&=\mathcal{P}_{\Delta}\, b^{-}\, m_{2}\otimes \bar m_{2}\big(\theta_{\mu}\bar\theta_{+}\, \lambda^{\mu}_{1},\theta_{\nu}\bar\theta_{+}\, \lambda^{\nu}_{2}\big)\\
    &=\mathcal{P}_{\Delta}\, b^{-}\,\mu \Big[\hat{m}_{2}(\theta_{\mu},\theta_{\nu})\, \hat{\bar m}_{2}(\theta_{+},\theta_{+})\, (\lambda_{1}^{\mu}\otimes \lambda_{2}^{\nu}) \Big]\\
    &=\cP_{\Delta}\, b ^{-}\, \mu\Bigg\{\Big[c\,\theta_\nu\Big[\big(\del_\mu\otimes \mathds{1}\big)+2\,\big(\mathds{1}\otimes\del_\mu\big)\Big]-c\,\theta_\mu\Big[\big(\mathds{1}\otimes\del_\nu\big)+2\,\big(\del_\nu\otimes \mathds{1}\big)\Big]\\
&+c\,\theta_\rho\,\eta_{\mu\nu}\,\Big[\big(\del^\rho\otimes \mathds{1}\big)-\big(\mathds{1}\otimes\del^\rho\big)\Big]\Big]\, \bar\theta_{+}\, \big(\mathds{1}\otimes \mathds{1}\big)\, ]\big( \lambda^{\mu}_{1}\otimes \lambda^{\nu}_{2} \big)\Bigg\}\\
 &=\cP_{\Delta}\, b^{-}c\, \theta_\rho\, \bar\theta_{+} \, \Big(\del\cdot \lambda_1\,\lambda_2^\rho+2\,\lambda_1\cdot\del \lambda_2^\rho+\del^\rho \lambda_1\cdot \lambda_2-(1\leftrightarrow2)\Big)  \\
    &=\tfrac{1}{2}\,\mathcal{P}_{\Delta}\,\theta_\rho\, \bar\theta_{+} \, \Big(\del\cdot \lambda_1\,\lambda_2^\rho+2\,\lambda_1\cdot\del \lambda_2^\rho+\del^\rho \lambda_1\cdot \lambda_2-(1\leftrightarrow2)\Big)\\
    &\equiv \tfrac{1}{2}\,\mathcal{P}_{\Delta} \,\theta_\rho\bar\theta_{+}\, \big( \lambda_{1}\bullet \lambda_{2} \big)^{\rho}\; \in \mathcal{V}_{-1}\; ,
\end{split}
\end{equation}
and we used the component form of $\hat m_{2}(\theta_{\mu},\theta_{\nu})$  and $\hat{\bar m}_{2}(\bar\theta_{+},\bar \theta_{+})$, which can be found in the appendix in equation \eqref{m2hat}. Using the above expression for the two-bracket $B_{2}$ we obtain the following Jacobiator:
\begin{equation}\label{jacobo2}
\begin{split}
        \text{Jac}(\Lambda_{1},\Lambda_{2},\Lambda_{3})\stackrel{[123]}{=}-\tfrac{3}{2}\, \mathcal{P}_{\Delta}\, \theta_{\mu}\bar\theta_{+}\, &\Big[\del^{\mu}(\lambda_{1\, \rho}\, \del^{\rho}\lambda_{2\, \nu}\, \lambda^{\nu}_{3})+2\, \del_{\rho}\lambda_{1\, \nu}\, \lambda_{2}^{\nu}\, \del^{\rho}\lambda_{3}^{\mu}+\Delta_{+} \lambda_{1\, \rho}\, \lambda^{\rho}_{2}\, \lambda_{3}^{\mu}\\
       &+ 2\, \del_{\rho}\lambda_{1}^{\rho}\, \lambda_{2}^{\nu}\, \del_{\nu}\lambda_{3}^{\mu} +\del_{\rho}\lambda_{1}^{\rho}\,\del^{\mu}\lambda_{2\, \nu}\, \lambda_{3}^{\nu}+\lambda_{2}^{\nu}\, \del_{\nu}\del_{\rho}\lambda^{\rho}_{1}\, \lambda_{3}^{\mu}\Big]\\
       +\tfrac{3}{4}\, \mathcal{P}_{\Delta}\, \theta_{\mu}\bar\theta_{+}\, &\Big[\Pperp{\lambda_{1}\bullet \lambda_{2}}\bullet \lambda_{3}\Big]^{\mu}\; ,
\end{split}
\end{equation}
where we use $\mathcal{P}_{\Delta}\, \Box=\mathcal{P}_{\Delta}\, \Delta_{+}$.

Having the Jacobiator \eqref{jacobo2} at our disposal, in order to verify the homotopy Jacobi relation \eqref{HomJacLam} we need the following components of $\cB_{3}$: first, $\cB_{3}$  on three gauge parameters, whose only non-trivial part can be found with the following computation:
\begin{align}
    \cB_{3}(\Lambda_{1},\Lambda_{2},\Lambda_{3})&=\mathcal{P}_{\Delta}\, B_{3}(\Lambda_{1},\Lambda_{2},\Lambda_{2})\nonumber\\
    &=-\tfrac{1}{2}\,\mathcal{P}_{\Delta}\, b^{-}\, \theta_{3\rm s}\otimes \bar m_{2}\bar m_{2}\, \Pi(\Lambda_{1},\Lambda_{2},\Lambda_{3})\nonumber\\
    &\stackrel{[123]}{=}\tfrac{1}{2}\,\mathcal{P}_{\Delta}\, b^{-}\, \theta_{3\rm s}\otimes \bar m_{2}\bar m_{2}\, (\theta_{\mu}\bar\theta_{+}\, \lambda^{\mu}_{1},\theta_{\nu}\bar\theta_{+}\, \lambda^{\nu}_{2},\theta_{\rho}\bar\theta_{+}\, \lambda^{\rho}_{3}\nonumber\\
    &\stackrel{[123]}{=}\tfrac{1}{2}\, \mathcal{P}_{\Delta}\, b^{-}\, \mu \Big[\hat\theta_{3\rm s}(\theta_{\mu},\theta_{\nu},\theta_{\rho})\, \bar\theta_{+}\,  \big(\mathds{1}\otimes \mathds{1}\otimes \mathds{1}\big) \, \big(\lambda_{1}^{\mu}\otimes \lambda_{2}^{\nu}\otimes \lambda_{3}^{\rho}\big)\Big]\\
 &\stackrel{[123]}{=}\tfrac{1}{2}\, \cP_{\Delta}\, b^{-}\, \mu \Big\{c\,\theta_+\Big[\eta_{\mu\nu}\,\big(\del_\rho\otimes\mathds{1}\otimes\mathds{1}\big)-\eta_{\mu\nu}\,\big(\mathds{1}\otimes\del_\rho\otimes\mathds{1}\big)+\eta_{\nu\rho}\,\big(\mathds{1}\otimes\del_\mu\otimes\mathds{1}\big)\nonumber\\
&-\eta_{\nu\rho}\,\big(\mathds{1}\otimes\mathds{1}\otimes\del_\mu\big)+\eta_{\mu\rho}\,\big(\mathds{1}\otimes\mathds{1}\otimes\del_\nu\big)-\eta_{\mu\rho}\,\big(\del_\nu\otimes\mathds{1}\otimes\mathds{1}\big)\Big]\, \bar\theta_{+}\,\big(\lambda_{1}^{\mu}\otimes \lambda_{2}^{\nu}\otimes \lambda_{3}^{\rho}\big)\Big\}\nonumber\\
&\stackrel{[123]}{=}3\, \cP_{\Delta}\, b^{-}\, c\,\theta_{+}\bar\theta_{+}\, \big\{ \lambda_{1\, \rho}\, \del^{\rho}\lambda_{2\, \nu}\, \lambda_{3}^{\nu} \big\}\nonumber\\
    &\stackrel{[123]}{=}\tfrac{3}{2}\, \mathcal{P}_{\Delta}\,  \theta_{+}\bar \theta_{+}\, \big\{ \lambda_{1\, \rho}\, \del^{\rho}\lambda_{2\, \nu}\, \lambda_{3}^{\nu} \big\} \in \mathcal{V}_{-2}\nonumber \; ,
\end{align}
where we used the component form $\hat{\theta}_{3{\rm s}}(\theta_{\mu},\theta_{\nu},\theta_{\rho})$ shown in equation \eqref{thetahat3A}. 
Second, we need to find $\mathcal{B}_{3}(\Lambda_{1},\Lambda_{2},\Psi)$. From a computation analogous to the above, we find
\begin{equation}\label{B32gauge1field}
\begin{split}
\mathcal{B}_{3}(\Lambda_{1},\Lambda_{2},\Psi)\stackrel{[12]}{=}&-\tfrac{1}{2}\, \mathcal{P}_{\Delta} \, \theta_{\mu}\bar\theta_{+}\, \Big[ 2\, f_{\rho}\, \lambda_{1}^{\rho}\, \lambda_{2}^{\mu}+4e\, \lambda_{1}^{\nu}\, \del_{\nu}\lambda_{2}^{\mu}+2\lambda_{1}^{\nu}\, \del_{\nu}e\, \lambda^{\mu}_{2}+2e\, \del^{\mu}\lambda_{1\, \nu}\lambda_{2}^{\nu}\\
&-\, e^{\mu\bar\nu}\, \bar\del_{\bar\nu}\lambda_{1\, \rho}\, \lambda_{2}^{\rho}+\, \bar\del_{\bar\nu}\lambda_{1}^{\mu}\, e^{\rho\bar\nu}\, \lambda_{2\, \rho}-\big( 2\, f^{\rho}+\bar\del_{\bar\nu} e^{\rho\bar\nu} \big)\, \lambda_{1\, \rho}\, \lambda_{2}^{\mu}\Big]\\
&+\mathcal{P}_{\Delta}\, \tfrac{1}{\Delta_{+}}\,\theta_{\mu}\bar\theta_{+}\, \bar\del^{\bar\nu}\Big[ \del_{\rho}\lambda_{1}^{\mu}\, \del^{\rho}\lambda_{2}^{\nu}\, e_{\nu\bar\nu}+\lambda^{\nu}_{1}\, \del_{\rho}\lambda_{2}^{\mu}\, \del^{\rho}e_{\nu\bar\nu}+\del_{\rho}e^{\mu}{}_{\bar\nu}\, \del^{\rho}\lambda_{1\, \nu}\, \lambda_{2}^{\nu} \\
&-\bar \del_{\bar\rho}\lambda_{1}^{\mu}\, \bar\del^{\bar\rho}\lambda_{2}^{\nu}\, e_{\nu\bar\nu}-\lambda^{\nu}_{1}\, \bar\del_{\bar\rho}\lambda_{2}^{\mu}\, \bar\del^{\bar\rho}e_{\nu\bar\nu}-\bar\del_{\bar\rho}e^{\mu}{}_{\bar\nu}\, \bar\del^{\bar\rho}\lambda_{1\, \nu}\, \lambda_{2}^{\nu}\Big]\\
&-\tfrac{1}{4}\, \mathcal{P}_{\Delta}\, \tfrac{1}{\Delta_{+}}\, \theta_{\mu}\bar\theta_{+}\, \bar\del^{\bar\nu}\Big[\Pperp{\lambda_{1}\bullet \lambda_{2}}\bullet e_{\bar\nu}+2\, \lambda_{2}\bullet \Pperp{\lambda_{1}\bullet e_{\bar \nu}}\Big]^{\mu}\\
&-\tfrac{1}{2}\, \mathcal{P}_{\Delta} \,\theta_{+}\bar\theta_{\bar\nu}\, \Big[\lambda_{1}^{\rho}\, \del_{\rho}\lambda_{2\, \nu}\, e^{\nu\bar\nu}+\lambda_{2}^{\rho}\, \del_{\rho}e^{\nu\bar\nu}\, \lambda_{1\, \nu}+e^{\nu\bar\nu}\, \del_{\nu}\lambda_{1\, \rho}\, \lambda_{2}^{\rho} \Big]\\
&-\tfrac{1}{2}\,  \mathcal{P}_{\Delta} \, c^{+}\theta_{+}\bar\theta_{+}\, \bar\del_{\bar\nu}\Big[\lambda_{1}^{\rho}\, \del_{\rho}\lambda_{2\, \nu}\, e^{\nu\bar\nu}+\lambda_{2}^{\rho}\, \del_{\rho}e^{\nu\bar\nu}\, \lambda_{1\, \nu}+e^{\nu\bar\nu}\, \del_{\nu}\lambda_{1\, \rho}\, \lambda_{2}^{\rho} \Big]\; \in \mathcal{V}_{-1}\; .
\end{split}
\end{equation}
From this expression one infers by inspection of the third to fifth line that the non-locality inherent in  $\tfrac{1}{\Delta_{+}}$ is 
unavoidable: there is no overall $\Delta_+$ that can be factored out to cancel it, as $\bar\del^{\bar\nu}$ is contracted 
with $e_{\mu\bar{\nu}}$ and not with a derivative. This changes after 
replacing the field  in \eqref{B32gauge1field} by ${\cal B}_1(\Lambda)$, which is the next step in order to verify the homotopy Jacobi relation.  
For instance, in  the last line in equation \eqref{B32gauge1field} one obtains 
\begin{equation}
\begin{split}
    & \mathcal{P}_{\Delta}\, \tfrac{1}{\Delta_{+}} \bar\del^{\bar\nu}\big\{\Pperp{\lambda_{[1}\bullet \lambda_{2}}\bullet \bar\del_{\bar\nu}\lambda_{3]}+2\, \lambda_{[2}\bullet \Pperp{\lambda_{1}\bullet \bar\del_{\bar\nu}\lambda_{3]}}\big\}^{\mu} \\ 
  & \ = \    \mathcal{P}_{\Delta}\, \tfrac{1}{\Delta_{+}}\, \bar\del^{\bar\nu}\bar \del_{\bar\nu}\big\{\Pperp{\lambda_{[1}\bullet \lambda_{2}}\bullet \lambda_{3]}\big\}^{\mu}\; ,
\end{split} 
\end{equation}
where the equality follows using the Leibniz rule and  the antisymmetry of the labels. 
Under the projector $\mathcal{P}_{\Delta}$ we can then use 
the weak constraint $\bar\del_{\bar\nu}\bar\del^{\bar\nu}\equiv \bar\square=\square$ together with $\mathcal{P}_{\Delta}\B=\mathcal{P}_{\Delta}\Delta_{+}$ to cancel $\tfrac{1}{\Delta_{+}}$. Doing so for the other terms and appropriate antisymmetrizations of the inputs leads to
\begin{equation}\label{B3del}
\begin{split}
    3\, \mathcal{B}_{3}(\Lambda_{[1},\Lambda_{2},\cB_{1}(\Lambda_{3]}))\stackrel{[123]}{=}&\tfrac{3}{2}\, \mathcal{P}_{\Delta}\, \theta_{\mu}\bar\theta_{+}\Big\{ \Delta_{+} \lambda_{3\, \rho}\, \lambda_{1}^{\rho}\, \lambda_{2}^{\mu}+2\, \del_{\rho}\lambda_{3}^{\rho}\, \lambda^{\nu}_{1}\, \del_{\nu}\lambda_{2}^{\mu}+\lambda_{1}^{\nu}\, \del_{\nu}\del_{\rho}\lambda_{3}^{\rho}\, \lambda_{2}^{\mu}\\
    +&\del_{\rho}\lambda_{3}^{\rho}\, \del^{\mu}\lambda_{1\, \nu}\, \lambda_{2}^{\nu}
    +2\, \del_{\rho}\lambda_{1}^{\nu}\, \lambda_{2\, \nu}\, \del^{\rho}\lambda_{3}^{\mu}\Big\}\\
    -&\tfrac{3}{4}\, \mathcal{P}_{\Delta}\, \theta_{\mu}\bar \theta_{+}\, \big\{ \Pperp{\lambda_{1}\bullet \lambda_{2}}\bullet \lambda_{3} \big\}^{\mu}\\
    -&\tfrac{3}{2}\, \mathcal{P}_{\Delta}\,  \theta_{+}\bar\theta_{\bar\nu}\, \bar\del^{\bar\nu}\, \Big\{ \lambda_{1\, \rho}\, \del^{\rho}\lambda_{2\, \nu}\, \lambda_{3}^{\nu} \Big\}\\
    -&\tfrac{3}{2}\, \mathcal{P}_{\Delta}\, c^{+}\, \theta_{+}\bar\theta_{+}\, \Delta_{+} \Big\{ \lambda_{1\, \rho}\, \del^{\rho}\lambda_{2\, \nu}\, \lambda_{3}^{\nu} \Big\}\; ,
\end{split}
\end{equation}
which has no non-localities. 

Next, we act with the differential on $\mathcal{B}_{3}(\Lambda_{1},\Lambda_{2},\Lambda_{3})$, which yields
\begin{equation}\label{delB3}
\begin{split}
    \cB_{1}\mathcal{B}_{3}(\Lambda_{1},\Lambda_{2},\Lambda_{3})\stackrel{[123]}{=}&\tfrac{3}{2}\, \mathcal{P}_{\Delta}\, \theta_{\mu}\bar \theta_{+}\, \del^{\mu}\big( \lambda_{1\, \rho}\, \del^{\rho}\lambda_{2\, \nu}\, \lambda_{3}^{\nu} \big)\\
    +&\tfrac{3}{2}\, \mathcal{P}_{\Delta}\, \theta_{+}\bar \theta_{\bar\nu}\, \bar\del^{\bar \nu}\big( \lambda_{1\, \rho}\, \del^{\rho}\lambda_{2\, \nu}\, \lambda_{3}^{\nu} \big)\\
    +&\tfrac{3}{2}\, \mathcal{P}_{\Delta}\, c^{+}\theta_{+}\bar \theta_{+}\, \Delta_{+} \big( \lambda_{1\, \rho}\, \del^{\rho}\lambda_{2\, \nu}\, \lambda_{3}^{\nu} \big)\; .
\end{split}
\end{equation}
Finally, adding up \eqref{jacobo2}, \eqref{delB3} and \eqref{B3del} one  verifies  the 
homotopy Jacobi relation \eqref{HomJacLam}.

\section{Conclusions and Outlook}

In this paper we have explicitly constructed weakly constrained double field theory to quartic order in fields, encoded 
in the three-brackets of the corresponding $L_{\infty}$ algebra. Due to the `weak constraint' originating from the level-matching 
constraints of string theory, the construction of such a theory is a highly non-trivial problem and requires an essential 
non-locality, which  is also present in the full string theory. Specifically, the weak constraint requires that all fields are annihilated by $\Delta$, 
the second-order Laplacian  w.r.t.~the flat metric of signature $(d,d)$. It is precisely the second-order character of $\Delta$  that 
complicates the construction of an algebra of fields, since the product of two fields satisfying the weak constraint in general 
does not satisfy  the weak  constraint. Rather, the naive product has to be modified by projecting the output to the $\Delta=0$ 
subspace, an operation that singles out certain Fourier modes and is hence non-local. 
Consequently, the resulting product is non-associative.

It is relatively straightforward to solve the resulting consistency problems to cubic order \cite{Hull:2009mi,Hohm:2022pfi}, which is essentially 
due to a `kinematical accident', but to quartic and higher order it is highly non-trivial to construct a consistent field theory 
(an $L_{\infty}$ algebra). In this paper we give the corresponding $L_{\infty}$ algebra up to and including three-brackets, 
constructed via a double copy procedure from Yang-Mills theory. 
Apart from the non-locality inherent in the weak constraint we found the need for additional non-localities in the 
form of inverses of $\Delta_+$, the positive definite flat-space Laplacian, but they only show up in terms where they 
are fully well-defined on the torus. We verified that these new non-localities are inevitable given the problem we set out 
to solve: finding the three-brackets $B_3$ so that the Jacobiator relation involving $B_2B_2$ is obeyed. 
Since $B_1$ and $B_2$ are fixed from the cubic theory of 
Hull and Zwiebach in \cite{Hull:2009mi},\footnote{A small caveat is that we do not include so-called cocylce factors, 
which are claimed to be necessary in the full string theory \cite{Hata:1986mz,Maeno:1989uc,Kugo:1992md}, 
see also sec.~3.3 in \cite{Hohm:2022pfi}, 
since in the present approach they appear
unnecessary.} 
which agrees with 
our double copy  \cite{Bonezzi:2022yuh}, the only freedom  is the definition 
of $B_3$ (which is only well-defined up to cohomologically trivial contributions that drop out from $[B_1,B_3]$).
We have verified for the gauge sector that the  inverses of $\Delta_+$ are essential.

The research presented here should be generalized in many directions, which include:

\begin{itemize}

\item So far we have given  the three-brackets only in the case that all dimensions are toroidal and hence 
Euclidean, with all coordinates being doubled. It remains to include an undoubled time coordinate or, 
more generally, an arbitrary number of dimensions for the `external' or non-compact space.   
Thus, the theory presented here should  be thought of as the `internal' sector of a split (or Hamiltonian-type) 
formulation as in \cite{Hohm:2013nja,Naseer:2015fba} for double field theory (or in \cite{Hohm:2013jma,Hohm:2013pua,Hohm:2013vpa} 
for the  closely related U-duality invariant `exceptional field theory'). 
It would be interesting to see whether such split formulations can be interpreted as a tensor product 
between `internal' and `external' algebras along the lines of \cite{Greitz:2013pua,Bonezzi:2019bek}.

\item One of the potentially most important applications, and one of strongest original motivations, 
 of a weakly constrained double field theory is in  the realm of cosmology. One may imagine 
  massive string modes being excited in the very early universe that  leave  an imprint on the 
  cosmic microwave background (CMB). For instance, 
  the string gas cosmology proposal of Brandenberger-Vafa invokes the winding modes that 
  must be present if some of the spatial dimensions in cosmology are toroidal \cite{Brandenberger:1988aj}, 
  see \cite{Brandenberger:2023ver} for 
  a recent review. Generalizing the previous item, it remains to find a weakly constrained double field 
  theory on time-dependent Friedmann-Robertson-Walker backgrounds, generalizing the cubic perturbation theory 
  of \cite{Hohm:2022pfi}  to quartic and higher order.

\item 
Independent of the inclusion of non-compact dimensions, the arguably most important outstanding problem is 
to generalize the construction to higher order in fields, even just for the `internal' or compact dimensions. 
Since the quartic theory exists, it is virtually certain that the theory exists to all orders, but since the detailed 
construction is already quite involved for the three-brackets we need a more efficient formulation for the 
kinematic BV$_{\infty}^{\B}$ structures that are present in Yang-Mills theory proper in order to display the algebra 
and its double copy to all orders.\footnote{In order to describe these structures to all orders it would 
be helpful to understand BV$_{\infty}^{\B}$ algebras in terms of so-called operads. 
We thank Bruno Vallette for explaining to us how a strict BV$^{\B}$-algebra 
can indeed be described in the language of differential graded operads.} 
It is intriguing that this is a problem already in pure Yang-Mills theory, 
which thus displays a complexity comparable to that of   gravity. 

\item Our double copy procedure developed in \cite{Bonezzi:2022yuh,Bonezzi:2022bse}, 
which is based on the additional structures involving the `$b$-ghost', 
may appear rather special and only applicable to peculiar formulations, but this is not so. 
We hope to be able to illustrate this with further examples in the future and to develop the 
general theory further. Notably, in this paper we have been cavalier about the cyclic structure of 
the $L_{\infty}$ algebra, which is needed in order to write an action. Thus, the results presented here 
are strictly applicable to the equations of motion only. We leave the detailed construction of 
the cyclic $L_{\infty}$ brackets, which might differ from the ones presented here 
by cohomologically trivial 
shifts (that, however, in the language of BV,  are not symplectomorphisms) to future work.

\item The weakly constrained double field theory constructed here to quartic order is quite complicated
and non-local. While above we have emphasized that the non-localities are  inevitable given the fixed starting 
point encoded in $B_1$ and $B_2$ of the cubic theory of Hull and Zwiebach, it is conceivable that there 
are simpler versions that carry more propagating fields, which would manifest themselves already 
to quadratic order, and that are effectively integrated out in the theory encountered here. 
One may wonder if there are versions with weaker constraints or perhaps even no level-matching constraints, 
as recently explored in string field theory \cite{Erbin:2022cyb,Okawa:2022mos}.

\item 
It would be very interesting to generalize weakly constrained double field theory to theories 
including massive `M-theory states' of the kind required by U-duality invariance. 
The strongly constrained versions are known as 
exceptional field theory  (see, e.g., \cite{Hillmann:2009ci,Berman:2010is,Hohm:2013jma,Hohm:2013pua,Hohm:2013vpa}). 
One of the challenges here is that there is  no immediate analogue of the double copy construction from 
Yang-Mills theory, but one may speculate that there are exotic field theories waiting to be constructed 
who could serve as similar building blocks \cite{Anastasiou:2013hba}.

\end{itemize}

\subsection*{Acknowledgements} 

We would like to thank Alex  Arvanitakis, Chris Hull, Victor Lekeu, Bruno Vallette, Ashoke Sen, and Barton Zwiebach for discussions and collaborations on 
related topics. O.H.~and R.B.~thank Barton Zwiebach and MIT for hospitality during the final stages of this project. 
O.H.~thanks Dennis Sullivan for discussions  and the CUNY Graduate Center New York for hospitality.

This work  is supported by the European Research Council (ERC) under the European Union's Horizon 2020 research and innovation program (grant agreement No 771862). The work of R.B.~ is funded by the Deutsche Forschungsgemeinschaft (DFG, German Research Foundation)–Projektnummer 524744955. 
F.D.~is supported by the Deutsche Forschungsgemeinschaft (DFG, German Research Foundation) - Projektnummer 417533893/GRK2575 ``Rethinking Quantum Field Theory".

%\bigskip 

\appendix

\section{Yang-Mills maps and operators}\label{App:YM}

In this appendix we collect all the relevant operators associated to the maps of the $\BV^\B$ algebra of $\cK$. We start from the operators $\hat m_1, \hat m_2$ and $\hat m_{3\rm h}$   in  \eqref{mhats} corresponding to the $C_\infty$ subalgebra. The differential $m_1$ is related to $\hat m_1$ via
\begin{equation}
m_1(\psi)=\hat m_1(Z_A)\,\psi^A(x)\;,    
\end{equation}
where the complete list of operators $\hat m_1(Z_A)$ is given by
\begin{equation}
\begin{split}
\hat m_1(\theta_+)&=\theta_\mu\,\del^\mu+c\,\theta_+\,\B\;,\quad \hat m_1(c\,\theta_+)=-c\,\theta_\mu\,\del^\mu-\theta_-\;,\\
\hat m_1(\theta_\mu)&=c\,\theta_\mu\,\B+\theta_-\,\del_\mu\;,\quad\,\hat m_1(c\,\theta_\mu)=-c\,\theta_-\,\del_\mu\;,\\
\hat m_1(\theta_-)&=c\,\theta_-\,\B\;,\quad\hspace{13mm} \hat m_1(c\,\theta_-)=0\;.
\end{split}    
\end{equation}
We continue with the two-product $m_2$, which acts as
\begin{equation}
m_2(\psi_1,\psi_2)=\mu\left[\hat m_2(Z_A,Z_B)\,\Big(\psi_1^A(x)\otimes\psi_2^B(x)\Big)\right]\;, \end{equation}
and the non-vanishing bidifferential operators $\hat m_2(Z_A,Z_B)$ read
\begin{equation}\label{m2hat}
\begin{split}
\hat m_2(\theta_+,\theta_+)&=\theta_+\big(\mathds{1}\otimes\mathds{1}\big)\;,\\ 
\hat  m_2(\theta_\mu,\theta_\nu)&=c\,\theta_\nu\Big[\big(\del_\mu\otimes \mathds{1}\big)+2\,\big(\mathds{1}\otimes\del_\mu\big)\Big]-c\,\theta_\mu\Big[\big(\mathds{1}\otimes\del_\nu\big)+2\,\big(\del_\nu\otimes \mathds{1}\big)\Big]\\
&+c\,\theta_\rho\,\eta_{\mu\nu}\,\Big[\big(\del^\rho\otimes \mathds{1}\big)-\big(\mathds{1}\otimes\del^\rho\big)\Big]\;,
\end{split}    
\end{equation}
for ``diagonal'' $(Z_A,Z_B)$, while for ``off-diagonal'' ones we give both orderings explicitly:
\begin{align}
\hat m_2(\theta_\mu,\theta_+)&=\theta_\mu\big(\mathds{1}\otimes\mathds{1}\big)+c\,\theta_+\big(\del_\mu\otimes\mathds{1}+\mathds{1}\otimes\del_\mu\big)\;,&\hat m_2(\theta_+,\theta_\mu)&=\hat m_2(\theta_\mu,\theta_+)\;,\nonumber\\
\hat m_2(\theta_+,c\,\theta_\mu)&=c\,\theta_\mu\big(\mathds{1}\otimes\mathds{1}\big)\;,& \hat m_2(c\,\theta_\mu, \theta_+)&=\hat m_2(\theta_+,c\,\theta_\mu)\;,\nonumber\\
\hat m_2(\theta_+,\theta_-)&=-c\,\theta_\mu\big(\mathds{1}\otimes\del^\mu\big)\;,&\hat m_2(\theta_-,\theta_+)&=-c\,\theta_\mu\big(\del^\mu\otimes\mathds{1}\big)\;,\nonumber\\
\hat m_2(\theta_\mu,c\,\theta_\nu)&=-c\,\theta_-\eta_{\mu\nu}\big(\mathds{1}\otimes\mathds{1}\big)\;,&\hat m_2(c\,\theta_\nu,\theta_\mu)&=\hat m_2(\theta_\mu,c\,\theta_\nu)\;,\nonumber\\
\hat m_2(\theta_\mu,\theta_-)&=c\,\theta_-\big(\mathds{1}\otimes\del_\mu\big)\;,&\hat m_2(\theta_-,\theta_\mu)&=c\,\theta_-\big(\del_\mu\otimes\mathds{1}\big)\;,\nonumber\\
\hat m_2(\theta_+,c\,\theta_-)&=c\,\theta_-\big(\mathds{1}\otimes\mathds{1}\big)\;, &\hat m_2(c\,\theta_-, \theta_+)&=\hat m_2(\theta_+,c\,\theta_-)\;,
\end{align}
which enforce graded symmetry of the map $m_2$.
The $C_\infty$ maps are exhausted with the three-product
\begin{equation}
m_{3\rm h}(\psi_1,\psi_2,\psi_3)=\mu\left[\hat m_{3\rm h}(Z_A,Z_B,Z_C)\,\Big(\psi_1^A(x)\otimes\psi^B_2(x)\otimes\psi^C_3(x)\Big)\right]\;,
\end{equation}
whose only non-vanishing component is associated to the operator
\begin{equation}
\hat m_{3\rm h}(\theta_\mu,\theta_\nu,\theta_\rho)=\Big(c\,\theta_\mu\,\eta_{\nu\rho}-c\,\theta_\nu\,\eta_{\mu\rho}\Big)\,\big(\mathds{1}\otimes\mathds{1}\otimes\mathds{1}\big)\;.   
\end{equation}

Coming to the $\BV^\B$ structure, the two-bracket $b_2$ is associated to a bidifferential operator $\hat b_2$ exactly as in \eqref{m2hat}:
\begin{equation}
b_2(\psi_1,\psi_2)=\mu\left[\hat b_2(Z_A,Z_B)\,\Big(\psi_1^A(x)\otimes\psi_2^B(x)\Big)\right]\;, \end{equation}
but we do not give the explicit form of the operators $\hat b_2(Z_A,Z_B)$, since they can be straightforwardly derived from $b_2=[b,m_2]$.
The homotopy Poisson map $\theta_3$ is related to tridifferential operators $\hat\theta_3(Z_A,Z_B,Z_C)$ by
\begin{equation}
\theta_{3}(\psi_1,\psi_2,\psi_3)=\mu\left[\hat \theta_{3}(Z_A,Z_B,Z_C)\,\Big(\psi_1^A(x)\otimes\psi^B_2(x)\otimes\psi^C_3(x)\Big)\right]\;.
\end{equation}
The following operators correspond to totally graded symmetric maps:
\begin{equation}\label{theta3sym}
\begin{split}
\hat\theta_3(\theta_+,\theta_+,\theta_-)&=\theta_+\big(\mathds{1}\otimes\mathds{1}\otimes\mathds{1}\big)\;,\\
\hat\theta_3(\theta_+,\theta_+,c\,\theta_-)&=-c\,\theta_+\big(\mathds{1}\otimes\mathds{1}\otimes\mathds{1}\big)\;,\\
\hat\theta_3(\theta_+,\theta_\mu,c\,\theta_\nu)&=c\,\theta_+\eta_{\mu\nu}\big(\mathds{1}\otimes\mathds{1}\otimes\mathds{1}\big)\;,\\
\hat\theta_3(\theta_+,\theta_\mu,\theta_-)&=\theta_\mu\big(\mathds{1}\otimes\mathds{1}\otimes\mathds{1}\big)+c\,\theta_+\big(\del_\mu\otimes\mathds{1}\otimes\mathds{1}\big)\;,\\
\hat\theta_3(\theta_+,c\,\theta_+,\theta_-)&=c\,\theta_+\big(\mathds{1}\otimes\mathds{1}\otimes\mathds{1}\big)\;,\\
\hat\theta_3(\theta_+,\theta_-,\theta_-)&=\theta_-\big(\mathds{1}\otimes\mathds{1}\otimes\mathds{1}\big)\;,\\
\hat\theta_3(\theta_+,\theta_-,c\,\theta_\mu)&=c\,\theta_\mu\big(\mathds{1}\otimes\mathds{1}\otimes\mathds{1}\big)\;,\\
\hat\theta_3(\theta_+,\theta_-,c\,\theta_-)&=c\,\theta_-\big(\mathds{1}\otimes\mathds{1}\otimes\mathds{1}\big)\;,\\
\hat\theta_3(\theta_\mu,\theta_-,\theta_-)&=2\,c\,\theta_-\big(\mathds{1}\otimes\del_\mu\otimes\mathds{1}+\mathds{1}\otimes\mathds{1}\otimes\del_\mu\big)\;,\\
\hat\theta_3(\theta_\mu,c\,\theta_\nu,\theta_-)&=-c\,\theta_-\eta_{\mu\nu}\big(\mathds{1}\otimes\mathds{1}\otimes\mathds{1}\big)\;,\\
\hat\theta_3(c\,\theta_+,\theta_-,\theta_-)&=c\,\theta_-\big(\mathds{1}\otimes\mathds{1}\otimes\mathds{1}\big)\;,
\end{split}    
\end{equation}
Given the ordering of $(Z_A,Z_B,Z_C)$ above, the operator corresponding to the exchange of the first two $Z$'s, \emph{i.e.} $\hat\theta_3(Z_B,Z_A,Z_C)$, is obtained by just exchanging the first two factors in $\big(\cO_1\otimes\cO_2\otimes\cO_3\big)$, since the sign $(-1)^{Z_AZ_B}$ in all these cases is $+1$. For instance, given the above expression for $\hat\theta_3(\theta_+,\theta_\mu,\theta_-)$, one has
\begin{equation}
\hat\theta_3(\theta_\mu,\theta_+,\theta_-)=\theta_\mu\big(\mathds{1}\otimes\mathds{1}\otimes\mathds{1}\big)+c\,\theta_+\big(\mathds{1}\otimes\del_\mu\otimes\mathds{1}\big)\;.    
\end{equation}
This ensures the graded symmetry of the map $\theta_3(\psi_1,\psi_2,\psi_3)$ in the first two arguments. Since all the maps associated with \eqref{theta3sym} are totally graded symmetric, the corresponding operators obey $\hat\theta_3=\hat\theta_{3\rm s}=\widehat{\theta_3\pi}$. 
The remaining permutations of the arguments $(Z_A,Z_B,Z_C)$ can then be recovered from \eqref{Tpiproperty}. The next $\hat\theta_3$ operators have both a totally graded symmetric part $\hat\theta_{3\rm s}=\widehat{\theta_3\pi}$ and a hook part $\hat\theta_{3\rm h}=\hat\theta_3-\widehat{\theta_3\pi}$, which we give separately:
\begin{equation}\label{thetahat3A}
\begin{split}
\hat\theta_{3\rm s}(\theta_\mu,\theta_\nu,\theta_\rho)&=c\,\theta_+\Big[\eta_{\mu\nu}\,\big(\del_\rho\otimes\mathds{1}\otimes\mathds{1}\big)-\eta_{\mu\nu}\,\big(\mathds{1}\otimes\del_\rho\otimes\mathds{1}\big)+\eta_{\nu\rho}\,\big(\mathds{1}\otimes\del_\mu\otimes\mathds{1}\big)\\
&\hspace{13mm}-\eta_{\nu\rho}\,\big(\mathds{1}\otimes\mathds{1}\otimes\del_\mu\big)+\eta_{\mu\rho}\,\big(\mathds{1}\otimes\mathds{1}\otimes\del_\nu\big)-\eta_{\mu\rho}\,\big(\del_\nu\otimes\mathds{1}\otimes\mathds{1}\big)\Big]\;,\\
\hat\theta_{3\rm h}(\theta_\mu,\theta_\nu,\theta_\rho)&=\Big(\theta_\nu\,\eta_{\mu\rho}-\theta_\mu\,\eta_{\nu\rho}\Big)\,\big(\mathds{1}\otimes\mathds{1}\otimes\mathds{1}\big)\;,
\end{split}    
\end{equation}
and one can see that they are antisymmetric in the simultaneous exchange of $\mu\leftrightarrow\nu$ and $\cO_1\leftrightarrow\cO_2$ in the factors $\cO_1\otimes\cO_2\otimes\cO_3$. The last group of non-vanishing $\hat\theta_3$ also has totally graded symmetric and hook components, given by
\begin{equation}
\begin{split}
\hat\theta_{3\rm s}(\theta_\mu, \theta_\nu, c\,\theta_\rho)&=\Big(c\,\theta_\nu\,\eta_{\mu\rho}-c\,\theta_\mu\,\eta_{\nu\rho}\Big)\big(\mathds{1}\otimes\mathds{1}\otimes\mathds{1}\big)\;,\\
\hat\theta_{3\rm s}(\theta_\mu, \theta_\nu, \theta_-)&=c\,\theta_\nu\big(\mathds{1}\otimes\mathds{1}\otimes\del_\mu\big)-c\,\theta_\mu\big(\mathds{1}\otimes\mathds{1}\otimes\del_\nu\big)+2\,c\,\theta_\nu\big(\mathds{1}\otimes\del_\mu\otimes\mathds{1}\big)-2\,c\,\theta_\mu\big(\del_\nu\otimes \mathds{1}\otimes\mathds{1}\big)\\
&\hspace{5mm}+c\,\theta_\rho\,\eta_{\mu\nu}\,\Big[\big(\del^\rho\otimes \mathds{1}\otimes \mathds{1}\big)-\big(\mathds{1}\otimes\del^\rho\otimes \mathds{1}\big)\Big]\;,\\
\hat\theta_{3\rm s}(\theta_\mu, c\,\theta_+, \theta_-)&=-\hat\theta_{3\rm s}(c\,\theta_+,\theta_\mu,  \theta_-)=-c\,\theta_\mu\big(\mathds{1}\otimes \mathds{1}\otimes \mathds{1}\big)\;,\\
\hat\theta_{3\rm h}(\theta_\mu, \theta_\nu, c\,\theta_\rho)&=\Big(c\,\theta_\nu\,\eta_{\mu\rho}-c\,\theta_\mu\,\eta_{\nu\rho}\Big)\big(\mathds{1}\otimes\mathds{1}\otimes\mathds{1}\big)\;,\\
\hat\theta_{3\rm h}(c\,\theta_\rho,\theta_\mu, \theta_\nu)&=\hat\theta_{3\rm h}(\theta_\mu,c\,\theta_\rho, \theta_\nu)=\Big(c\,\theta_\mu\,\eta_{\nu\rho}-c\,\theta_\rho\,\eta_{\mu\nu}\Big)\big(\mathds{1}\otimes\mathds{1}\otimes\mathds{1}\big)\;.
\end{split}    
\end{equation}
As for the two-bracket $b_2$, we do not give the explicit form of the operators $\hat b_3$ corresponding to the three-bracket, since they can be derived from $b_3=-[b,\theta_{3\rm s}]$. 

\section{Derivation of $\Theta_3$ and $B_3$}\label{sec:Theta3}

In this appendix  we compute the symmetric projection of the Poisson relation \eqref{Input free Poisson}, in order to determine the symmetric part of the homotopy $\Theta_3$. We then use this to compute the jacobiator of the bracket $B_2$, yielding the deformed homotopy Jacobi identity.

We begin by writing the maps in terms of their Yang-Mills building blocks:
\begin{equation}
\begin{split}
&\Big\{[b^-,M_2M_2]-3\,M_2B_2\,\Pi\Big\}\Pi\\
&=\frac12\,\Big\{\big[b-\bar b,(m_2\otimes\bar m_2)\,(m_2\otimes\bar m_2)\big]-3\,(m_2\otimes\bar m_2)\, (b_2\otimes\bar m_2-m_2\otimes\bar b_2)\Big\}\,\Pi\\
&=\frac12\,\Big\{\big[b-\bar b,m_2m_2\otimes\bar m_2\bar m_2\big]-3\,\big(m_2b_2\otimes\bar m_2\bar m_2-m_2m_2\otimes\bar m_2\bar b_2\big)\Big\}\,\Pi\\
&=\frac12\,\Big\{\Big([b,m_2m_2]-3\,m_2b_2\Big)\otimes\bar m_2\bar m_2-m_2m_2\otimes\Big([\bar b,\bar m_2\bar  m_2]-3\,\bar m_2\bar b_2\Big)\Big\}\,\Pi\;.
\end{split}    
\end{equation}
We continue by substituting the Poisson relation \eqref{Poiss reminder} for $[b,m_2m_2]$ and its barred counterpart:
\begin{equation}\label{Poiss mess}
\begin{split}
&\Big\{[b^-,M_2M_2]-3\,M_2B_2\,\Pi\Big\}\Pi\\
&=\frac12\,\Big\{\Big([m_1,\theta_3]+m_{3\rm h}(d_\B-3\,d_s\,\pi)-3\,m_2b_2\,(1-\pi)\Big)\otimes\bar m_2\bar m_2\\
&\hspace{10mm}-m_2m_2\otimes\Big([\bar m_1,\bar\theta_3]+\bar m_{3\rm h}(\bar d_{\bar\B}-3\,\bar d_s\,\bar\pi)-3\,\bar m_2\bar b_2\,(1-\bar\pi)\Big)\Big\}\,\Pi\\
&=\big[M_1, \tfrac12(\theta_3\otimes\bar m_2\bar m_2-m_2m_2\otimes\bar\theta_3)\Pi\big]+\frac12\,\Big\{\Big(m_{3\rm h}(d_\B-3\,d_s\,\pi)-3\,m_2b_2\,(1-\pi)\Big)\otimes\bar m_2\bar m_2\\
&\hspace{5mm}-m_2m_2\otimes\Big(\bar m_{3\rm h}(\bar d_{\bar\B}-3\,\bar d_s\,\bar\pi)-3\,\bar m_2\bar b_2\,(1-\bar\pi)\Big)\Big\}\,\Pi\;,
\end{split}    
\end{equation}
where we have used $[m_1,m_2m_2]=0$ to extract a total differential $M_1$, which gives the first contribution to $\Theta_{3\rm s}$. For the next steps we will repeatedly use the projector relations $\pi\Pi=\bar\pi\Pi$ and $(1-\pi)\Pi=(1-\bar\pi)\Pi$. The terms involving $m_2b_2$ and $\bar m_2\bar b_2$ in \eqref{Poiss mess} can be further manipulated as
\begin{equation}\label{Poiss mess 2}
\begin{split}
&-\frac32\,\Big\{m_2b_2\,(1-\pi)\otimes\bar m_2\bar m_2-m_2m_2\otimes\bar m_2\bar b_2\,(1-\bar\pi)\Big\}\,\Pi\\
&=-\frac32\,\Big\{m_2b_2\otimes\bar m_2\bar m_2\,(1-\bar\pi)-m_2m_2\,(1-\pi)\otimes\bar m_2\bar b_2\Big\}\,\Pi\\
&=-\frac32\,\Big\{m_2b_2\otimes[\bar m_1,\bar m_{3\rm h}]-[m_1,m_{3\rm h}]\otimes\bar m_2\bar b_2\Big\}\,\Pi\\
&=\frac32\,\big[M_1,(m_2b_2\otimes\bar m_{3\rm h}+m_{3\rm h}\otimes\bar m_2\bar b_2)\Pi\big]-\frac32\,\Big\{[m_1,m_2b_2]\otimes\bar m_{3\rm h}-m_{3\rm h}\otimes[\bar m_1,\bar m_2\bar b_2]\Big\}\,\Pi\\
&=\frac32\,\big[M_1,(m_2b_2\otimes\bar m_{3\rm h}+m_{3\rm h}\otimes\bar m_2\bar b_2)\Pi\big]-\frac32\,\Big\{m_2[\B,m_2]\otimes\bar m_{3\rm h}-m_{3\rm h}\otimes\bar m_2[\bar\B,\bar m_2]\Big\}\,\Pi\;,
\end{split}    
\end{equation}
with the total $M_1-$commutator contributing to $\Theta_{3\rm s}$. Let us now consider the above terms containing $\B-$commutators, together with the $m_{3\rm h}\,d_s\,\pi$ terms from \eqref{Poiss mess}. Using the projector relations and rewriting $d_s=\frac12(d_s+\bar d_s)+\frac12(d_s-\bar d_s)$, $\bar d_s=\frac12(d_s+\bar d_s)-\frac12(d_s-\bar d_s)$, we obtain 
\begin{align}\label{Poiss mess 3}
&-\frac32\,\Big\{m_{3\rm h}\,d_s\,\pi\otimes\bar m_2\bar m_2-m_2m_2\otimes\bar m_{3\rm h}\,\bar d_s\,\bar\pi+m_2[\B,m_2]\otimes\bar m_{3\rm h}-m_{3\rm h}\otimes\bar m_2[\bar\B,\bar m_2]\Big\}\,\Pi\nonumber\\[2mm]
&=-\frac32\,\Big\{m_{3\rm h}\,d_s\,\pi\otimes\bar m_2\bar m_2-m_2m_2\otimes\bar m_{3\rm h}\,\bar d_s\,\bar\pi+m_2m_2d_s\otimes\bar m_{3\rm h}-m_{3\rm h}\otimes\bar m_2\bar m_2\bar d_s\Big\}\,\Pi\nonumber\\[2mm]
&=-\frac32\,\Big\{m_{3\rm h}\,d_s\otimes\bar m_2\bar m_2\bar\pi-m_2m_2\pi\otimes\bar m_{3\rm h}\,\bar d_s+m_2m_2d_s\otimes\bar m_{3\rm h}-m_{3\rm h}\otimes\bar m_2\bar m_2\bar d_s\Big\}\,\Pi\nonumber\\[2mm]
&=-\frac34\,\Big\{m_2m_2(1-\pi)\otimes\bar m_{3\rm h}\,\bar d_s+m_2m_2(1-\pi) d_s\otimes\bar m_{3\rm h}\nonumber\\
&\hspace{15mm}-m_{3\rm h}\otimes\bar m_2\bar m_2(1-\bar\pi)\bar d_s-m_{3\rm h}d_s\otimes\bar m_2\bar m_2(1-\bar\pi)\Big\}\,\Pi\\
&\hspace{3mm}-\frac34\,\Big\{m_{3\rm h}\,d_s\otimes\bar m_2\bar m_2\bar\pi-m_{3\rm h}\otimes\bar m_2\bar m_2\bar\pi \bar d_s+m_2m_2\pi d_s\otimes\bar m_{3\rm h}-m_2m_2\pi\otimes\bar m_{3\rm h}\,\bar d_s\nonumber\\
&\hspace{15mm}+m_2m_2d_s\otimes\bar m_{3\rm h}-m_2m_2\otimes\bar m_{3\rm h}\bar d_s+m_{3\rm h}d_s\otimes\bar m_2\bar m_2-m_{3\rm h}\otimes\bar m_2\bar m_2\bar d_s\Big\}\,\Pi\nonumber\\[2mm]
&=-\frac34\,\Big\{[m_1,m_{3\rm h}]\otimes\bar m_{3\rm h}\,(d_s+\bar d_s)-m_{3\rm h}\otimes[\bar m_1,\bar m_{3\rm h}]\,(d_s+\bar d_s)\Big\}\,\Pi\nonumber\\
&\hspace{3mm}-\frac32\,\Big\{m_{3\rm h}\otimes\bar m_2\bar m_2\bar\pi\,D_s+m_2m_2\pi \otimes\bar m_{3\rm h}\,D_s+m_2m_2\otimes\bar m_{3\rm h}\,D_s+m_{3\rm h}\otimes\bar m_2\bar m_2\,D_s\Big\}\,\Pi\nonumber\\[2mm]
&=-\frac34\,\big[M_1,(m_{3\rm h}\otimes\bar m_{3\rm h})\,(d_s+\bar d_s)\Pi\big]-\frac32\,\Big\{m_{3\rm h}\otimes\bar m_2\bar m_2\,(1+\bar\pi)+m_2m_2\,(1+\pi) \otimes\bar m_{3\rm h}\Big\}\,D_s\,\Pi\nonumber\;,  
\end{align}
with a new contribution to $\Theta_{3\rm s}$. We now combine the obstructions above, proportional to $D_s$, with the remaining $m_{3\rm h}\,d_\B$ terms from \eqref{Poiss mess}, and rewrite $d_\B=\frac12\,(d_\B+\bar d_{\bar\B})+\frac12\,(d_\B-\bar d_{\bar\B})$ and $\bar d_{\bar\B}=\frac12\,(d_\B+\bar d_{\bar\B})-\frac12\,(d_\B-\bar d_{\bar\B})$. Using the definition $D_\Delta=\frac12\,(d_\B-\bar d_{\bar\B})$, this yields
\begin{equation}\label{dboxmess}
\begin{split}
&\frac12\,\Big\{m_{3\rm h}\,d_\B\otimes\bar m_2\bar m_2-m_2m_2\otimes\bar m_{3\rm h}\,\bar d_{\bar\B}\Big\}\,\Pi\\
&-\frac32\,\Big\{m_{3\rm h}\otimes\bar m_2\bar m_2\,(1+\bar\pi)+m_2m_2\,(1+\pi) \otimes\bar m_{3\rm h}\Big\}\,D_s\,\Pi\\[2mm]
&=\frac14\,\Big\{m_{3\rm h}\,d_\B\otimes\bar m_2\bar m_2+m_{3\rm h}\otimes\bar m_2\bar m_2\,\bar d_{\bar\B}-m_2m_2\otimes\bar m_{3\rm h}\,\bar d_{\bar\B}-m_2m_2\,d_\B\otimes\bar m_{3\rm h}\Big\}\,\Pi\\
&+\frac12\,\Big\{m_{3\rm h}\otimes\bar m_2\bar m_2\,D_\Delta+m_2m_2\otimes\bar m_{3\rm h}\,D_\Delta\Big\}\,\Pi\\
&-\frac32\,\Big\{m_{3\rm h}\otimes\bar m_2\bar m_2\,(1+\bar\pi)+m_2m_2\,(1+\pi) \otimes\bar m_{3\rm h}\Big\}(1-\Pi+\Pi)\,D_s\,\Pi\\[2mm]
&=\frac14\,\Big\{m_{3\rm h}\,d_\B\otimes\bar m_2\bar m_2+m_{3\rm h}\otimes\bar m_2\bar m_2\,\bar d_{\bar\B}-m_2m_2\otimes\bar m_{3\rm h}\,\bar d_{\bar\B}-m_2m_2\,d_\B\otimes\bar m_{3\rm h}\Big\}\,\Pi\\
&+\frac12\,\Big\{m_{3\rm h}\otimes\bar m_2\bar m_2\,D_\Delta+m_2m_2\otimes\bar m_{3\rm h}\,D_\Delta\Big\}\,\Pi\\
&-\frac12\,\Big\{m_{3\rm h}\otimes\bar m_2\bar m_2\,(1+\bar\pi)+m_2m_2\,(1+\pi) \otimes\bar m_{3\rm h}\Big\}\,D_\Delta\,\Pi-3\,M_{3\rm h}\,D_s\,\Pi\;,
\end{split}    
\end{equation}
where in the last line we used \eqref{Ddeltarel} and recognized $M_{3\rm h}$ from \eqref{M3h}.
One can now use the implicit projection $m_{3\rm h}=m_{3\rm h}\,(1-\pi)$ to see that $m_{3\rm h}\otimes\bar m_2\bar m_2\,\bar\pi\,\Pi=0$ and $m_2m_2\,\pi\otimes\bar m_{3\rm h}\,\Pi=0$ 
in the last line above. From \eqref{dboxmess} we are thus left with
\begin{equation}\label{final mess}
\begin{split}
&\frac14\,\Big\{m_{3\rm h}\,d_\B\otimes\bar m_2\bar m_2+m_{3\rm h}\otimes\bar m_2\bar m_2\,\bar d_{\bar\B}-m_2m_2\otimes\bar m_{3\rm h}\,\bar d_{\bar\B}-m_2m_2\,d_\B\otimes\bar m_{3\rm h}\Big\}\,\Pi-3\,M_{3\rm h}\,D_s\,\Pi\\[2mm]
&=\frac14\,\Big[\B+\bar\B\,,m_{3\rm h}\otimes\bar m_2\bar m_2-m_2m_2\otimes\bar m_{3\rm h}\Big]\,\Pi-3\,M_{3\rm h}\,D_s\,\Pi\\[2mm]
&=\frac12\,\Big[\big[M_1,b^+\big]\,,m_{3\rm h}\otimes\bar m_2\bar m_2-m_2m_2\otimes\bar m_{3\rm h}\Big]\,\Pi-3\,M_{3\rm h}\,D_s\,\Pi\\[2mm]
&=\frac12\,\Big[M_1\,,\big[b^+,m_{3\rm h}\otimes\bar m_2\bar m_2-m_2m_2\otimes\bar m_{3\rm h}\big]\Big]\,\Pi\\
&+\frac12\,\Big[b^+\,,\big[M_1,m_{3\rm h}\otimes\bar m_2\bar m_2-m_2m_2\otimes\bar m_{3\rm h}\big]\Big]\,\Pi-3\,M_{3\rm h}\,D_s\,\Pi\;.
\end{split}    
\end{equation}
We now use $m_{3\rm h}\otimes\bar m_2\bar m_2\,\Pi=m_{3\rm h}\otimes\bar m_2\bar m_2\,(1-\bar\pi)\,\Pi$ to show that the $b^+-$commutator above vanishes:
\begin{equation}
\begin{split}
\Big(m_{3\rm h}\otimes\bar m_2\bar m_2-m_2m_2\otimes\bar m_{3\rm h}\Big)\,\Pi&=\Big(m_{3\rm h}\otimes\bar m_2\bar m_2\,(1-\bar\pi)-m_2m_2\,(1-\pi)\otimes\bar m_{3\rm h}\Big)\,\Pi\\
&=\Big(m_{3\rm h}\otimes[\bar m_1, \bar m_{3\rm h}]-[m_1, m_{3\rm h}]\otimes\bar m_{3\rm h}\Big)\,\Pi\\
&=-\big[M_1\,,m_{3\rm h}\otimes\bar m_{3\rm h}\big]\,\Pi\;,
\end{split}    
\end{equation}
thus reducing \eqref{final mess} to
\begin{equation}\label{Poiss mess 4}
\frac12\,\Big[M_1\,,\big[b^+,m_{3\rm h}\otimes\bar m_2\bar m_2-m_2m_2\otimes\bar m_{3\rm h}\big]\Big]\,\Pi-3\,M_{3\rm h}\,D_s\,\Pi\;.    
\end{equation}
Collecting all the $M_1-$commutators from \eqref{Poiss mess}, \eqref{Poiss mess 2}, \eqref{Poiss mess 3} and \eqref{Poiss mess 4}, we finally find that the symmetric projection of \eqref{Input free Poisson} obeys
\begin{equation}\label{Psym}
\Big\{[b^-,M_2M_2]-3\,M_2B_2\,\Pi\Big\}\Pi=[M_1,\Theta_{3\rm s}]-3\,M_{3\rm h}\,D_s\,\Pi\;, \end{equation}
which has the same structure as the symmetric projection of \eqref{Poiss reminder}, where the symmetric part of $\Theta_3$ is given by
\begin{equation}\label{Theta3s initial}
\begin{split}
\Theta_{3\rm s}&=\frac12\,\Big\{\theta_3\otimes\bar m_2\bar m_2-m_2m_2\otimes\bar\theta_3+3\,m_2b_2\otimes\bar m_{3\rm h}+3\,m_{3\rm h}\otimes\bar m_2\bar b_2-\tfrac32\,m_{3\rm h}\otimes\bar m_{3\rm h}\,(d_s+\bar d_s)\\
&\hspace{10mm}+\big[b^+,m_{3\rm h}\otimes\bar m_2\bar m_2-m_2m_2\otimes\bar m_{3\rm h}\big]\Big\}\,\Pi\;.    
\end{split}    
\end{equation}
The constributions to $\Theta_{3\rm s}$ involving $\theta_3$ and $\bar\theta_3$ are not separately projected in their single copy constituents. We do so by splitting $\theta_3=\theta_{3\rm s}+\theta_{3\rm h}=\theta_{3\rm s}-[b,m_{3\rm h}]$. This produces
\begin{equation}
\theta_3\otimes\bar m_2\bar m_2-m_2m_2\otimes\bar\theta_3=\theta_{3\rm s}\otimes\bar m_2\bar m_2-m_2m_2\otimes\bar\theta_{3\rm s}-[b,m_{3\rm h}\otimes\bar m_2\bar m_2]+[\bar b,m_2m_2\otimes\bar m_{3\rm h}]\;,    
\end{equation}
which, inserted in \eqref{Theta3s initial}, gives the final expression for $\Theta_{3\rm s}$:
\begin{equation}\label{Theta3s}
\begin{split}
\Theta_{3\rm s}&=\frac12\,\Big\{\theta_{3\rm s}\otimes\bar m_2\bar m_2-m_2m_2\otimes\bar\theta_{3\rm s}+3\,m_2b_2\otimes\bar m_{3\rm h}+3\,m_{3\rm h}\otimes\bar m_2\bar b_2-\tfrac32\,m_{3\rm h}\otimes\bar m_{3\rm h}\,(d_s+\bar d_s)\\
&\hspace{10mm}-\big[b^-,m_{3\rm h}\otimes\bar m_2\bar m_2+m_2m_2\otimes\bar m_{3\rm h}\big]\Big\}\,\Pi\;.    
\end{split}    
\end{equation}
Collecting the two projections \eqref{Phook} and \eqref{Psym}, we find the complete homotopy Poisson relation:
\begin{equation}\label{Poissfinal}
[b^-,M_2M_2]-3\,M_2B_2\,\Pi=\big[M_1,\Theta_3\big]+M_{3\rm h}\big(D_\Delta-3\,D_s\,\Pi\big)\;,    
\end{equation}
analogous to the single copy one \eqref{Poiss reminder}.

Given the Poisson relation, the homotopy Jacobi identity of $B_2$ is fixed by taking a $b^--$commutator of \eqref{Poissfinal}:
\begin{equation}
\begin{split}
0&=\Big[b^-,3\,M_2B_2\,\Pi-[b^-,M_2M_2]+\big[M_1,\Theta_3\big]+M_{3\rm h}\big(D_\Delta-3\,D_s\,\Pi\big)\Big] \\
&=3\,B_2B_2\,\Pi+\big[b^-,[M_1,\Theta_3]\big]+\big[b^-,M_{3\rm h}\big]\big(D_\Delta-3\,D_s\,\Pi\big)\\
&=3\,B_2B_2\,\Pi-\big[M_1,[b^-,\Theta_3]\big]+\Theta_3\,D_\Delta-\Theta_{3\rm h}\big(D_\Delta-3\,D_s\,\Pi\big)\;.
\end{split}    
\end{equation}
Upon decomposing $\Theta_3=\Theta_3\,(\Pi+1-\Pi)$ and using $3\,\Pi\,D_s\,\Pi=\Pi\,D_\Delta$, we can further manipulate the above expression as follows:
\begin{equation}\label{JacX}
\begin{split}
0&=3\,B_2B_2\,\Pi-\big[M_1,[b^-,\Theta_3]\big]+\Theta_3\,D_\Delta-\Theta_{3\rm h}\big(D_\Delta-3\,D_s\,\Pi\big)\\
&=3\,B_2B_2\,\Pi-\big[M_1,[b^-,\Theta_{3\rm s}]\big]+\Theta_3\,\Pi\,D_\Delta+3\,\Theta_{3}\,(1-\Pi)\,D_s\,\Pi\\
&=3\,B_2B_2\,\Pi+\big[M_1,B_3\big]+3\,\Theta_{3}\,D_s\,\Pi\;,
\end{split}    
\end{equation}
which is the deformed homotopy Jacobi relation. Above we have defined the three-bracket $B_3$ as
$B_3:=-[b^-,\Theta_{3}]=-[b^-,\Theta_{3\rm s}]$, where the last equality follows from $\Theta_{3\rm h}=-[b^-, M_{3\rm h}]$. This exhausts the relations of the $\BV^\Delta$ algebra up to trilinear maps.

\bibliography{WeakDFT-DC.bib}

\providecommand{\href}[2]{#2}\begingroup\raggedright\begin{thebibliography}{10}

\bibitem{Erbin:2021smf}
H.~Erbin, \href{http://dx.doi.org/10.1007/978-3-030-65321-7}{{\em {String Field
  Theory: A Modern Introduction}}}, vol.~980 of {\em Lecture Notes in Physics}.
\newblock 3, 2021.
\newblock \href{http://arxiv.org/abs/2301.01686}{{\ttfamily arXiv:2301.01686
  [hep-th]}}.

\bibitem{Doubek:2020rbg}
M.~Doubek, B.~Jur\v{c}o, M.~Markl, and I.~Sachs,
  \href{http://dx.doi.org/10.1007/978-3-030-53056-3}{{\em {Algebraic Structure
  of String Field Theory}}}, vol.~973 of {\em Lecture Notes in Physics}.
\newblock 11, 2020.

\bibitem{Hull:2009mi}
C.~Hull and B.~Zwiebach, ``{Double Field Theory},''
  \href{http://dx.doi.org/10.1088/1126-6708/2009/09/099}{{\em JHEP} {\bfseries
  09} (2009) 099}, \href{http://arxiv.org/abs/0904.4664}{{\ttfamily
  arXiv:0904.4664 [hep-th]}}.

\bibitem{Kawai:1985xq}
H.~Kawai, D.~C. Lewellen, and S.~H.~H. Tye, ``{A Relation Between Tree
  Amplitudes of Closed and Open Strings},''
  \href{http://dx.doi.org/10.1016/0550-3213(86)90362-7}{{\em Nucl. Phys. B}
  {\bfseries 269} (1986) 1--23}.

\bibitem{Bern:2008qj}
Z.~Bern, J.~J.~M. Carrasco, and H.~Johansson, ``{New Relations for Gauge-Theory
  Amplitudes},'' \href{http://dx.doi.org/10.1103/PhysRevD.78.085011}{{\em Phys.
  Rev. D} {\bfseries 78} (2008) 085011},
  \href{http://arxiv.org/abs/0805.3993}{{\ttfamily arXiv:0805.3993 [hep-ph]}}.

\bibitem{Bern:2019prr}
Z.~Bern, J.~J. Carrasco, M.~Chiodaroli, H.~Johansson, and R.~Roiban, ``{The
  Duality Between Color and Kinematics and its Applications},''
  \href{http://arxiv.org/abs/1909.01358}{{\ttfamily arXiv:1909.01358
  [hep-th]}}.

\bibitem{Siegel:1993th}
W.~Siegel, ``{Superspace duality in low-energy superstrings},''
  \href{http://dx.doi.org/10.1103/PhysRevD.48.2826}{{\em Phys. Rev. D}
  {\bfseries 48} (1993) 2826--2837},
  \href{http://arxiv.org/abs/hep-th/9305073}{{\ttfamily arXiv:hep-th/9305073}}.

\bibitem{Hohm:2010jy}
O.~Hohm, C.~Hull, and B.~Zwiebach, ``{Background independent action for double
  field theory},'' \href{http://dx.doi.org/10.1007/JHEP07(2010)016}{{\em JHEP}
  {\bfseries 07} (2010) 016}, \href{http://arxiv.org/abs/1003.5027}{{\ttfamily
  arXiv:1003.5027 [hep-th]}}.

\bibitem{Hohm:2010pp}
O.~Hohm, C.~Hull, and B.~Zwiebach, ``{Generalized metric formulation of double
  field theory},'' \href{http://dx.doi.org/10.1007/JHEP08(2010)008}{{\em JHEP}
  {\bfseries 08} (2010) 008}, \href{http://arxiv.org/abs/1006.4823}{{\ttfamily
  arXiv:1006.4823 [hep-th]}}.

\bibitem{Hohm:2011dz}
O.~Hohm, ``{On factorizations in perturbative quantum gravity},''
  \href{http://dx.doi.org/10.1007/JHEP04(2011)103}{{\em JHEP} {\bfseries 04}
  (2011) 103}, \href{http://arxiv.org/abs/1103.0032}{{\ttfamily arXiv:1103.0032
  [hep-th]}}.

\bibitem{Sen:2016qap}
A.~Sen, ``{Wilsonian Effective Action of Superstring Theory},''
  \href{http://dx.doi.org/10.1007/JHEP01(2017)108}{{\em JHEP} {\bfseries 01}
  (2017) 108}, \href{http://arxiv.org/abs/1609.00459}{{\ttfamily
  arXiv:1609.00459 [hep-th]}}.

\bibitem{Arvanitakis:2020rrk}
A.~S. Arvanitakis, O.~Hohm, C.~Hull, and V.~Lekeu, ``{Homotopy Transfer and
  Effective Field Theory I: Tree-level},''
  \href{http://arxiv.org/abs/2007.07942}{{\ttfamily arXiv:2007.07942
  [hep-th]}}.

\bibitem{Arvanitakis:2021ecw}
A.~S. Arvanitakis, O.~Hohm, C.~Hull, and V.~Lekeu, ``{Homotopy Transfer and
  Effective Field Theory II: Strings and Double Field Theory},''
  \href{http://arxiv.org/abs/2106.08343}{{\ttfamily arXiv:2106.08343
  [hep-th]}}.

\bibitem{Hohm:2013jaa}
O.~Hohm, W.~Siegel, and B.~Zwiebach, ``{Doubled $\alpha'$-geometry},''
  \href{http://dx.doi.org/10.1007/JHEP02(2014)065}{{\em JHEP} {\bfseries 02}
  (2014) 065}, \href{http://arxiv.org/abs/1306.2970}{{\ttfamily arXiv:1306.2970
  [hep-th]}}.

\bibitem{Hohm:2014xsa}
O.~Hohm and B.~Zwiebach, ``{Double field theory at order $\alpha'$},''
  \href{http://dx.doi.org/10.1007/JHEP11(2014)075}{{\em JHEP} {\bfseries 11}
  (2014) 075}, \href{http://arxiv.org/abs/1407.3803}{{\ttfamily arXiv:1407.3803
  [hep-th]}}.

\bibitem{Bonezzi:2023ced}
R.~Bonezzi, C.~Chiaffrino, F.~Diaz-Jaramillo, and O.~Hohm, ``{Weakly
  constrained double field theory: the quartic theory},''
  \href{http://arxiv.org/abs/2306.00609}{{\ttfamily arXiv:2306.00609
  [hep-th]}}.

\bibitem{Zwiebach:1992ie}
B.~Zwiebach, ``{Closed string field theory: Quantum action and the B-V master
  equation},'' \href{http://dx.doi.org/10.1016/0550-3213(93)90388-6}{{\em Nucl.
  Phys. B} {\bfseries 390} (1993) 33--152},
  \href{http://arxiv.org/abs/hep-th/9206084}{{\ttfamily arXiv:hep-th/9206084}}.

\bibitem{Zeitlin:2007vv}
A.~M. Zeitlin, ``{Homotopy Lie Superalgebra in Yang-Mills Theory},''
  \href{http://dx.doi.org/10.1088/1126-6708/2007/09/068}{{\em JHEP} {\bfseries
  09} (2007) 068}, \href{http://arxiv.org/abs/0708.1773}{{\ttfamily
  arXiv:0708.1773 [hep-th]}}.

\bibitem{Zeitlin:2008cc}
A.~M. Zeitlin, ``{Conformal Field Theory and Algebraic Structure of Gauge
  Theory},'' \href{http://dx.doi.org/10.1007/JHEP03(2010)056}{{\em JHEP}
  {\bfseries 03} (2010) 056}, \href{http://arxiv.org/abs/0812.1840}{{\ttfamily
  arXiv:0812.1840 [hep-th]}}.

\bibitem{Zeitlin:2009tj}
A.~M. Zeitlin, ``{Quasiclassical Lian-Zuckerman Homotopy Algebras, Courant
  Algebroids and Gauge Theory},''
  \href{http://dx.doi.org/10.1007/s00220-011-1206-0}{{\em Commun. Math. Phys.}
  {\bfseries 303} (2011) 331--359},
  \href{http://arxiv.org/abs/0910.3652}{{\ttfamily arXiv:0910.3652 [math.QA]}}.

\bibitem{Zeitlin:2014xma}
A.~M. Zeitlin, ``{Beltrami-Courant differentials and $G_{\infty}$-algebras},''
  \href{http://dx.doi.org/10.4310/ATMP.2015.v19.n6.a3}{{\em Adv. Theor. Math.
  Phys.} {\bfseries 19} (2015) 1249--1275},
  \href{http://arxiv.org/abs/1404.3069}{{\ttfamily arXiv:1404.3069 [math.QA]}}.

\bibitem{costellorenormalization}
K.~Costello, {\em Renormalization and Effective Field Theory}.
\newblock Mathematical surveys and monographs. American Mathematical Soc.
\newblock \url{https://books.google.de/books?id=9iM3NSy\_xCUC}.

\bibitem{Costello:2016vjw}
K.~Costello and O.~Gwilliam, {\em {Factorization Algebras in Quantum Field
  Theory}}.
\newblock Cambridge University Press, 12, 2016.

\bibitem{Hohm:2017pnh}
O.~Hohm and B.~Zwiebach, ``{$L_{\infty}$ Algebras and Field Theory},''
  \href{http://dx.doi.org/10.1002/prop.201700014}{{\em Fortsch. Phys.}
  {\bfseries 65} no.~3-4, (2017) 1700014},
  \href{http://arxiv.org/abs/1701.08824}{{\ttfamily arXiv:1701.08824
  [hep-th]}}.

\bibitem{Reiterer:2019dys}
M.~Reiterer, ``{A homotopy BV algebra for Yang-Mills and color-kinematics},''
  \href{http://arxiv.org/abs/1912.03110}{{\ttfamily arXiv:1912.03110
  [math-ph]}}.

\bibitem{Bonezzi:2022bse}
R.~Bonezzi, C.~Chiaffrino, F.~Diaz-Jaramillo, and O.~Hohm, ``{Gauge invariant
  double copy of Yang-Mills theory: The quartic theory},''
  \href{http://dx.doi.org/10.1103/PhysRevD.107.126015}{{\em Phys. Rev. D}
  {\bfseries 107} no.~12, (2023) 126015},
  \href{http://arxiv.org/abs/2212.04513}{{\ttfamily arXiv:2212.04513
  [hep-th]}}.

\bibitem{Bonezzi:2023ciu}
R.~Bonezzi, C.~Chiaffrino, F.~Diaz-Jaramillo, and O.~Hohm, ``{Gravity =
  Yang-Mills},''
\newblock 6, 2023.
\newblock \href{http://arxiv.org/abs/2306.14788}{{\ttfamily arXiv:2306.14788
  [hep-th]}}.

\bibitem{Diaz-Jaramillo:2021wtl}
F.~Diaz-Jaramillo, O.~Hohm, and J.~Plefka, ``{Double field theory as the double
  copy of Yang-Mills theory},''
  \href{http://dx.doi.org/10.1103/PhysRevD.105.045012}{{\em Phys. Rev. D}
  {\bfseries 105} no.~4, (2022) 045012},
  \href{http://arxiv.org/abs/2109.01153}{{\ttfamily arXiv:2109.01153
  [hep-th]}}.

\bibitem{Lee:2018gxc}
K.~Lee, ``{Kerr-Schild Double Field Theory and Classical Double Copy},''
  \href{http://dx.doi.org/10.1007/JHEP10(2018)027}{{\em JHEP} {\bfseries 10}
  (2018) 027}, \href{http://arxiv.org/abs/1807.08443}{{\ttfamily
  arXiv:1807.08443 [hep-th]}}.

\bibitem{Cho:2019ype}
W.~Cho and K.~Lee, ``{Heterotic Kerr-Schild Double Field Theory and Classical
  Double Copy},'' \href{http://dx.doi.org/10.1007/JHEP07(2019)030}{{\em JHEP}
  {\bfseries 07} (2019) 030}, \href{http://arxiv.org/abs/1904.11650}{{\ttfamily
  arXiv:1904.11650 [hep-th]}}.

\bibitem{Berman:2020xvs}
D.~S. Berman, K.~Kim, and K.~Lee, ``{The classical double copy for M-theory
  from a Kerr-Schild ansatz for exceptional field theory},''
  \href{http://dx.doi.org/10.1007/JHEP04(2021)071}{{\em JHEP} {\bfseries 04}
  (2021) 071}, \href{http://arxiv.org/abs/2010.08255}{{\ttfamily
  arXiv:2010.08255 [hep-th]}}.

\bibitem{Lescano:2022nhp}
E.~Lescano and S.~Roychowdhury, ``{Heterotic Kerr-Schild Double Field Theory
  and its double Yang-Mills formulation},''
  \href{http://dx.doi.org/10.1007/JHEP04(2022)090}{{\em JHEP} {\bfseries 04}
  (2022) 090}, \href{http://arxiv.org/abs/2201.09364}{{\ttfamily
  arXiv:2201.09364 [hep-th]}}.

\bibitem{Lescano:2023pai}
E.~Lescano, G.~Menezes, and J.~A. Rodr\'\i{}guez, ``{Aspects of Conformal
  Gravity and Double Field Theory from a Double Copy Map},''
  \href{http://arxiv.org/abs/2307.14538}{{\ttfamily arXiv:2307.14538
  [hep-th]}}.

\bibitem{Bonezzi:2023pox}
R.~Bonezzi, F.~Diaz-Jaramillo, and S.~Nagy, ``{Gauge Independent Kinematic
  Algebra of Self-Dual Yang-Mills},''
  \href{http://arxiv.org/abs/2306.08558}{{\ttfamily arXiv:2306.08558
  [hep-th]}}.

\bibitem{Borsten:2021hua}
L.~Borsten, H.~Kim, B.~Jur\v{c}o, T.~Macrelli, C.~Saemann, and M.~Wolf,
  ``{Double Copy from Homotopy Algebras},''
  \href{http://dx.doi.org/10.1002/prop.202100075}{{\em Fortsch. Phys.}
  {\bfseries 69} no.~8-9, (2021) 2100075},
  \href{http://arxiv.org/abs/2102.11390}{{\ttfamily arXiv:2102.11390
  [hep-th]}}.

\bibitem{Borsten:2022vtg}
L.~Borsten, B.~Jurco, H.~Kim, T.~Macrelli, C.~Saemann, and M.~Wolf,
  ``{Kinematic Lie Algebras From Twistor Spaces},''
  \href{http://arxiv.org/abs/2211.13261}{{\ttfamily arXiv:2211.13261
  [hep-th]}}.

\bibitem{Borsten:2023reb}
L.~Borsten, B.~Jurco, H.~Kim, T.~Macrelli, C.~Saemann, and M.~Wolf,
  ``{Tree-Level Color-Kinematics Duality from Pure Spinor Actions},''
  \href{http://arxiv.org/abs/2303.13596}{{\ttfamily arXiv:2303.13596
  [hep-th]}}.

\bibitem{Borsten:2023ned}
L.~Borsten, B.~Jurco, H.~Kim, T.~Macrelli, C.~Saemann, and M.~Wolf, ``{Double
  Copy from Tensor Products of Metric BV${}^{\color{gray}
  \blacksquare}$-algebras},'' \href{http://arxiv.org/abs/2307.02563}{{\ttfamily
  arXiv:2307.02563 [hep-th]}}.

\bibitem{Borsten:2023paw}
L.~Borsten, B.~Jurco, H.~Kim, T.~Macrelli, C.~Saemann, and M.~Wolf,
  ``{Double-Copying Self-Dual Yang-Mills Theory to Self-Dual Gravity on Twistor
  Space},'' \href{http://arxiv.org/abs/2307.10383}{{\ttfamily arXiv:2307.10383
  [hep-th]}}.

\bibitem{Bonezzi:2022yuh}
R.~Bonezzi, F.~Diaz-Jaramillo, and O.~Hohm, ``{The gauge structure of double
  field theory follows from Yang-Mills theory},''
  \href{http://dx.doi.org/10.1103/PhysRevD.106.026004}{{\em Phys. Rev. D}
  {\bfseries 106} no.~2, (2022) 026004},
  \href{http://arxiv.org/abs/2203.07397}{{\ttfamily arXiv:2203.07397
  [hep-th]}}.

\bibitem{Barnich:2003wj}
G.~Barnich and M.~Grigoriev, ``{Hamiltonian BRST and Batalin-Vilkovisky
  formalisms for second quantization of gauge theories},''
  \href{http://dx.doi.org/10.1007/s00220-004-1275-4}{{\em Commun. Math. Phys.}
  {\bfseries 254} (2005) 581--601},
  \href{http://arxiv.org/abs/hep-th/0310083}{{\ttfamily arXiv:hep-th/0310083}}.

\bibitem{Barnich:2004cr}
G.~Barnich, M.~Grigoriev, A.~Semikhatov, and I.~Tipunin, ``{Parent field theory
  and unfolding in BRST first-quantized terms},''
  \href{http://dx.doi.org/10.1007/s00220-005-1408-4}{{\em Commun. Math. Phys.}
  {\bfseries 260} (2005) 147--181},
  \href{http://arxiv.org/abs/hep-th/0406192}{{\ttfamily arXiv:hep-th/0406192}}.

\bibitem{Grigoriev:2023lcc}
M.~Grigoriev and D.~Rudinsky, ``{Notes on the $L_\infty$-approach to local
  gauge field theories},'' \href{http://arxiv.org/abs/2303.08990}{{\ttfamily
  arXiv:2303.08990 [hep-th]}}.

\bibitem{Ben-Shahar:2021doh}
M.~Ben-Shahar and M.~Guillen, ``{10D super-Yang-Mills scattering amplitudes
  from its pure spinor action},''
  \href{http://dx.doi.org/10.1007/JHEP12(2021)014}{{\em JHEP} {\bfseries 12}
  (2021) 014}, \href{http://arxiv.org/abs/2108.11708}{{\ttfamily
  arXiv:2108.11708 [hep-th]}}.

\bibitem{Ben-Shahar:2021zww}
M.~Ben-Shahar and H.~Johansson, ``{Off-Shell Color-Kinematics Duality for
  Chern-Simons},'' \href{http://arxiv.org/abs/2112.11452}{{\ttfamily
  arXiv:2112.11452 [hep-th]}}.

\bibitem{Hohm:2022pfi}
O.~Hohm and A.~F. Pinto, ``{Cosmological Perturbations in Double Field
  Theory},'' \href{http://dx.doi.org/10.1007/JHEP04(2023)073}{{\em JHEP}
  {\bfseries 04} (2023) 073}, \href{http://arxiv.org/abs/2207.14788}{{\ttfamily
  arXiv:2207.14788 [hep-th]}}.

\bibitem{Hata:1986mz}
H.~Hata, K.~Itoh, T.~Kugo, H.~Kunitomo, and K.~Ogawa, ``{Gauge String Field
  Theory for Torus Compactified Closed String},''
  \href{http://dx.doi.org/10.1143/PTP.77.443}{{\em Prog. Theor. Phys.}
  {\bfseries 77} (1987) 443}.

\bibitem{Maeno:1989uc}
M.~Maeno and H.~Takano, ``{Derivation of the cocycle factor of vertex in closed
  bosonic string field theory on torus},''
  \href{http://dx.doi.org/10.1143/PTP.82.829}{{\em Prog. Theor. Phys.}
  {\bfseries 82} (1989) 829}.

\bibitem{Kugo:1992md}
T.~Kugo and B.~Zwiebach, ``{Target space duality as a symmetry of string field
  theory},'' \href{http://dx.doi.org/10.1143/ptp/87.4.801}{{\em Prog. Theor.
  Phys.} {\bfseries 87} (1992) 801--860},
  \href{http://arxiv.org/abs/hep-th/9201040}{{\ttfamily arXiv:hep-th/9201040}}.

\bibitem{Hohm:2013nja}
O.~Hohm and H.~Samtleben, ``{Gauge theory of Kaluza-Klein and winding modes},''
  \href{http://dx.doi.org/10.1103/PhysRevD.88.085005}{{\em Phys. Rev. D}
  {\bfseries 88} (2013) 085005},
  \href{http://arxiv.org/abs/1307.0039}{{\ttfamily arXiv:1307.0039 [hep-th]}}.

\bibitem{Naseer:2015fba}
U.~Naseer, ``{Canonical formulation and conserved charges of double field
  theory},'' \href{http://dx.doi.org/10.1007/JHEP10(2015)158}{{\em JHEP}
  {\bfseries 10} (2015) 158}, \href{http://arxiv.org/abs/1508.00844}{{\ttfamily
  arXiv:1508.00844 [hep-th]}}.

\bibitem{Hohm:2013jma}
O.~Hohm and H.~Samtleben, ``{U-duality covariant gravity},''
  \href{http://dx.doi.org/10.1007/JHEP09(2013)080}{{\em JHEP} {\bfseries 09}
  (2013) 080}, \href{http://arxiv.org/abs/1307.0509}{{\ttfamily arXiv:1307.0509
  [hep-th]}}.

\bibitem{Hohm:2013pua}
O.~Hohm and H.~Samtleben, ``{Exceptional Form of D=11 Supergravity},''
  \href{http://dx.doi.org/10.1103/PhysRevLett.111.231601}{{\em Phys. Rev.
  Lett.} {\bfseries 111} (2013) 231601},
  \href{http://arxiv.org/abs/1308.1673}{{\ttfamily arXiv:1308.1673 [hep-th]}}.

\bibitem{Hohm:2013vpa}
O.~Hohm and H.~Samtleben, ``{Exceptional Field Theory I: $E_{6(6)}$ covariant
  Form of M-Theory and Type IIB},''
  \href{http://dx.doi.org/10.1103/PhysRevD.89.066016}{{\em Phys. Rev. D}
  {\bfseries 89} no.~6, (2014) 066016},
  \href{http://arxiv.org/abs/1312.0614}{{\ttfamily arXiv:1312.0614 [hep-th]}}.

\bibitem{Greitz:2013pua}
J.~Greitz, P.~Howe, and J.~Palmkvist, ``{The tensor hierarchy simplified},''
  \href{http://dx.doi.org/10.1088/0264-9381/31/8/087001}{{\em Class. Quant.
  Grav.} {\bfseries 31} (2014) 087001},
  \href{http://arxiv.org/abs/1308.4972}{{\ttfamily arXiv:1308.4972 [hep-th]}}.

\bibitem{Bonezzi:2019bek}
R.~Bonezzi and O.~Hohm, ``{Duality Hierarchies and Differential Graded Lie
  Algebras},'' \href{http://dx.doi.org/10.1007/s00220-021-03973-8}{{\em Commun.
  Math. Phys.} {\bfseries 382} no.~1, (2021) 277--315},
  \href{http://arxiv.org/abs/1910.10399}{{\ttfamily arXiv:1910.10399
  [hep-th]}}.

\bibitem{Brandenberger:1988aj}
R.~H. Brandenberger and C.~Vafa, ``{Superstrings in the Early Universe},''
  \href{http://dx.doi.org/10.1016/0550-3213(89)90037-0}{{\em Nucl. Phys. B}
  {\bfseries 316} (1989) 391--410}.

\bibitem{Brandenberger:2023ver}
R.~Brandenberger, ``{Superstring Cosmology -- A Complementary Review},''
  \href{http://arxiv.org/abs/2306.12458}{{\ttfamily arXiv:2306.12458
  [hep-th]}}.

\bibitem{Erbin:2022cyb}
H.~Erbin and M.~M\'edevielle, ``{Closed string theory without level-matching at
  the free level},'' \href{http://dx.doi.org/10.1007/JHEP03(2023)091}{{\em
  JHEP} {\bfseries 03} (2023) 091},
  \href{http://arxiv.org/abs/2209.05585}{{\ttfamily arXiv:2209.05585
  [hep-th]}}.

\bibitem{Okawa:2022mos}
Y.~Okawa and R.~Sakaguchi, ``{Closed string field theory without the
  level-matching condition},''
  \href{http://arxiv.org/abs/2209.06173}{{\ttfamily arXiv:2209.06173
  [hep-th]}}.

\bibitem{Hillmann:2009ci}
C.~Hillmann, ``{Generalized E(7(7)) coset dynamics and D=11 supergravity},''
  \href{http://dx.doi.org/10.1088/1126-6708/2009/03/135}{{\em JHEP} {\bfseries
  03} (2009) 135}, \href{http://arxiv.org/abs/0901.1581}{{\ttfamily
  arXiv:0901.1581 [hep-th]}}.

\bibitem{Berman:2010is}
D.~S. Berman and M.~J. Perry, ``{Generalized Geometry and M theory},''
  \href{http://dx.doi.org/10.1007/JHEP06(2011)074}{{\em JHEP} {\bfseries 06}
  (2011) 074}, \href{http://arxiv.org/abs/1008.1763}{{\ttfamily arXiv:1008.1763
  [hep-th]}}.

\bibitem{Anastasiou:2013hba}
A.~Anastasiou, L.~Borsten, M.~J. Duff, L.~J. Hughes, and S.~Nagy, ``{A magic
  pyramid of supergravities},''
  \href{http://dx.doi.org/10.1007/JHEP04(2014)178}{{\em JHEP} {\bfseries 04}
  (2014) 178}, \href{http://arxiv.org/abs/1312.6523}{{\ttfamily arXiv:1312.6523
  [hep-th]}}.

\end{thebibliography}\endgroup
\bibliographystyle{utphys}

\end{document}